\documentclass[lineno]{JFM-FLM_Au}
\usepackage{amsmath}
\usepackage{etoolbox}
\usepackage{subcaption}
\usepackage[T1]{fontenc}
\usepackage{textcomp}

\newcommand*\patchAmsMathEnvironmentForLineno[1]{%
  \expandafter\pretocmd\csname #1\endcsname{\linenomath}{}{}%
  \expandafter\pretocmd\csname end#1\endcsname{\endlinenomath}{}{}%
}
\newcommand*\patchBothAmsMathEnvironmentsForLineno[1]{%
  \patchAmsMathEnvironmentForLineno{#1}%
  \patchAmsMathEnvironmentForLineno{#1*}%
}
\AtBeginDocument{%
  \patchBothAmsMathEnvironmentsForLineno{equation}%
  \patchBothAmsMathEnvironmentsForLineno{align}%
  \patchBothAmsMathEnvironmentsForLineno{flalign}%
  \patchBothAmsMathEnvironmentsForLineno{alignat}%
  \patchBothAmsMathEnvironmentsForLineno{gather}%
  \patchBothAmsMathEnvironmentsForLineno{multline}%
}

\newlength{\panelheight}
\AtBeginDocument{\setlength{\panelheight}{0.38\textwidth}}

\newcommand{\panel}[3][1]{%
  \begin{subfigure}[b]{0.49\textwidth}\centering
    \includegraphics[height=#1\panelheight]{#3}%
    \caption{}\label{#2}%
  \end{subfigure}}

\newcommand{\bp}{\boldsymbol{p}}
\newcommand{\br}{\boldsymbol{r}}
\newcommand{\bx}{\boldsymbol{x}}
\newcommand{\bg}{\hat{\boldsymbol{g}}}

\newcommand{\bnu}{\hat{\boldsymbol{n}}}
\newcommand{\dU}{\Delta\boldsymbol{U}}
\newcommand{\tI}{\mathsfbi{I}}
\newcommand{\tJ}{\mathsfbi{J}}
\newcommand{\tD}{\mathsfbi{D}}
\newcommand{\Gpa}{G_\parallel}
\newcommand{\Gpe}{G_\perp}
\newcommand{\bpa}{\beta_\parallel}
\newcommand{\bpe}{\beta_\perp}
\newcommand{\lpe}{\lambda_\perp^{*}}
\newcommand{\lpa}{\lambda_\parallel^{*}}
\newcommand{\Kz}{K^{(0)}}
\newcommand{\Kb}{K^{(0)}_{b}}
\newcommand{\Kbar}{\bar{K}}
\newcommand{\Iint}[2]{\mathcal{I}_{#1,#2}}
\newcommand{\avg}[1]{\left\langle #1 \right\rangle}
\newcommand{\Rr}{\mathcal{R}}
\newcommand{\Wf}{\boldsymbol{\mathcal{W}}}
\newcommand{\dd}{\mathrm{d}}

\newcommand{\XX}[1]{\ifmmode\mathbf{[#1]}\else\textbf{[#1]}\fi}

\lefttitle{A. Choudhury and A. Roy}
\righttitle{Journal of Fluid Mechanics}

\title{Collision of a sphere and a prolate spheroid settling in a quiescent ambient}

\author{Arnab Choudhury\aff{1} \and Anubhab Roy\aff{1}}

\affiliation{\aff{1}Department of Applied Mechanics and Biomedical Engineering, Indian Institute of Technology Madras, Chennai, Tamil Nadu 600036, India}

\corresau{Anubhab Roy, anubhab@smail.iitm.ac.in}

\begin{document}
\nolinenumbers
\maketitle

\begin{abstract}
The collision rate of a prolate spheroid and a sphere settling through a quiescent fluid at small Reynolds number is computed for randomly oriented spheroids and for spheroids turned broadside by weak fluid inertia. Without hydrodynamic interactions, the collision cross-section is the projection of the spheroid enlarged by the radius of the sphere. The ideal kernel then follows in closed form for aligned spheroids and from a one-dimensional quadrature for isotropic ones. Alignment raises the ideal kernel by at most a factor of $4/\upi$, but extinguishes it on the extinction locus, where an isolated broadside spheroid settles as fast as the sphere. Hydrodynamic interactions are computed from pair trajectories in a first-reflection slender-body model, in which the interaction velocity is proportional to the difference of the buoyant weights and only the flow of the sphere rotates the spheroid. A strongly aligned spheroid weathervanes, whereas a freely rotating one feathers and glides. Weathervaning lowers the collision efficiency, as does feathering far from the extinction locus, but gliding raises it near the locus. There, the collision rate of spheroids aligned at a finite rate lies far above both the strong-alignment result and the ideal isotropic kernel. Just above the extinction locus, where strongly aligned spheroids are blocked, collisions persist at all the finite alignment rates computed. A boundary-element solution shows that the model overestimates the rate at which the gap closes near contact. A lubrication film largely corrects this and reduces the collision efficiencies from their interception values. Simple fits for the ideal kernels, the efficiencies and the near-contact correction are provided for use in collision models.
\end{abstract}

\section{Introduction}\label{sec:intro}
Particles that settle at different speeds collide, and these collisions control the growth of aggregates in sedimenting suspensions, in the ocean and in the atmosphere. In mixed-phase clouds, ice crystals collide with supercooled droplets, which freeze on contact. This process, riming, turns crystals into graupel and helps to start precipitation \citep{PruppacherKlett1997,LambVerlinde2011}. For the collision of spheres the theory is mature. The ideal kernel of \citet{Smoluchowski1917} assumes that each particle moves with its isolated settling velocity. Hydrodynamic interactions lower this rate, because each particle is carried by the flow disturbance of the other \citep{Batchelor1982,BatchelorWen1982,Davis1984}. Near contact, the breakdown of continuum lubrication in a gas and van der Waals attraction decide the outcome \citep{Sundararajakumar1996,Dhanasekaran2021a,Patra2022}. Electric charge \citep{Patra2023,Warrier2026} and turbulence \citep{SaffmanTurner1956,GrabowskiWang2013,Dhanasekaran2021b,Patra2026} also change the collision rate.

Ice crystals are rarely spherical. Columns and needles are close to prolate spheroids, and plates to oblate ones \citep{PruppacherKlett1997}. The collision rate of such a particle depends on its orientation, through both its settling velocity and its cross-section, and the orientation in turn depends on the flow. Two routes have been followed. The first computes the flow around a crystal of fixed orientation and integrates droplet trajectories through it \citep{Schlamp1975,WangJi2000}. The second neglects hydrodynamic interactions and computes geometric kernels, in turbulence \citep{Siewert2014,Jucha2018,Naso2018,Sheikh2022,Sheikh2024} or in a quiescent fluid \citep{SlomkaStocker2020a,SlomkaStocker2020b,GavzeKhain2022,Jiang2024}. \citet{GavzeKhain2022} built the collecting area of two spheroids by moving the projection of one around the projection of the other, on the plane normal to their relative velocity, and computed swept volumes for many shapes and orientations. Identical spheroids also settle relative to one another when their orientations differ, because the settling velocity depends on orientation. For two identical slender fibres, \citet{KochShaqfeh1989} showed that the relative settling velocity is proportional to $(\bp_1\bp_1-\bp_2\bp_2)\bcdot\bg$, the difference between the orientation dyads of the two fibres contracted with gravity, where $\bp_1$ and $\bp_2$ are their orientation vectors, and $\bg$ is the unit vector along the gravity. It therefore vanishes only when the two fibres are parallel or antiparallel, or when both lie broadside to gravity. \citet{SlomkaStocker2020a} computed the resulting collision rate of identical rods. For crystals and droplets in turbulence, \citet{Naso2018} noted that the effect of hydrodynamic interactions is difficult to estimate from first principles, and \citet{Sheikh2022} left the collisions of elongated crystals as an open question.

Hydrodynamic interactions between spheroids and spheres have been studied mainly in suspension mechanics. \citet{Kim1985} computed the sedimentation of two spheroids by the method of reflections, and \citet{ClaeysBrady1993} built Stokesian dynamics for prolate spheroids. \citet{KochShaqfeh1989} showed that the disturbance of a settling fibre rotates a neighbour at a rate that decays as the inverse square of the distance, and \citet{ShaqfehKoch1988,ShaqfehKoch1990} described how repeated encounters disperse the orientations of fibres. Closest to the present problem, \citet{Harlen1999} simulated a sphere settling through a suspension of fibres, using slender-body theory for the fibres and Fax\'en's law for the sphere. They also pointed out that the lubrication force between a sphere and a thin fibre is much weaker than that between two spheres, so that the gap closes quickly and the two make contact. None of these studies computed a collision rate.

Orientation is the second ingredient. A settling spheroid keeps its orientation in Stokes flow, but weak fluid inertia turns a prolate spheroid broadside to gravity (\S\,\ref{sec:orientation}). This is seen for falling cylinders and fibres \citep{JayaweeraMason1965,Roy2019} and helps to explain why ice crystals in clouds fall with their long axis horizontal \citep{Klett1995}. In turbulence the same torque competes with the velocity gradients \citep{Gustavsson2019,Sheikh2020,Roy2026orientation}. An isotropic population and a fully broadside one are therefore the two limiting cases that a collision theory must treat.

In this paper we study the simplest problem with both ingredients: a rigid prolate spheroid and a rigid sphere that settle through a quiescent fluid at small Reynolds number. We make four contributions. First, we obtain the ideal kernel geometrically. Without interactions, the collision cross-section is the projection of the spheroid, enlarged by the radius of the sphere, on the plane normal to their relative velocity, which is an ellipse enlarged by a disc. For a sphere and a spheroid this gives the collecting area of \citet{GavzeKhain2022} in closed form. The kernel of an aligned population then follows in closed form, and that of an isotropic population as a one-dimensional quadrature. We find a locus in the plane of aspect ratio and size ratio on which the aligned collision rate vanishes, a closed-form floor of the isotropic collision rate on that locus, and a bound of $4/\upi$ on the gain from alignment. Second, we compute collision efficiencies with hydrodynamic interactions from pair trajectories, using a first-reflection slender-body model of the kind used by \citet{Harlen1999} and letting the spheroid rotate. Third, we add the inertial torque at finite alignment rates and explain the results with a simple model of forced relaxation. Near the extinction locus we find that slow alignment raises the collision rate far above the strong-alignment result and the ideal isotropic kernel. Fourth, we test the model against an exact first reflection and a boundary-element solution, add a lubrication film and a capture gap, and give simple formulas and tables for use in collision models.

The paper is organised as follows. In \S\,\ref{sec:formulation} the problem is formulated, including the orientation distributions of the two populations considered and the slender-body description of the hydrodynamic interactions. In \S\,\ref{sec:ideal} the ideal collision rate, in which interactions are neglected, is derived in closed form for both populations, and the extinction locus, on which the relative velocity of an aligned spheroid and the sphere vanishes, is identified. In \S\,\ref{sec:interacting} the collision rate with hydrodynamic interactions is computed by integrating the pair trajectories, both away from and near the extinction locus, for a finite alignment rate, and with an assessment of the accuracy of the model and of the effect of the gap between the surfaces. Section~\ref{sec:practical} collects fitted formulas for practical use, together with a dimensional example, and \S\,\ref{sec:conclusions} summarises the main results.

\section{Problem formulation}\label{sec:formulation}

\subsection{The pair and its isolated settling}\label{sec:isolated}

\begin{figure}
\vspace{5mm}
  \centerline{\includegraphics[width=0.85\textwidth]{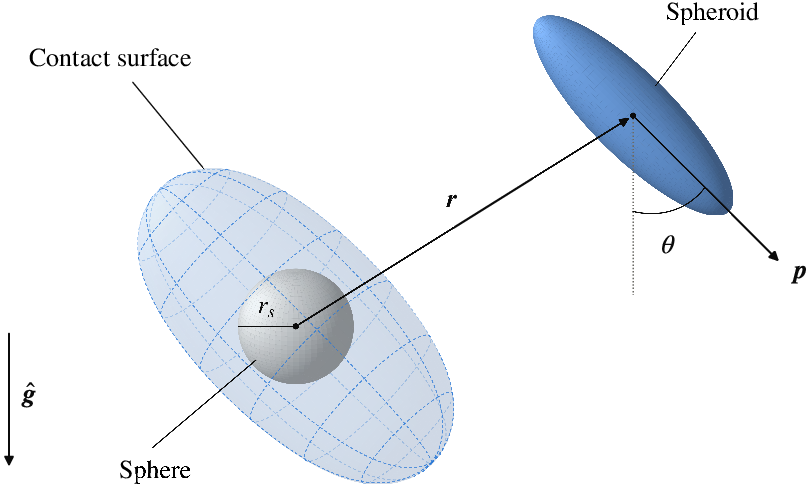}}
  \caption{Schematic diagram of a sphere settling along with a spheroid in a quiescent ambient. The axis $\bp$ of the spheroid makes the angle $\theta$ with gravity $\bg$, and the separation $\br$ points from the centre of the sphere (radius $r_s$) to the centroid of the spheroid. The translucent surface is the contact surface for the orientation shown: the positions of the centroid at which the two particles touch.}
\label{fig:pair}
\end{figure}

A dilute suspension of rigid prolate spheroids and rigid spheres settles under gravity through a quiescent fluid of density $\rho_f$, dynamic viscosity $\mu_f$ and kinematic viscosity $\nu=\mu_f/\rho_f$ (figure~\ref{fig:pair}). The spheroid has length $L$, symmetry semi-axis $c=L/2$, transverse semi-axis $a=L/(2\kappa)$ with aspect ratio $\kappa=c/a>1$, and its eccentricity is $e=(1-\kappa^{-2})^{1/2}$. Its axis points along the unit vector $\bp$. The sphere has radius $r_s$, and the size ratio is $\lambda=c/r_s=L/(2r_s)$, the ratio of the half-length of the spheroid to the radius of the sphere, so that $\lambda$ reduces to the ratio of the two radii in the spherical limit $\kappa\to1$. The spheroid and the sphere have densities $\rho_c$ and $\rho_s$, respectively, both larger than $\rho_f$, and we write $\gamma=(\rho_s-\rho_f)/(\rho_c-\rho_f)$. Gravity acts along the unit vector $\bg$, $\theta$ is the angle between $\bp$ and $\bg$, and $\mu=\bp\bcdot\bg=\cos\theta$. The number densities of spheroids and spheres are $n_1$ and $n_2$, respectively. The suspension is assumed sufficiently dilute that encounters involving three or more particles are rare, and the analysis is restricted to binary interactions.

Three effects determine the range of validity of the present model: the inertia of the particles, their Brownian diffusion, and the inertia of the surrounding fluid. The first two are neglected throughout, as justified below. The third, when present, is taken to be weak, and its effect on the orientation dynamics of the spheroid is discussed separately in \S\,\ref{sec:orientation}. The inertia of each particle is measured by a Stokes number, the ratio of its viscous relaxation time to the time in which it settles through its own size. For the spheroid, $St_c=\tau_cU_\perp/L$, where $U_\perp$ is the broadside settling speed of the spheroid, and $\tau_c=2\rho_ca^{2}/(9\mu_f X)$ is its relaxation time for translation along its axis, with $X$ the axial resistance function defined in (\ref{eq:XY}). For the sphere, $St_s=\tau_sU_s/(2r_s)$, where $U_s$ is the settling speed of the sphere and $\tau_s=2\rho_sr_s^{2}/(9\mu_f)$ is its relaxation time. Both Stokes numbers are taken to be small, $St_c\ll1$ and $St_s\ll1$, so that the inertia of the particles is negligible. The intensity of Brownian motion is measured in the same way by two P\'eclet numbers, each the ratio of the time in which a particle diffuses through its own size to the time in which it settles through it: $Pe_c=U_\perp L/D_c$ for the spheroid and $Pe_s=2U_sr_s/D_s$ for the sphere. Here $D_c=k_BT/(6\upi\mu_f cX)$ is the translational diffusivity of the spheroid along its axis, $D_s=k_BT/(6\upi\mu_f r_s)$ is that of the sphere, $k_B$ is the Boltzmann constant and $T$ is the absolute temperature. Both P\'eclet numbers are taken to be large, so that Brownian motion is negligible.

In the subsequent formulation, the lengths are non-dimensionalised by the length of the spheroid, $L$, velocities by
\begin{equation}
U^{*}=\frac{(\rho_c-\rho_f)\,g\,L^{2}}{18\mu_f},
\label{eq:Ustar}
\end{equation}
the Stokes settling speed of a sphere of diameter $L$, and time by $L/U^{*}$. In these units $c=1/2$, $a=1/(2\kappa)$ and $r_s=1/(2\lambda)$. The drag on a prolate spheroid translating with velocity $\boldsymbol{U}$ through a quiescent fluid is $-6\upi\mu_f c\,[X\bp\bp+Y(\tI-\bp\bp)]\bcdot\boldsymbol{U}$, where $\tI$ denotes the identity tensor, and the axial and transverse resistance functions, normalised by the Stokes drag on a sphere of radius $c$, are \citep{KimKarrila1991}
\begin{equation}
X=\frac{8e^{3}}{3\left[(1+e^{2})L_e-2e\right]},\qquad
Y=\frac{16e^{3}}{3\left[(3e^{2}-1)L_e+2e\right]}.
\label{eq:XY}
\end{equation}
Here, $L_e=\ln[(1+e)/(1-e)]$. Gravity exerts no torque on the spheroid, so its isolated settling velocity is
\begin{equation}
\boldsymbol{U}_c=\Gpe\,\bg+(\Gpa-\Gpe)\,\mu\,\bp,
\label{eq:Uc}
\end{equation}
where $\Gpa=1/(\kappa^{2}X)$ and $\Gpe=1/(\kappa^{2}Y)$ are its settling speeds when the symmetry axis of the spheroid is kept vertical and horizontal (broadside), respectively. Since $X<Y$, a prolate spheroid falls fastest when vertical, and a tilted one also drifts sideways. For a slender spheroid ($\kappa\gg1$),
\begin{equation}
\Gpa\simeq\frac{3}{2\kappa^{2}}\left(\ln2\kappa-\frac12\right),\qquad
\Gpe\simeq\frac{3}{4\kappa^{2}}\left(\ln2\kappa+\frac12\right).
\label{eq:Gslender}
\end{equation}
The sphere settles with $\boldsymbol{U}_s=S\,\bg$, where $S=\gamma/\lambda^{2}$. The relative velocity of the spheroid with respect to the sphere is
\begin{equation}
\dU(\bp)=\boldsymbol{U}_c-\boldsymbol{U}_s=\bpe\,\bg+(\Gpa-\Gpe)\,\mu\,\bp,
\label{eq:dU}
\end{equation}
where $\bpe=\Gpe-S$ is the speed mismatch between a broadside oriented spheroid and a sphere. Similarly, the speed mismatch between a vertical spheroid and a sphere is denoted by $\bpa = \Gpa-S$. The components of $\dU$ along and across the symmetry axis of the spheroid $\boldsymbol{p}$ are $\bpa\mu\,\bp$ and $\bpe(\bg-\mu\bp)$, so that
\begin{equation}
|\dU|^{2}=\bpa^{2}\mu^{2}+\bpe^{2}(1-\mu^{2}).
\label{eq:dUmag}
\end{equation}
Now, if the angle between $\dU$ and the symmetry axis of the spheroid $\bp$ is $\chi\in[0,\upi/2]$, then 
\begin{equation}
    |\dU|\sin\chi=|\bpe|\sin\theta,\qquad |\dU|\cos\chi=|\bpa|\,|\mu|.
    \label{eq:dUmag_Orientation}
\end{equation}
The conditions $\bpe=0$ and $\bpa=0$ define two loci in the $(\kappa,\lambda)$ plane that organise the ideal collision rate (\S\,\ref{sec:locus}). In the spherical limit $\dU=(1-S)\bg$.

\subsection{Orientation of the spheroid and the two populations}\label{sec:orientation}
In Stokes flow a settling spheroid feels no torque, because its resistance does not couple translation and rotation \citep{KimKarrila1991}. Its orientation is then inherited from its history. The orientation distribution of the spheroids far from any sphere, $\psi_\infty(\bp)$, is a probability density on the unit sphere of orientations $S^{2}=\{\bp:|\bp|=1\}$, normalised as $\int_{S^{2}}\psi_\infty(\bp)\,\dd^{2}\bp=1$, where $\dd^{2}\bp$ is the element of solid angle. It is not determined by the settling motion and is taken to be isotropic, $\psi_\infty(\bp)=1/(4\upi)$. This is the distribution attained by small spheroids whose orientations have been randomised by turbulence in the absence of fluid inertia \citep{Gustavsson2019,Sheikh2020}. We call this the Stokes population.

Weak fluid inertia changes this picture. At small particle Reynolds number $Re_p=U_\perp L/\nu$, with $U_\perp=\Gpe U^{*}$ the broadside settling speed, the disturbance flow exerts a torque on the spheroid \citep{Cox1965,KhayatCox1989,Dabade2015,Jiang2021}. Divided by the rotational resistance, it turns the axis at the rate
\begin{equation}
\dot{\bp}_I=-\,\frac{F_0\,U^{*}L}{\nu}\,(\boldsymbol{U}_c\bcdot\bp)\,(\tI-\bp\bp)\bcdot\boldsymbol{U}_c,
\label{eq:omegaI}
\end{equation}
with $F_0(\kappa)>0$ the torque coefficient, written as in \citet{Roy2026orientation}. It is the closed-form $O(Re)$ torque of \citet[][]{Dabade2015}.
Vertical orientations are unstable, and every broadside orientation, the girdle $\mu=0$, is stable but neutral in azimuth. Near broadside the tilt relaxes as
\begin{equation}
\dot\mu=-a_I\,\mu,\qquad a_I=F_b\,\frac{U_\perp^{2}\,L}{\nu\, U^{*}} =F_b\,\Gpe\,Re_p,\qquad F_b=F_0\,\frac{\Gpa}{\Gpe},
\label{eq:aI}
\end{equation}
in units of $U^{*}/L$, or $a_I=F_bU_\perp^{2}/\nu$ in dimensional form, with $F_0=0.138$, $0.131$ and $0.119$, and hence $F_b=0.183$, $0.189$ and $0.182$, for $\kappa=5$, $10$ and $20$, respectively \citep{Dabade2015}.
The slender-body form $F_b\simeq\tfrac58(\Gpa/\Gpe)/\ln2\kappa$ \citep{KhayatCox1989,Shin2009} is $1.6$ times larger at $\kappa=10$. Our results are expressed through $a_I$, so $F_b$ enters only their conversion to $Re_p$. By (\ref{eq:Uc}), the components of $\boldsymbol{U}_c$ along and across $\bp$ can be determined exactly, so that (\ref{eq:omegaI}) reduces to
$\dot{\bp}_I=-a_I\,\mu\,(\bg-\mu\bp)$, and $\dot\mu=-a_I\,\mu\,(1-\mu^{2})$ \citep{Roy2026orientation}. The amplitude $a_I=F_0\Gpa\Gpe U^{*}L/\nu$ is therefore independent of the tilt at $O(Re_p)$, and the linearisation in (\ref{eq:aI}) neglects only the factor $(1-\mu^{2})$.

Three conditions are required for the population to be fully broadside at each encounter and for its pair dynamics to be described as above. First, the alignment time must be short compared with the time between encounters with spheres; their ratio is $O(\lambda(\lambda+1)^{2}\phi_s/(F_bRe_p))$, where $\phi_s$ is the volume fraction of spheres. Second, rotational diffusion must be slow compared with alignment, $D_r\ll a_I\,U^{*}/L$, where $D_r=3k_BT\ln(2\kappa)/(\upi\mu_f L^{3})$ at leading order, so that the r.m.s.\ tilt $\{D_r\,L/(a_I\,U^{*})\}^{1/2}$ is small; any ambient velocity gradient, such as the shear of turbulence, must likewise be weak compared with $a_I\,U^{*}/L$. Third, $Re_p$ must be small enough that the translation of both particles obeys the Stokes equations, since inertia changes the velocities only at $O(Re_p)$ \citep{ProudmanPearson1957,KhayatCox1989}. Under these conditions the orientations far from any sphere are distributed as
\begin{equation}
\psi_\infty^{\,b}(\bp)=\frac{\delta_D(\mu)}{2\upi},
\label{eq:psib}
\end{equation}
with $\delta_D$ the Dirac delta function. We call this the inertial population. Section~\ref{sec:example} gives an indicative dimensional example regarding it.

During an encounter the inertial torque competes with the hydrodynamic torque from the disturbance flow of the sphere. By (\ref{eq:pdoth}) below, that flow rotates the spheroid at a rate of order $\Omega_h=9\gamma/(2\lambda^{3})$ (in units of $U^{*}/L$). This rate is set by the weight of the sphere alone, and it does not vanish when the two particles settle at the same speed. The relative strength of the two torques is measured by the ratio of the inertial alignment rate (\ref{eq:aI}) to $\Omega_h$,
\begin{equation*}
\frac{a_I}{\Omega_h}=\frac{2F_b\,Re_p\Gpe\lambda^{3}}{9\gamma}.
\end{equation*}
This ratio grows with the size ratio $\lambda$. On the extinction locus $\lambda=\lpe$, the isolated settling speed of a broadside spheroid and a sphere becomes equal, i.e., $\Gpe=S$ (\S\,\ref{sec:locus}), which gives $\Gpe{\lpe}^{2}=\gamma$ and $\lpe=\kappa(\gamma Y)^{1/2}$. For $\lambda\le\lpe$, the ratio is therefore bounded by $a_I/\Omega_h\le\tfrac29F_b\lpe Re_p$, which does not exceed $0.5\,Re_p$ for $\kappa\le20$ when $\gamma=1$. In the regime $Re_p\ll1$, the inertial torque is thus weak compared with the hydrodynamic torque during most of an encounter. It still matters far upstream, where the interaction is weak, and when the relative velocity is small (\S\,\ref{sec:finiteRe}). We therefore use four encounter models. (i) Frozen orientation: $\bp$ is fixed. This is not a physical limit, but it isolates the translational interaction. (ii) Strong alignment, $a_I\to\infty$: the tilt is held at zero, while the azimuth of $\bp$ is free, since the inertial torque has no azimuthal component. (iii) Free rotation, $a_I\to0$: the spheroid turns under the hydrodynamic torque alone, but initially the spheroids are aligned in broadside orientation. (iv) Finite $a_I$: the inertial rotation (\ref{eq:omegaI}) is added to the hydrodynamic one. The Stokes population always rotates freely.

\subsection{Hydrodynamic interactions}\label{sec:sbt}

We compute the interactions at first reflection, i.e., each particle moves in the flow that the other would generate if it settled alone. The flow of the sphere is exact. The spheroid is treated by slender-body theory at leading order in $1/\ln2\kappa$ \citep{Batchelor1970,Cox1970}, as in the sphere--fibre simulations of \citet{Harlen1999} and the fibre suspensions of \citet{KochShaqfeh1989}. The isolated velocities (\ref{eq:Uc})--(\ref{eq:dU}) keep the exact resistance functions, so the loci of \S\,\ref{sec:locus} are placed exactly; the slender-body approximation enters only in the interaction terms. Positions are measured from the centre of the sphere. The centroid of the spheroid is at $\br$, and a point on its axis is at $\bx(s)=\br+s\bp$, with $-\tfrac12\le s\le\tfrac12$. The motion of the two particles is determined using the following three steps:

(i) \emph{The disturbance flow of the sphere.} The disturbance field of a sphere of radius $r_s = 1/(2\lambda)$ settling in a quiescent ambient with a velocity $\boldsymbol{U}_s = S\boldsymbol{\hat{g}}$ consists of a Stokeslet and potential-doublet field
\begin{equation}
\boldsymbol{u}_s'(\bx)=\frac{3\gamma}{8\lambda^3}\left[\tJ(\bx)+\frac{1}{12\lambda^2}\tD(\bx)\right]\bcdot\boldsymbol{\hat{g}},
\label{eq:usphere}
\end{equation}
where, 
\begin{equation}
    \tJ=\frac{\tI}{|\bx|}+\frac{\bx\bx}{|\bx|^{3}},\qquad
    \tD=\frac{\tI}{|\bx|^{3}}-\frac{3\bx\bx}{|\bx|^{5}},
    \label{eq:vec_J_D}
\end{equation}
which satisfies no slip on its surface.

(ii) \emph{The response of the spheroid in the sphere's disturbance field.} At leading order in $1/\ln2\kappa$, the force per unit length exerted by the disturbance field (\ref{eq:usphere}) on the element of the spheroid at $\br+s\bp$ is local and, in units of $\mu_fU^{*}$, reads $(4\upi/\ln2\kappa)(\tI-\tfrac12\bp\bp)\bcdot[\boldsymbol{u}_s'(\br+s\bp)-\boldsymbol{U}-s\dot{\bp}]$. Here $\boldsymbol{U}$ is the translational velocity of the centroid of the spheroid and $\dot{\bp}$ is its rotation rate, both induced by the disturbance field of the sphere. Since the spheroid is free of force and torque, these give
\begin{eqnarray}
\boldsymbol{U}&=&\int_{-1/2}^{1/2}\boldsymbol{u}_s'(\br+s\bp)\,\dd s,    \label{eq:sbtresponse_u} \\
\dot{\bp}&=&12\,(\tI-\bp\bp)\bcdot\int_{-1/2}^{1/2}s\,\boldsymbol{u}_s'(\br+s\bp)\,\dd s.     \label{eq:sbtresponse_p}
\end{eqnarray}
The spheroid thus translates with the mean of the disturbance velocity along its axis and rotates with its first moment \citep{Jeffery1922}; the coefficient $1/\ln2\kappa$ cancels from both relations.

(iii) \emph{The disturbance flow of the spheroid and the response of the sphere.} At the same order in $1/\ln2\kappa$, the settling spheroid disturbs the fluid as a line of Stokeslets distributed uniformly along its axis, whose total strength is its buoyant weight $F_c\bg$, with $F_c=(\rho_c-\rho_f)\,g\,V_c$ and $V_c=\tfrac43\upi ca^{2}=\upi L^{3}/(6\kappa^{2})$ the volume of the spheroid. Its disturbance field is
\begin{eqnarray}
\boldsymbol{u}_c'(\bx)&=&\frac{3}{8\kappa^{2}}\int_{-1/2}^{1/2}\tJ(\bx-\br-s\bp)\bcdot\bg\;\dd s,
\label{eq:uspheroid}
\end{eqnarray}
where $\bx-\br-s\bp$ is the position of the field point $\bx$ relative to the element of the spheroid axis at $\br+s\bp$. The sphere located at $\boldsymbol{x}=0$, free of force and torque in this field, translates with the velocity $\boldsymbol{w}_s$ given by Fax\'en's law,
\begin{equation}
\boldsymbol{w}_s=\left[\left(1+\frac{1}{24\lambda^{2}}\nabla^{2}\right)\boldsymbol{u}_c'(\bx)\right]_{\bx=\boldsymbol{0}}.
\label{eq:faxen_sphere}
\end{equation}
Since $\nabla^{2}\tJ=2\tD$, Fax\'en's operator acting on a Stokeslet gives the same kernel as (\ref{eq:usphere}), a consequence of the reciprocal theorem. With $\tJ$ and $\tD$ even in $\bx$, the velocity $\boldsymbol{U}$ of the spheroid, and the velocity $\boldsymbol{w}_s$ of the sphere are therefore both proportional to the single vector field
\begin{equation}
\Wf(\br,\bp)=\int_{-1/2}^{1/2}\left[\tJ(\bx)+\frac{1}{12\lambda^{2}}\tD(\bx)\right]_{\bx=\br+s\bp}\bcdot\bg\;\dd s,
\label{eq:W}
\end{equation}
as
\begin{equation}
\boldsymbol{U}=\frac{3\gamma}{8\lambda^{3}}\,\Wf(\br,\bp),\qquad
\boldsymbol{w}_s=\frac{3}{8\kappa^{2}}\,\Wf(\br,\bp).
\label{eq:wcws}
\end{equation}
The relative interaction velocity is then
\begin{equation}
\boldsymbol{U}-\boldsymbol{w}_s=\delta\,\Wf(\br,\bp),
\label{eq:wrel}
\end{equation}
where 
\begin{equation}
    \delta=\frac38\left(\frac{\gamma}{\lambda^{3}}-\frac{1}{\kappa^{2}}\right)=\frac{F_s-F_c}{8\upi\mu_fLU^{*}},
    \label{eq:delta}
\end{equation}
where $F_s=(\rho_s-\rho_f)\,g\,V_s$, with $V_s=\tfrac43\upi r_s^{3}$, is the buoyant weight of the sphere. The relative interaction velocity is thus proportional to the difference between the buoyant weights of the sphere and the spheroid. It therefore vanishes, for every configuration, when the two weights are equal,
\begin{equation}
\lambda=\lambda_w=\gamma^{1/3}\kappa^{2/3},
\label{eq:lamw}
\end{equation}
and changes sign there. For $\gamma=1$, $\lambda_w$ marks equal volumes. This cancellation holds only at first reflection: the reflected flows, which are omitted in the model, do not cancel at $\lambda_w$, and the boundary-element solution of \S\,\ref{sec:accuracy} shows that the exact interaction, although weak, lowers the closing rate of a head-on approach by $8.5\,\%$ at $h=4a$ for $\kappa=10$. Far from the spheroid, $|\br|\gg1$, $\Wf\to\tJ(\br)\bcdot\bg$, the familiar Stokeslet interaction of two settling particles.
 
The relative velocity of the pair is the sum of the relative settling velocity (\ref{eq:dU}) and the relative interaction velocity (\ref{eq:wrel}), and the rotation of the spheroid follows from (\ref{eq:sbtresponse_p}) with (\ref{eq:usphere}). The pair dynamics is thus
\begin{eqnarray}
\dot{\br}&=&\bpe\,\bg+(\Gpa-\Gpe)\,\mu\,\bp+\delta\,\Wf(\br,\bp),
\label{eq:rdot}\\
\dot{\bp}&=&\dot{\bp}_h+\dot{\bp}_I,
\label{eq:pdot_h_I}
\end{eqnarray}
where
\begin{equation}
    \dot{\bp}_h=\frac{9\gamma}{2\lambda^{3}}\,(\tI-\bp\bp)\bcdot\int_{-1/2}^{1/2}s\left[\tJ(\bx)+\frac{1}{12\lambda^{2}}\tD(\bx)\right]_{\bx=\br+s\bp}\bcdot\bg\;\dd s.
    \label{eq:pdoth}
\end{equation}
Here $\dot{\bp}_h$ is the hydrodynamic rotation rate, that is $\dot{\bp}$ of (\ref{eq:sbtresponse_p}), and $\dot{\bp}_I$ is the inertial rotation (\ref{eq:omegaI}). All the integrals reduce to elementary line integrals, given in closed form in Appendix~\ref{app:lineint}. Each term of $\Wf$ is a line integral of a Stokes flow, and is therefore divergence-free. Hence $\nabla_{\br}\bcdot\Wf=0$, where $\nabla_{\br}$ is the gradient with respect to the separation $\br$ at fixed orientation $\bp$, and the relative translational velocity field is solenoidal at fixed orientation. This is a property of the model rather than of the exact pair hydrodynamics, whose mobilities are compressible \citep{Batchelor1982}.

The model has two kinds of error. The first is the slender-body closure. It replaces the exact singularity solution of the spheroid \citep{ChwangWu1975} and its exact Fax\'en laws \citep{Kim1985} by their slender limits. The strength of the line of Stokeslets is fixed by the buoyant weight, and the coefficient $1/\ln2\kappa$ cancels from (\ref{eq:sbtresponse_u}) and (\ref{eq:sbtresponse_p}), so no error of order $1/\ln2\kappa$ arises. The resulting errors are of relative order $\kappa^{-2}$, from the finite length of the Stokeslet line and from approximating the Bretherton constant $\mathcal{B}=(\kappa^{2}-1)/(\kappa^{2}+1)$ by unity \citep{Jeffery1922, Bretherton1962}, and of order $a^{2}/\rho^{2}$ at a distance $\rho$ from the axis, from the neglected potential doublets. The second is the reflection truncation, which omits the response of each particle to the flow reflected by the other. Both errors grow near contact. The model also has no lubrication resistance, so the particles touch in finite time. We count a collision when the surfaces touch, and we call the resulting values interception efficiencies. Section~\ref{sec:accuracy} tests them against an exact first reflection and a boundary-element solution, and adds a lubrication film and a capture gap.

\subsection{The collision rate}\label{sec:rate}

The state of a pair is described by the separation $\br$ and the orientation $\bp$ of the spheroid. Let $P(\br,\bp)$ be the pair probability density, defined such that $n_1n_2P(\br,\bp)\,\dd^{3}\br\,\dd^{2}\bp$ is the number of sphere--spheroid pairs per unit volume of suspension whose separation lies within $\dd^{3}\br$ of $\br$ and whose orientation lies within the solid angle $\dd^{2}\bp$ of $\bp$. Far from the sphere the two particles are uncorrelated and $P\to\psi_\infty(\bp)$. Since Brownian motion is negligible, $P$ is advected by the pair dynamics (\ref{eq:rdot})--(\ref{eq:pdoth}), and in the steady state it satisfies the conservation equation
\begin{equation}
\nabla_{\br}\bcdot(\dot{\br}P)+\nabla_{\bp}\bcdot(\dot{\bp}P)=0,
\label{eq:Liouville}
\end{equation}
where $\nabla_{\br}$ is the gradient with respect to $\br$ at fixed $\bp$, and $\nabla_{\bp}=(\tI-\bp\bp)\bcdot\partial/\partial\bp$ is the surface gradient on the unit sphere $|\bp|=1$, the projection of the gradient onto the plane tangent to the sphere at $\bp$, so that $\nabla_{\bp}\bcdot$ is the surface divergence. Contact is described by the gap $h(\br,\bp)$: the distance from the centre of the sphere to the surface of the spheroid, minus $r_s$. For fixed $\bp$ the set $h\le0$ consists of all points within a distance $r_s$ of the spheroid, that is, the spheroid enlarged by the radius of the sphere (figure~\ref{fig:pair}). If $\bx^{*}$ is the point of the spheroid surface nearest to the centre of the sphere, and $\bnu$ is the outward unit normal to the spheroid at $\bx^{*}$, which points from the spheroid towards the sphere, the gap changes at the rate
\begin{equation}
\dot{h}=-\bnu\bcdot\left[\dot{\br}+(\bp\times\dot{\bp})\times(\bx^{*}-\br)\right].
\label{eq:hdot}
\end{equation}
The first term in the brackets is the translation of the centroid relative to the sphere, as for a pair of spheres. The second is the additional motion of $\bx^{*}$ due to the rotation of the spheroid, at angular velocity $\bp\times\dot{\bp}$. The collision rate $K_{12}$ is the number of pairs per unit volume per unit time that reach contact, that is, the flux of pairs into the contact manifold $h=0$ of the joint space of $\br$ and $\bp$. For fixed $\bp$, the contact surface $\mathcal{S}(\bp)=\{\br:h(\br,\bp)=0\}$ is the set of positions of the centroid of the spheroid at which the two particles touch (figure~\ref{fig:pair}), and a pair crosses it inwards when $\dot{h}<0$. The collision rate in terms of $\dot{h}$ can be expressed as 
\begin{equation}
K_{12}=n_1n_2U^{*}L^{2}\int_{S^{2}}\dd^{2}\bp\oint_{\mathcal{S}(\bp)}P\,\max(-\dot{h},0)\,\dd A,
\label{eq:K12flux}
\end{equation}
where $\max(-\dot{h},0)$ equals $-\dot{h}$ where the gap closes ($\dot{h}<0$) and vanishes where it opens, so that only inward crossings contribute, and $\dd A$ is the dimensionless area element of $\mathcal{S}(\bp)$. Outside the contact manifold the phase-space current $(\dot{\br}P,\dot{\bp}P)$ is divergence-free by (\ref{eq:Liouville}), so (\ref{eq:K12flux}) equals the flux through any cross-section of the tube of trajectories that reach contact. Pairs that come from infinity but neither reach contact nor separate, because the model holds them in a steady configuration or keeps them in a bound orbit (\S\,\ref{sec:accuracy} and Appendix~\ref{app:numerics}), are not counted: $K_{12}$ is the rate of collisions in a single encounter, from which such trapped pairs are excluded. Every other trajectory that reaches contact comes from infinity, where the interaction has decayed: there the orientation is a constant $\bp_0$, $P=\psi_\infty(\bp_0)$ and the relative velocity is $\dU(\bp_0)$. Evaluating the flux on a cross-section far upstream, normal to $\dU(\bp_0)$, gives
\begin{equation}
K_{12}=n_1n_2U^{*}L^{2}\int_{S^{2}}\psi_\infty(\bp_0)\,|\dU(\bp_0)|\,A_c(\bp_0)\,\dd^{2}\bp_0,
\label{eq:K12capture}
\end{equation}
where $A_c(\bp_0)$ is the dimensionless capture area: the area of the set of upstream impact parameters, in the plane normal to $\dU(\bp_0)$, whose trajectories reach contact. This extends the construction of \citet{Patra2022} to the joint space of position and orientation. The collision efficiency is the ratio of $K_{12}$ to the ideal rate $\Kz$ of the same population (\S\,\ref{sec:ideal}),
\begin{equation}
E=K_{12}/\Kz.
\label{eq:E12}
\end{equation}

\section{Ideal collision rate}\label{sec:ideal}

The ideal collision rate $\Kz$ is the collision rate in the absence of hydrodynamic interactions, when the two particles approach each other along straight lines at their isolated relative velocity. Without interactions the spheroid keeps its orientation and moves relative to the sphere in a straight line with the velocity $\dU(\bp)$. To evaluate $\Kz$ from (\ref{eq:K12capture}), the capture area $A_c(\bp)$ must therefore be determined. Since every trajectory is a straight line parallel to $\dU(\bp)$, a pair reaches contact if and only if this line intersects the region enclosed by the contact surface $\mathcal{S}(\bp)$, and every point of $\mathcal{S}(\bp)$ that faces the oncoming pairs is reached directly from infinity, where $P=\psi_\infty(\bp)$. This region is the Minkowski sum of the spheroid and a sphere of radius $r_s$, that is, the set of all points $\boldsymbol{y}+\boldsymbol{z}$ with $\boldsymbol{y}$ in the spheroid and $|\boldsymbol{z}|\le r_s$ \citep{Schneider2014}. The capture area is thus the projected area $A_\perp(\bp)$ of this Minkowski sum on the plane normal to $\dU(\bp)$. For spheroids of fixed orientation $\bp$, the ideal collision rate is then
\begin{equation}
\mathcal{K}^{(0)}(\bp)=n_1n_2U^{*}L^{2}\,|\dU(\bp)|\,A_\perp(\bp),
\label{eq:K0p}
\end{equation}
and for a population with orientation distribution $\psi_\infty (\bp)$ it is the orientational average
\begin{equation}
\Kz=\int_{S^{2}}\psi_\infty(\bp)\,\mathcal{K}^{(0)}(\bp)\,\dd^{2}\bp.
\label{eq:K0}
\end{equation}
The downstream shadow of the sphere is depleted of pairs, but it does not enter (\ref{eq:K0p}).

\subsection{The collision cross-section}\label{sec:projection}

\begin{figure}
  \centerline{\includegraphics[width=0.5\textwidth]{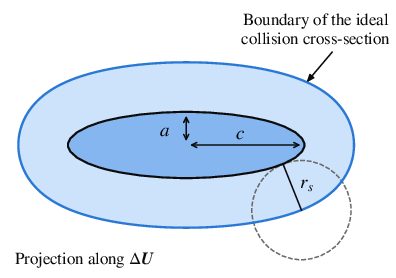}}
  \caption{Projection of the region enclosed by the contact surface on the plane normal to the relative velocity $\dU$: an ellipse with semi-axes $a_e$ and $a$ fattened by a disc of radius $r_s$. Its area is the ideal collision cross-section (\ref{eq:steiner}).}
\label{fig:projection}
\end{figure}

Since projection onto a plane is a linear map, it preserves this enlargement \citep{Schneider2014}. Hence, the projection of the region enclosed by $\mathcal{S}(\bp)$ is the projection of the spheroid enlarged by the projection of the sphere, as illustrated in figure~\ref{fig:projection}. On the plane normal to $\dU(\bp)$, the spheroid projects onto an ellipse with semi-axes
\begin{equation}
a_e=\left(c^{2}\sin^{2}\chi+a^{2}\cos^{2}\chi\right)^{1/2}\quad\text{and}\quad a
\label{eq:ae}
\end{equation}
(Appendix~\ref{app:shadow}), and the sphere onto a disc of radius $r_s$. By Steiner's formula \citep{Schneider2014}, the area of an ellipse fattened by a disc is the sum of the area of the ellipse, its perimeter $P_e$ times $r_s$, and the area of the disc:
\begin{equation}
A_\perp=\upi a_e\,a+P_er_s+\upi r_s^{2},\qquad P_e=4a_e\mathcal{E}(e_\perp),
\label{eq:steiner}
\end{equation}
where $\mathcal{E}(e_\perp)=\int_0^{\upi/2}(1-e_\perp^{2}\sin^{2}\vartheta)^{1/2}\dd\vartheta$ is the complete elliptic integral of the second kind, and $e_\perp=\sqrt{1-a^2/a_e^2}$. This is the collecting area that the construction of \citet{GavzeKhain2022} gives when one particle is a sphere, and (\ref{eq:steiner}) gives it in closed form. Monte Carlo sampling of straight lines past the exact spheroid agrees with (\ref{eq:steiner}) within the sampling error. As a check on the projection, the mean of (\ref{eq:steiner}) over viewing directions distributed uniformly and independently of the orientation of the spheroid equals one quarter of the area of $\mathcal{S}(\bp)$, as required by Cauchy's formula for the mean projected area of a convex body. This is not the average over the isotropic population, in which the viewing direction $\dU(\bp)$ depends on the orientation (\S\,\ref{sec:populations}).

We normalise the ideal collision rate $K^{(0)}$ by the Smoluchowski kernel of two spheres of radii $L/2$ and $r_s$, with the densities of the spheroid and of the sphere,
\begin{equation}
K_S=\frac{\upi n_1n_2 U^{*}L^{2}\,|1-S|}{4}\left(1+\frac1\lambda\right)^{2},
\label{eq:KS}
\end{equation}
and write $\Kbar=\Kz/K_S$. In the spherical limit $\mathcal{K}^{(0)}(\bp)=K_S$ for every orientation. Since $K_S$ vanishes at $S=1$, results are reported for $\lambda\ge1.5$.

\subsection{The two populations}\label{sec:populations}

For the inertial population (\ref{eq:psib}), $\mu=0$ and $\dU=\bpe\bg$ is vertical: a broadside spheroid does not drift. Then $\chi=\upi/2$, $a_e=c$ and $e_\perp=e$, for every azimuth, and the kernel follows in closed form,
\begin{equation}
\Kb=n_1n_2U^{*}L^{2}\,|\bpe|\,A_b,
\label{eq:Kb}
\end{equation}
where $A_b$ is the broadside cross-section in units of $L^{2}$, and expressed as 
\begin{equation}
    A_b=\frac{\upi}{4\kappa}+\frac{\mathcal{E}(e)}{\lambda}+\frac{\upi}{4\lambda^{2}}.
    \label{eq:Ab_ICR}
\end{equation}
We can observe that the ideal kernel of the aligned population, $\Kb$ vanishes only where $\bpe=0$.

For the Stokes population, $\psi_\infty=1/(4\upi)$ in (\ref{eq:K0}), and since $\mathcal{K}^{(0)}(\bp)$ depends on the orientation only through $|\mu|$, the orientational average reduces to an integral of (\ref{eq:K0p}) over $|\mu|\in[0,1]$. Expressing $|\dU|$ and the semi-axis $a_e$ in terms of $\mu$, each of the three contributions to the cross-section (\ref{eq:steiner}) yields one term of the ideal collision rate
\begin{equation}
\Kz=n_1n_2U^{*}L^{2}\left(T_1+T_2+T_3\right),
\label{eq:Kiso}
\end{equation}
where
\begin{equation}
\left.
\begin{aligned}
T_1&=\frac{\upi}{4\kappa}\,\Phi\!\left(\bpe^{2},\,\kappa^{-2}\bpa^{2}-\bpe^{2}\right),\\
T_2&=\frac1\lambda\int_0^{1}\left[\bpe^{2}(1-\mu^{2})+\kappa^{-2}\bpa^{2}\mu^{2}\right]^{1/2}\mathcal{E}\big(e_\perp(\mu)\big)\,\dd\mu,\\
T_3&=\frac{\upi}{4\lambda^{2}}\,\Phi\!\left(\bpe^{2},\,\bpa^{2}-\bpe^{2}\right),
\end{aligned}
\right\}
\label{eq:T123}
\end{equation}
and $\Phi(\beta_\perp^{2},m)=\int_0^{1}(\beta_\perp^{2}+m\mu^{2})^{1/2}\dd\mu$ is elementary (Appendix~\ref{app:shadow}). Only $T_2$ requires numerical quadrature. For a slender spheroid, $a\ll a_e$ except near $\chi=0$, so that $e_\perp\simeq1$ and $\mathcal{E}(e_\perp)\simeq1$ in $T_2$. Dropping in addition the terms $\kappa^{-2}\bpa^{2}$ in $T_1$ and $T_2$, which arise from the end-on projection of the spheroid, gives $T_1+T_2\simeq\tfrac{\upi}{4}|\bpe|(1/\lambda+\upi/4\kappa)$, while $T_3$ is unchanged:
\begin{equation}
\Kz\simeq n_1n_2U^{*}L^{2}\left[\frac{\upi}{4}|\bpe|\left(\frac1\lambda+\frac{\upi}{4\kappa}\right)+\frac{\upi}{4\lambda^{2}}\,\Phi\!\left(\bpe^{2},\,\bpa^{2}-\bpe^{2}\right)\right].
\label{eq:Kisoslender}
\end{equation}
This form is not used for the results, all of which are computed from (\ref{eq:Kiso}) and (\ref{eq:T123}), but it exposes the structure of the isotropic kernel. Its first two terms are $\upi/4=\avg{\sin\theta}$ times the corresponding terms of the broadside kernel (\ref{eq:Kb}), the origin of the factor $4/\upi$ in \S\,\ref{sec:compare}. The slender form (\ref{eq:Kisoslender}) is accurate when the broadside speed mismatch $|\bpe|$ is large, but it deteriorates as $|\bpe|$ approaches zero, where the neglected terms $\kappa^{-2}\bpa^{2}$ become comparable to $\bpe^{2}$ and sustain the floor of the isotropic kernel discussed in \S\,\ref{sec:locus}. For instance, its error is $0.6\,\%$ at $(\kappa,\lambda)=(20,4)$, far from the extinction locus, but $50\,\%$ at $(10,6.2)$, close to $\lpe=6.17$.

\subsection{The extinction locus}\label{sec:locus}

The broadside-aligned kernel (\ref{eq:Kb}) vanishes on the extinction locus
\begin{equation}
\lpe=\kappa\,(\gamma Y)^{1/2},\qquad\text{equivalently}\qquad r_{s\perp}^{*}=\frac{a}{(\gamma Y)^{1/2}},
\label{eq:lperp}
\end{equation}
the radius of the sphere, on which a broadside aligned spheroid and the sphere settle at the same velocity. Differential settling is then extinguished for every aligned pair. The matched radius ($r_{s\perp}^*$) depends only on $a$, $\kappa$ and $\gamma$.
For a slender spheroid,
\begin{equation}
\lambda_{sl\perp}^*\simeq2\kappa\left[\frac{2\gamma}{3(2\ln2\kappa+1)}\right]^{1/2}.
\label{eq:lperpslender}
\end{equation}
For equal densities, i.e., $\gamma=1$, the radius of the matched sphere is always larger than the semi-minor axis of the spheroid. 
Near the spherical limit $\lpe=1+\tfrac{7}{10}(\kappa-1)+O((\kappa-1)^{2})$ for $\gamma=1$, so the locus starts at the point where two equal spheres settle together.

\begin{figure}
  \centering
  \panel[1.18]{fig:locus_a}{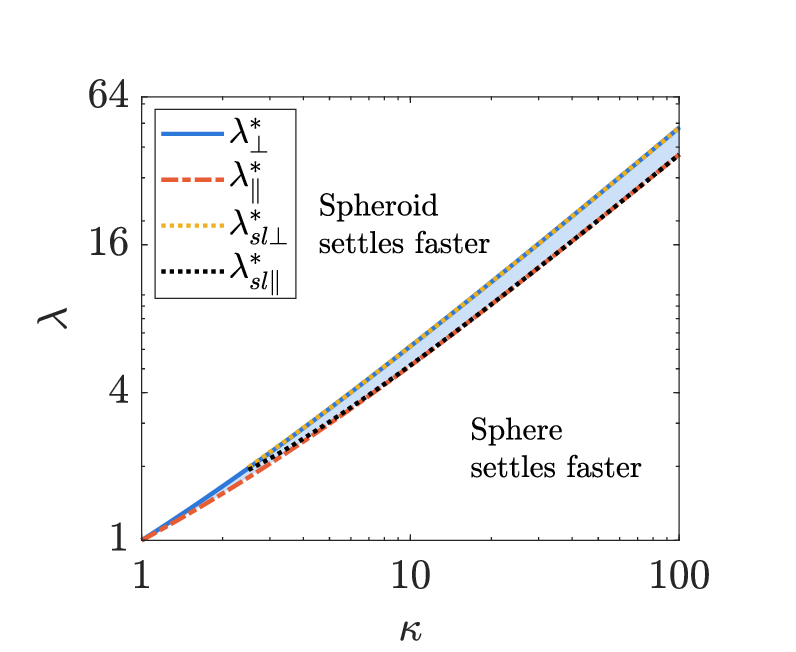} \hfill 
  \panel[1.18]{fig:locus_b}{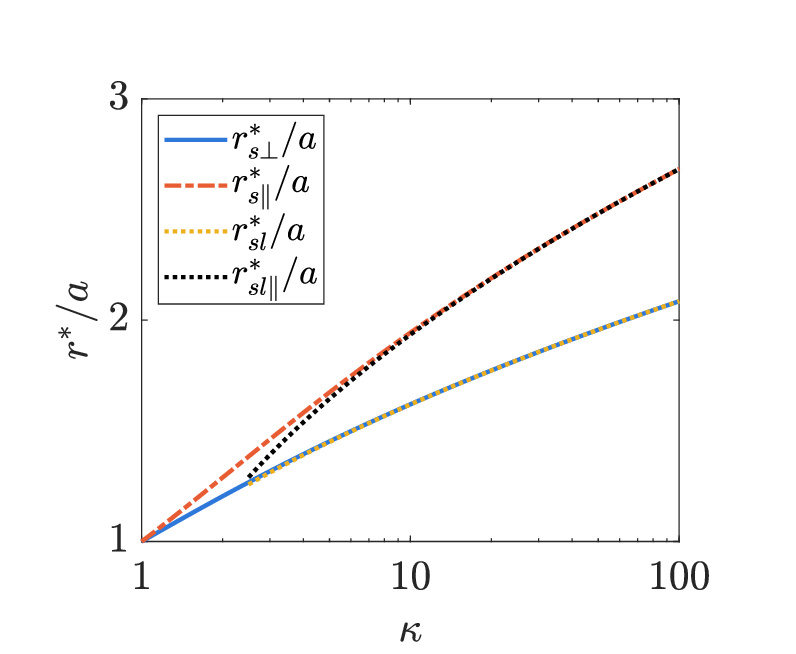}
  \caption{(\textit{a}) The variation of the extinction locus $\lpe$, the axial locus $\lpa$, and their slender counterparts $\lambda_{sl\perp}^*$ and $\lambda_{sl\parallel}^*$ with respect to the aspect ratio $\kappa$ for $\gamma=1$. In the shaded region bounded by $\lambda_{\perp}^*$ and $\lambda_\parallel^*$, any of the particles may settle faster depending upon the spheroid's orientation. (\textit{b}) The variation of the matched sphere radii $r_{s\perp}^*$, $r_{s\parallel}^*$, and their slender counterparts $r_{sl\perp}^*$ and $r_{sl\parallel}^*$ in the units of transverse semi-axis $a$, with respect to $\kappa$ for $\gamma=1$.}
\label{fig:locus}
\end{figure}

In a similar manner, the axial locus is defined by $\bpa=0$,
\begin{equation}
\lpa=\kappa\,(\gamma X)^{1/2},  \qquad\text{equivalently}\qquad r_{s\parallel}^{*}=\frac{a}{(\gamma X)^{1/2}},
\label{eq:lpar}
\end{equation}
the radius of the sphere, on which a spheroid with its axis vertical and the sphere settle at the same velocity. For a slender spheroid, the slender form (\ref{eq:Gslender}) of $\Gpa$ gives
\begin{equation}
\lambda_{sl\parallel}^{*}\simeq2\kappa\left[\frac{\gamma}{3(2\ln2\kappa-1)}\right]^{1/2}.
\label{eq:lparslender}
\end{equation}
In figure~\ref{fig:locus_a} the variation of the extinction locus and the axial locus is shown along with their slender forms with respect to the aspect ratio of the spheroid. Since the settling velocity of a spheroid at vertical orientation is more than its broadside orientation, the sphere that matches its settling velocity must be larger. The axial locus therefore lies below the extinction locus, $\lpa<\lpe$ as shown in figure.~\ref{fig:locus_a}. Their ratio, $\lpa/\lpe=(X/Y)^{1/2}$, decreases from unity in the spherical limit $\kappa=1$ towards the slender-body value $1/\sqrt{2}$ as $\kappa\to\infty$, although the approach is logarithmically slow. The two loci divide the $(\kappa,\lambda)$ plane into three regions. For $\lambda>\lpe$ the settling velocity of the sphere is slower than the broadside-aligned spheroid, and hence irrespective of the orientation of the spheroid, the sphere settles slower. For $\lambda<\lpa$ the sphere settles faster than even a vertical spheroid, so that every spheroid settles more slowly. In the band $\lpa<\lambda<\lpe$, the vertical settling speed of the spheroid, $\Gpe+(\Gpa-\Gpe)\mu^{2}$, equals that of the sphere at the tilt $\theta_v$ given by $\cos^{2}\theta_v=(S-\Gpe)/(\Gpa-\Gpe)$ so that spheroids tilted by less than $\theta_v$ settle faster than the sphere, whereas those tilted by more lag behind it. Even at $\theta=\theta_v$, however, the relative velocity does not vanish, because the tilted spheroid still drifts sideways. By (\ref{eq:dUmag}), $|\dU|$ vanishes only for a broadside spheroid on $\lpe$ and for a vertical spheroid on $\lpa$.

Figure~\ref{fig:locus_b} shows the radii of the matched spheres, in units of the transverse semi-axis $a$ of the spheroid, on the two loci. On the extinction locus $r_{s\perp}^{*}/a=(\gamma Y)^{-1/2}$, by (\ref{eq:lperp}), and on the axial locus $r_{s\parallel}^{*}/a=(\gamma X)^{-1/2}$, by (\ref{eq:lpar}). Both radii equal $a$ in the spherical limit, where two equal spheres settle together, and both increase with $\kappa$. For a slender spheroid, (\ref{eq:lperpslender}) and (\ref{eq:lparslender}) give
\begin{equation}
\frac{r_{sl\perp}^{*}}{a}\simeq\left[\frac{3(2\ln2\kappa+1)}{8\gamma}\right]^{1/2},\qquad
\frac{r_{sl\parallel}^{*}}{a}\simeq\left[\frac{3(2\ln2\kappa-1)}{4\gamma}\right]^{1/2},
\label{eq:matched_radii_slender}
\end{equation}
where $r_{sl\perp}^{*}$ and $r_{sl\parallel}^{*}$ denote the matched radii of the sphere settling at the same velocity as the spheroid at $\lambda_\perp^*$ and $\lambda_\parallel^*$, respectively. The matched radii grow only as $(\ln2\kappa)^{1/2}$. This weak dependence reflects the balance of weight and drag: the weight of a slender spheroid scales as $a^{2}L$ and its drag as $L/\ln2\kappa$, so that its settling speed scales as $a^{2}\ln2\kappa$, whereas that of the sphere scales as $r_s^{2}$. For $\gamma=1$, the sphere matched to a broadside spheroid is therefore thicker than the spheroid, by a factor that increases slowly with the aspect ratio, yet remains much smaller than its half-length, since $r_{s\perp}^{*}/c=1/\lpe$ decreases as $\kappa$ increases. The sphere matched to a vertical slender spheroid is still larger, and the ratio $r_{s\parallel}^{*}/r_{s\perp}^{*}=(Y/X)^{1/2}$ increases slowly towards $\sqrt{2}$ as $\kappa\to\infty$.

Unlike the broadside-aligned kernel, the isotropic kernel does not vanish on the extinction locus. There $S=\Gpe$ as $\bpe=0$, resulting in $\bpa=\Gpa-\Gpe$, and the relative velocity (\ref{eq:dU}) reduces to $\dU=\bpa\mu\,\bp$. The broadside part of the settling velocity of every spheroid is thus matched exactly by the sphere, and only the excess axial speed of a tilted spheroid remains. Every spheroid, whatever its tilt with respect to gravity, therefore moves relative to the sphere along its own axis with speed $\bpa|\mu|$. Then $\chi=0$, and its cross-section (\ref{eq:steiner}) reduces to the end-on value $\upi(a+r_{s\perp}^{*})^{2}$, with $r_{s\perp}^{*}$ the matched radius (\ref{eq:lperp}). The kernel at fixed orientation (\ref{eq:K0p}) is then $n_1n_2U^{*}L^{2}\bpa|\mu|\,\upi(a+r_{s\perp}^{*})^{2}$, and averaging over the isotropic population, for which $|\mu|$ is uniform on $[0,1]$ with mean $\tfrac12$, gives
\begin{equation}
\Kz\big|_{\lambda=\lpe}=\frac{\upi}{8\kappa^{2}}\,n_1n_2U^{*}L^{2}\left(\Gpa-\Gpe\right)\left[1+\frac{1}{\sqrt{\gamma Y}}\right]^{2}.
\label{eq:Kisolocus}
\end{equation}
We call this value the floor of the isotropic kernel: whereas the broadside-aligned kernel falls to zero on the locus ($\lambda_\perp^*$), the isotropic kernel falls only to approximately (\ref{eq:Kisolocus}) due to the contribution of the tilted spheroids, which still settle faster than the sphere there. It exists only because the settling mobility of the spheroid is anisotropic, $\Gpa\neq\Gpe$, and it vanishes in the spherical limit, where $\Gpa=\Gpe$. Subsequently it is shown that the floor lies close to, but not exactly at, the minimum of the isotropic kernel.

A similar closed form holds on the axial locus. There $\bpa=0$, so that $\dU=\bpe(\bg-\mu\bp)$ is perpendicular to the axis of every spheroid, resulting in $\chi=\upi/2$. Therefore, at the axial locus every spheroid presents its broadside cross-section $A_b$, and $|\dU|=|\bpe|\sin\theta$. Averaging $\sin\theta$ over the isotropic population we get $\Kz=(\upi/4)\Kb$, so that on $\lpa$ alignment raises the ideal rate by the factor $4/\upi$ with respect to the broadside-aligned kernel (\S\,\ref{sec:compare}).
\begin{table}
  \begin{center}
\def~{\hphantom{0}}\small\setlength{\tabcolsep}{3.5pt}
  \begin{tabular}{rcccccccc}
      $\kappa$ & $\lpa$ & $\lpe$ & $r_{s\perp}^{*}/a$ & $\Kbar|_{\lpe}$ & $\lambda_{min}$ & $\Kbar_{min}$ & $\lambda_-$ & $\lambda_+$\\[3pt]
      ~2 & ~1.5517 & ~1.6607 & 1.2043 & $1.956\times10^{-2}$ & ~1.6515 & $1.901\times10^{-2}$ & ~1.6133 & ~1.9778\\
      ~3 & ~2.0527 & ~2.2766 & 1.3178 & $7.925\times10^{-3}$ & ~2.2670 & $7.819\times10^{-3}$ & ~2.1892 & ~2.6301\\
      ~5 & ~2.9873 & ~3.4431 & 1.4522 & $2.185\times10^{-3}$ & ~3.4361 & $2.175\times10^{-3}$ & ~3.2932 & ~3.8388\\
      10 & ~5.1450 & ~6.1738 & 1.6198 & $3.012\times10^{-4}$ & ~6.1704 & $3.010\times10^{-4}$ & ~5.9246 & ~6.6212\\
      20 & ~9.1341 & 11.2828 & 1.7726 & $3.376\times10^{-5}$ & 11.2814 & $3.375\times10^{-5}$ & 10.9320 & 11.7767\\
      50 & 20.1455 & 25.5523 & 1.9568 & $1.512\times10^{-6}$ & 25.5518 & $1.512\times10^{-6}$ & 25.0862 & 26.0988\\
  \end{tabular}
    \caption{Quantities on and near the extinction locus for $\gamma=1$. $\lpa$ and $\lpe$: the axial locus (\ref{eq:lpar}) and the extinction locus (\ref{eq:lperp}), the size ratios at which a vertical and a broadside spheroid, respectively, settle at the same speed as the sphere. $r_{s\perp}^{*}/a=(\gamma Y)^{-1/2}$: radius of the sphere matched on the extinction locus, in units of the transverse semi-axis $a$ of the spheroid. $\Kbar|_{\lpe}$: the isotropic floor (\ref{eq:Kisolocus}), normalised by the Smoluchowski kernel $K_S$ (\ref{eq:KS}). $\lambda_{min}$ and $\Kbar_{min}$: the size ratio at which the normalised isotropic kernel $\Kbar=\Kz/K_S$ is smallest at fixed $\kappa$, and its value there. $\lambda_-$ and $\lambda_+$: the lower and upper limits of the band $\lambda_-<\lambda<\lpe<\lambda_+$ around the extinction locus in which alignment lowers the ideal rate, $\Rr=\Kb/\Kz<1$ (\S\,\ref{sec:compare}); outside this band $\Rr>1$.}
  \label{tab:locus}
  \end{center}
\end{table}

Table~\ref{tab:locus} lists the two loci, the matched radius, the floor for isotropic kernel, the minimum value of the isotropic kernel near extinction locus and the corresponding size ratio, and the lower and upper limits of the band around the extinction locus in which alignment lowers the ideal rate, when $\gamma=1$. We observe that at a fixed $\kappa$, the normalised isotropic kernel $\Kbar$ attains a minimum $\Kbar_{min}$ at the size ratio $\lambda_{min}$, which is different from the extinction locus ($\lambda_\perp^*$). Because the floor itself varies with $\lambda$, through $\bpa$ and through the cross-section, the minimum lies slightly below the extinction locus, by less than $1\,\%$, and its depth $\Kbar_{min}$ is within $3\,\%$ of the normalised floor for $\kappa\ge2$. Both differences decrease rapidly as $\kappa$ increases. Such a minimum might be read as the coincidence of the mean settling speeds of the two species. It is instead a remnant of the extinction locus. The broadside part of the cross-section, which dominates for a slender body, is swept only by the component of $\dU$ normal to the axis, $\bpe(\bg-\mu\bp)$, and this component vanishes on $\lpe$ for every orientation. The same argument holds for a thin oblate plate, whose face is swept only by the component of $\dU$ along its normal, set by the face-down settling speed, as mentioned in \citet{Naso2018}. For thin plates (aspect ratio $0.01$, radius $150$\,\textmu m) with nearly random orientations in weak turbulence, they found a sharp dip of the collision kernel at a droplet radius of $15$\,\textmu m, with nearly equal values at $10$ and $20$\,\textmu m. With their parameters, a droplet of radius $15.6$\,\textmu m matches the face-down speed of the plate, while the mean settling speed of their plates, $4.9$\,cm\,s$^{-1}$, is matched by a droplet of $19$\,\textmu m. The dip sits at the face-down match.

Fluid inertia, by aligning spheroids broadside to gravity, turns the minimum of the Stokes kernel into a true zero. Broadside-aligned spheroids of transverse semi-axis $a$ do not collide by differential settling with spheres of radius $r_{s\perp}^{*}$, and collisions near the locus must come from effects that the ideal problem omits. Hydrodynamic interactions are one such effect, which is discussed in detail in \S\,\ref{sec:locusHI} near the extinction locus.

\subsection{Isotropic versus broadside-aligned populations}\label{sec:compare}
Figure~\ref{fig:lambda_a} shows the normalised kernels $\Kbar$ of the isotropic and broadside-aligned populations against the size ratio for different aspect ratios when $\gamma=1$. When the radius of the sphere is comparable to the half-length of the spheroid, $\lambda=O(1)$, the sphere settles much faster than the spheroid, so that $|\dU|\simeq S$ depends weakly on the orientation. This results in very close values of the two kernels near $\lambda=1$. As the size ratio increases, the broadside-aligned kernel falls to zero on the extinction locus $\lpe$, where the settling speed of every broadside spheroid matches that of the sphere, whereas the isotropic kernel falls only to the floor (\ref{eq:Kisolocus}). Beyond the locus both kernels rise again and approach constant values as the sphere becomes small and the spheroid settles faster than the sphere. Away from the locus the broadside-aligned kernel lies above the isotropic one for every $\kappa$. Figure~\ref{fig:lambda_b} shows the two kernels in the neighbourhood of the two loci for $\kappa=10$ and $\gamma=1$. The aligned kernel varies linearly with $|\bpe|$ on either side of $\lpe$, whereas the isotropic kernel passes through a smooth minimum close to it, and the two curves cross on either side of the locus. The closed forms of \S\,\ref{sec:locus}, on $\lpe$ and on $\lpa$, are marked by open circles.
\begin{figure}
  \centering
  \panel[1.1]{fig:lambda_a}{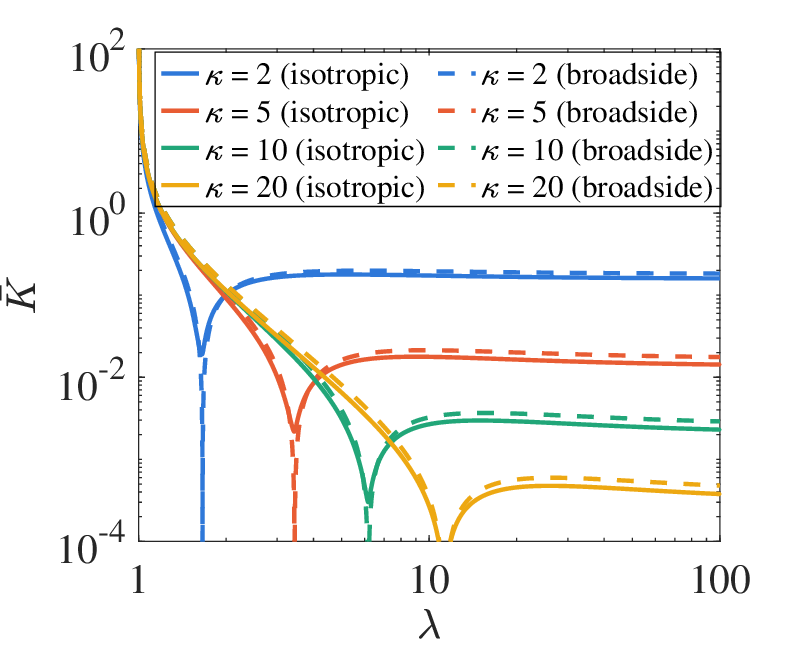}\hfill
  \panel[1.1]{fig:lambda_b}{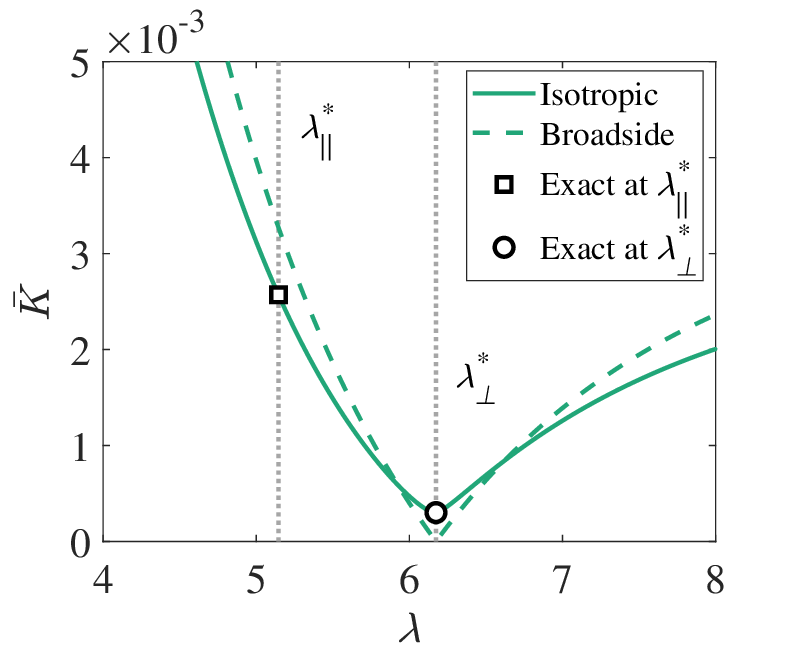}
  \caption{The variation of $\Kbar$, the ideal collision rate normalised by the Smoluchowski kernel (\ref{eq:KS}), with the size ratio $\lambda$ for $\gamma=1$. The solid lines represent the isotropic (Stokes) population and the dashed lines represent the broadside-aligned (inertial) population. (\textit{a}) Variation of $\Kbar$ against the size ratio $\lambda$ for $\kappa=2$, $5$, $10$ and $20$. (\textit{b}) Variation of $\bar{K}$ for $\kappa=10$ near the axial locus $\lpa$ and the extinction locus $\lpe$, marked by dotted vertical lines. The square symbol represents the exact value of $\bar{K}$ on $\lambda=\lpa$, and the circle symbol represents the exact value of $\bar{K}$ on $\lambda=\lpe$.}
\label{fig:lambda}
\end{figure}

In order to quantify the effect of alignment in the collision rate we define a ratio $\Rr=\Kb/\Kz$. Figure~\ref{fig:ratio_a} maps $\Rr$ in the $(\kappa,\lambda)$ plane. We observe that the broadside-alignment raises the ideal collision rate, $\Rr>1$, everywhere except in a narrow band $\lambda_-<\lambda<\lambda_+$ that follows the extinction locus (table~\ref{tab:locus}). The absolute width of this band increases with $\kappa$, but its width relative to the locus, $(\lambda_+-\lambda_-)/\lpe$, decreases from about $0.22$ at $\kappa=2$ to $0.04$ at $\kappa=50$. Figure~\ref{fig:ratio_b} shows the variation of $\Rr$ against $\lambda/\lpe$ for different aspect ratios. With this scaling the curves nearly collapse: $\Rr$ falls to zero at $\lambda=\lpe$, reaches its largest value, $4/\upi$, on the axial locus $\lambda=\lpa$, and approaches a value below $4/\upi$ for $\lambda\gg\lpe$. The gain from alignment never exceeds $4/\upi$, i.e., 
\begin{equation}
\Kb\le\frac{4}{\upi}\,\Kz,
\label{eq:bound}
\end{equation}
with equality only on $\lambda=\lpa$. The bound (\ref{eq:bound}) has a simple geometric origin. Whatever its orientation, a spheroid viewed along the relative velocity presents a cross-section (\ref{eq:steiner}) that is never smaller than its broadside cross-section reduced by the factor $\sin\chi$, where $\chi$ is the angle between $\dU$ and its axis $\bp$. By (\ref{eq:dUmag_Orientation}), the product of $|\dU|$ and $\sin\chi$ is the speed with which the broadside part of the relative velocity, $\bpe$, sweeps the spheroid at tilt $\theta$. The kernel at fixed orientation (\ref{eq:K0p}) is therefore never smaller than the aligned kernel multiplied by $\sin\theta$, and averaging $\sin\theta$ over an isotropic population gives the factor $\upi/4$. Equality would require every spheroid to move normal to its own axis, which by (\ref{eq:dU}) occurs only on the axial locus, where $\bpa=0$. Note that in figure~\ref{fig:ratio_b}, the variation of $\Rr$ is shown for $\lambda>1.05$, i.e. for spheres smaller than the half-length of the spheroid, so that for $\kappa=2$, where $\lpe=1.66$, the curve begins at $\lambda/\lpe\approx0.63$.

\begin{figure}
  \vspace{5mm}
  \centering
  \panel[1]{fig:ratio_a}{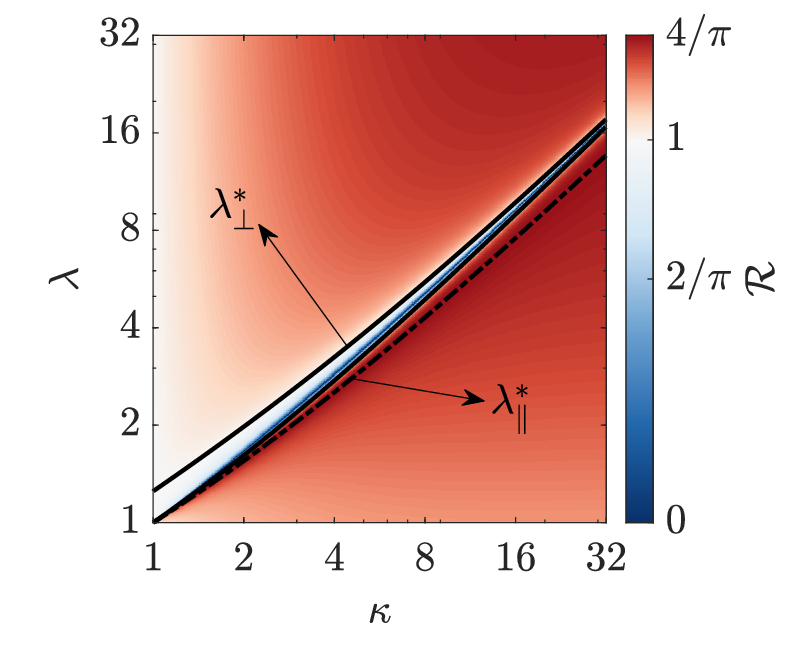}\hfill
  \panel[1]{fig:ratio_b}{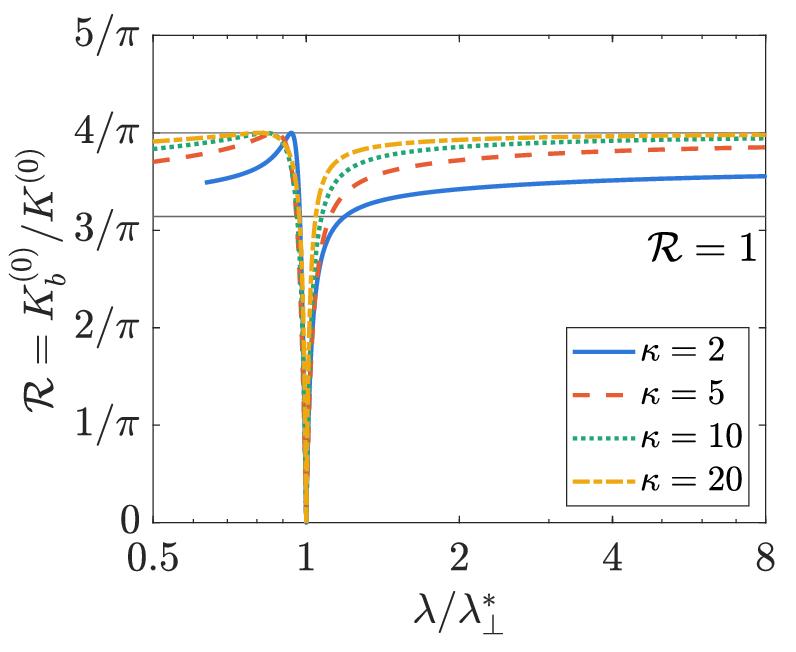}
    \caption{(\textit{a}) Contours of $\Rr=\Kb/\Kz$ in the $(\kappa,\lambda)$ plane, with the extinction locus $\lpe$ (\ref{eq:lperp}) (solid line), and the axial locus $\lpa$ (\ref{eq:lpar}) (dash-dotted line) for $\gamma=1$. The region $\Rr<1$ is confined to a narrow band around the extinction locus. (\textit{b}) Variation of $\Rr$ against the size ratio $\lambda$ scaled by the extinction locus $\lambda_\perp^*$ for $\kappa=2$, $5$, $10$ and $20$ when $\gamma=1$. The curves are shown for $\lambda>1.05$. The thin horizontal lines mark $\Rr=1$ and $\Rr=4/\upi$.}
\label{fig:ratio}
\end{figure}

The effect of alignment thus has two competing parts. The gain is geometric: an isotropic population presents, on average, no less than the fraction $\upi/4$ of the broadside cross-section, so aligning the spheroids raises the cross-section by at most $4/\upi$. The loss is kinematic: alignment places every spheroid in its slowest orientation, which a sphere of radius $r_{s\perp}^{*}$ matches exactly, so that near the extinction locus the relative velocity, and with it the aligned kernel, vanishes. Three limits complete the picture. When the sphere is much smaller than the spheroid, $\lambda\to\infty$, the cross-section reduces to that of the spheroid alone, and $\Rr$ approaches a constant that increases with the aspect ratio, and it tends to $4/\upi$ for a needle ($\kappa\to\infty$). For a needle at fixed $\lambda$, both particles move relative to each other essentially at the settling speed of the sphere, so that $\Rr$ depends only on the size ratio and lies between $4/\upi$ and unity. Near the spherical limit (Appendix~\ref{app:nearsphere}),
\begin{equation}
\Rr=1+e^{2}\,\frac{\lambda(4\lambda-5)}{30(\lambda^{2}-1)}+O(e^{4}),
\label{eq:Rnearsphere}
\end{equation}
so that for nearly spherical spheroids, alignment increases the ideal collision rate when $\lambda>5/4$ and decreases it when $1<\lambda<5/4$; this narrow interval is the one into which both loci contract as $\kappa\to1$.

\section{Collision rate with hydrodynamic interactions}\label{sec:interacting}

The interacting kernel follows from (\ref{eq:K12capture}), once the capture area $A_c(\bp_0)$ has been determined by integrating the pair dynamics (\ref{eq:rdot})--(\ref{eq:pdoth}). Far upstream, each trajectory is labelled by its impact parameter $\boldsymbol{b}=(b_1,b_2)$, its position on the plane normal to $\dU(\bp_0)$, measured from the line through the centre of the sphere parallel to $\dU(\bp_0)$. The component $b_1$ is taken along the projection of $\bp_0$ on this plane, and $b_2$ across it. The capture set is the set of impact parameters whose trajectories reach contact, and $A_c(\bp_0)$ is its area. Without interactions it reduces to the shadow of \S\,\ref{sec:ideal}, the projection of the region enclosed by the contact surface. Since the relative translational velocity is solenoidal in $\br$ for every orientation, the flux of pairs through any cross-section of the tube of colliding trajectories equals the flux far upstream, provided that the orientation along the tube has not yet departed from $\bp_0$.

\subsection{Methodology}\label{sec:method}

In order to identify the capture set and calculate the interacting kernel for a particular population of spheroid, we consider the trajectories that start on a plane normal to $\dU(\bp_0)$ and at a distance $Z$ upstream of the sphere. The pair dynamics between the spheroid and the sphere, i.e., the coupled relative velocity (\ref{eq:rdot}) and orientation rate of the spheroid (\ref{eq:pdot_h_I}) are integrated in time with an adaptive Runge--Kutta scheme. Contact between the two particles is detected according to the exact gap $h$ between their surfaces, and the integration time step is refined as the gap closes, so that grazing contacts are not stepped over (Appendix~\ref{app:numerics}). The set of colliding impact parameters is traced along rays in the impact plane. The capture area $A_c(\bp_0)$ is computed in the upstream plane normal to $\dU(\bp_0)$, by launching trajectories along rays that emanate from the point directly upstream of the sphere and are distributed uniformly in the polar angle of that plane. For a broadside spheroid the pair configuration is symmetric about two vertical mirror planes, the plane containing the axis of the spheroid and the plane normal to it, so that only one quadrant of the upstream plane needs to be traced. Under frozen orientation and strong alignment the capture set is star-shaped with respect to the centre of this plane. Here, each ray crosses its boundary only once, so that trajectories launched inside the crossing point reach contact and those launched outside it miss. The crossing point on each ray is then located by bisection between a colliding and a non-colliding launch, and the area follows from the boundary radii by quadrature over the polar angle. With free rotation of the spheroid the capture set need not be star-shaped, and a ray may cross its boundary several times. The outcome of a trajectory, collision or miss, is therefore first sampled at a sequence of points along each ray, every change of outcome between neighbouring points is refined by bisection, and the contribution of each ray to the area is assembled from all the crossings it contains. The kernel is evaluated as the flux of the actual relative velocity through the capture set on the starting plane,
\begin{equation}
K_{12}(\bp_0)=n_1n_2U^*L^2\int_{A_c(Z)}\dot{\br}\bcdot\hat{\boldsymbol{u}}\;\dd A,\qquad \hat{\boldsymbol{u}}=\frac{\dU(\bp_0)}{|\dU(\bp_0)|},
\label{eq:Kflux}
\end{equation}
which, since $\nabla_{\br}\bcdot\dot{\br}=0$ at fixed orientation, equals $n_1n_2U^{*}L^{2}|\dU(\bp_0)|A_c(\bp_0)$, the integrand of (\ref{eq:K12capture}) without the weight $\psi_\infty$, for any $Z$ upstream of which the orientation of the spheroid remains equal to $\bp_0$. If the upstream dynamics has already changed the orientation by the time the pair reaches the starting plane, the orientation there is correlated with the impact parameter, and the exact flux through the plane would have to be evaluated with the pair density transported from infinity by (\ref{eq:Liouville}), rather than with the upstream density $\psi_\infty(\bp_0)$. In the present computations every trajectory is instead released on the starting plane with the orientation $\bp_0$ and the weight $\psi_\infty(\bp_0)$. This finite release is a controlled approximation: its error is set by the rotation of the spheroid upstream of the plane, which is estimated below for each encounter model, and its convergence with $Z$ is verified in Appendix~\ref{app:numerics}. The choice of the starting plane depends on the encounter model and on the population, because the disturbance of the sphere decays slowly. Far upstream, the relative velocity of a broadside pair consists of the uniform stream $\bpe\bg$ and a Stokeslet (\ref{eq:farfield}); at the distance $Z$ on the vertical through the sphere, the contribution from Stokeslet is smaller than the stream by the factor $2|\delta|/(|\bpe|Z)$. This ratio equals unity at $r_0=2|\delta|/|\bpe|$, the distance at which the interaction velocity equals the difference of settling speeds (\S\,\ref{sec:locusHI}). Near the extinction locus $\bpe\to0$, and consequently the value $r_0$ increases significantly. The simpler estimate of the kernel, $|\dU(\bp_0)|$ times the area of the capture set on the starting plane, neglects this correction and is therefore inadequate (Appendix~\ref{app:numerics}), whereas the flux (\ref{eq:Kflux}) does not depend on $Z$ provided that the orientation does not change upstream of the starting plane. The starting plane is accordingly chosen as follows.

At frozen orientation the spheroid does not rotate, so that the flux (\ref{eq:Kflux}) is exactly independent of $Z$. Under strong alignment the inertial torque holds the tilt at zero but leaves the azimuth free, and the sphere turns the spheroid about the vertical even far upstream. For a sphere offset by $b_1$ along $\bp$ and by $b_2$ across it, at a vertical distance $z\gg1$, the far-field rotation (\ref{eq:pdoth_far}) changes the azimuth $\phi$ of the axis at the rate
\begin{equation}
\dot\phi\simeq\frac{9\gamma}{8\lambda^{3}}\,\frac{b_1b_2}{z^{4}}.
\label{eq:pji_dot}
\end{equation}
Since the pair approaches at the speed $|\bpe|$, the azimuthal rotation accumulated upstream of a plane at $z=Z/L$ is
\begin{equation}
\Delta\phi\simeq\frac{3\gamma}{8\lambda^{3}|\bpe|}\,b_1b_2\left(\frac{L}{Z}\right)^{3}.
\label{eq:Delta_phi}
\end{equation}
Under strong alignment the flux is therefore independent of $Z$ only up to terms of relative order $(L/Z)^{3}$. For the capture sets of figure~\ref{fig:capture} at $Z=6L$ these terms amount to at most a few times $10^{-4}$, consistent with the change of at most $0.12\,\%$ found on moving the starting plane (Appendix~\ref{app:numerics}). For both encounter models it therefore suffices that the starting plane lie beyond the region in which the interaction dominates the stream. The plane is placed at $Z=\max(6L,2r_0)$, at which the interaction is at most half the difference of settling speeds.
 
For the free rotation population, the sphere tries to tilt the spheroid even far upstream due to hydrodynamic interactions, so that the orientation on any finite starting plane differs from $\bp_0$, and the flux depends on $Z$. The spheroid is therefore released from broadside on the starting plane at $Z$, and the efficiency for release at infinity is recovered from the release law
\begin{equation}
E(Z)\simeq E(\infty)\,\exp(-\Lambda L/Z),\qquad \Lambda=\frac{3\gamma(\Gpa-\Gpe)}{2\lambda^{3}\bpe^{2}},
\label{eq:release}
\end{equation}
derived in Appendix~\ref{app:release}. The length $\Lambda$ characterises the combined effect, far upstream, of the tilt induced by the sphere and of the resulting glide of the spheroid along its axis, which together narrow the capture set along $\bp_0$. It increases as $\bpe^{-2}$ near the extinction locus, where the approach is slow and the spheroid has longer to tilt and glide. The free-rotation efficiencies reported below are computed with release at $Z=24L$ and corrected by (\ref{eq:release}), and they are omitted where $\Lambda>2.5$.
 
For the Stokes population, every spheroid rotates freely, from an upstream orientation drawn at random. The orientational average is evaluated by six-point Gauss--Legendre quadrature in $|\bp_0\bcdot\bg|$, with trajectories started at $Z=24L$. The release law (\ref{eq:release}) holds only for spheroids released from broadside, and no correction for the finite starting distance is made. Moving the starting plane from $24L$ to $48L$ changes the efficiency of the Stokes population by at most $1.1\,\%$, which bounds the error of the finite release for this population.

Appendix~\ref{app:numerics} describes the numerical checks. With interactions switched off, the traced areas reproduce (\ref{eq:steiner}) to within $0.3\,\%$ ($0.5\,\%$ for inclined spheroids at $\kappa=20$). At $\lambda=\lambda_w$, where trajectories at frozen orientation are straight, from the first-reflection model, the computed efficiency is approximately $1.0002$, $1.0003$ and $0.9991$ for $\kappa=5$, $10$ and $20$, respectively, against the exact value of unity. Moving the starting plane changes the efficiencies by less than $0.2\,\%$ ($1.1\,\%$ for the Stokes population), and tracing with a half or a quarter of the rays changes them by less than $2\,\%$, except where stated.

\subsection{Collision efficiencies away from and below the extinction locus}\label{sec:awayHI}

\begin{figure}
  \vspace{5mm}
  \centering
  \panel[1.12]{fig:eff_a}{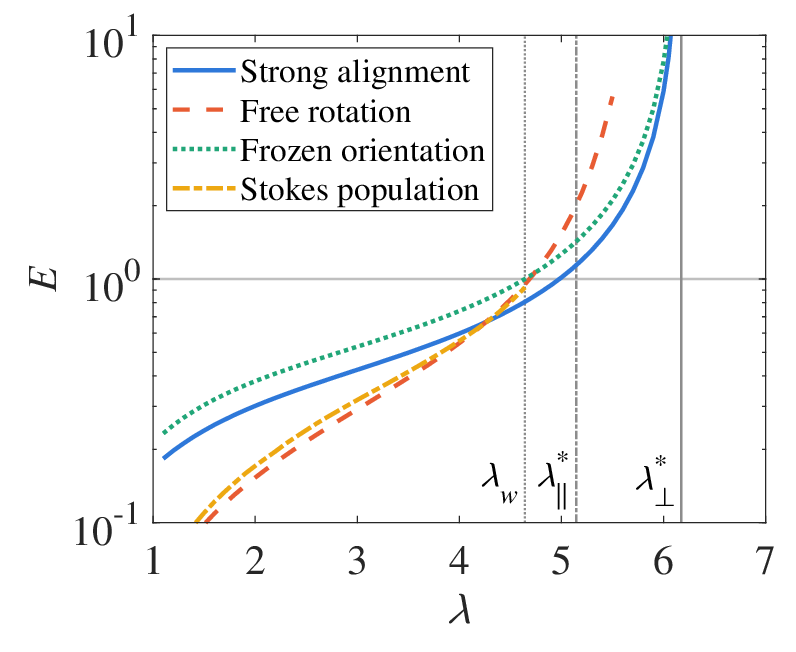}\hfill
  \panel[1.12]{fig:eff_b}{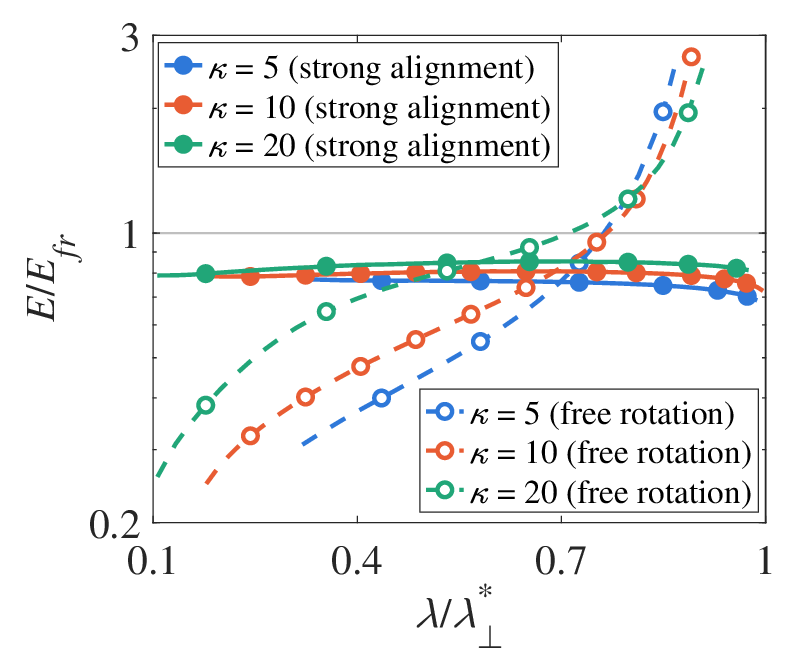}
  \caption{(\textit{a}) The variation of collision efficiency, $E$, with respect to the size ratio, $\lambda$, at $\kappa=10$ and $\gamma=1$ for the inertial population under strong alignment, free rotation extrapolated by (\ref{eq:release}) where $\Lambda\le2.5$, frozen orientation, and Stokes population. Vertical lines mark $\lambda_w$, $\lpa$ and $\lpe$. Between $\lambda\approx5.2$, just above $\lpa$, and the end of the free-rotation curve, the model has a stable steady configuration in which the spheroid settles vertically above the sphere (\S\,\ref{sec:accuracy}). The free-rotation efficiency there excludes the pairs trapped in this configuration, whose flux is $16\,\%$ of the colliding flux at $\lambda=5.5$ and reaches $94\,\%$ at $\lambda=5.2$. (\textit{b}) Ratio of collision efficiency, $E$, to its value at frozen orientation, $E_{fr}$, under strong alignment and with free rotation, for $\kappa=5$, $10$ and $20$ at $\gamma=1$.}
\label{fig:eff}
\end{figure}

\begin{figure}
  \centering
  \panel[1.2]{fig:capture_a}{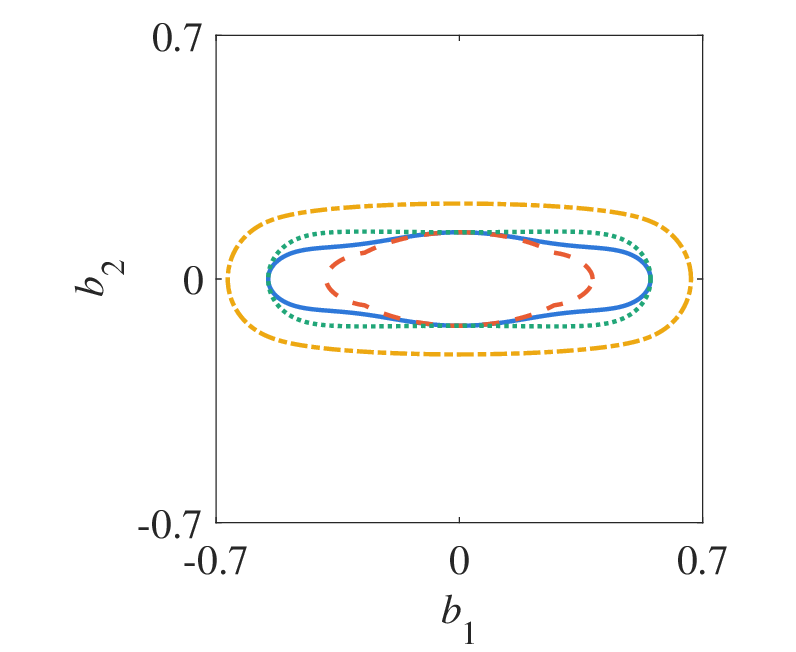}\hfill
  \panel[1.2]{fig:capture_b}{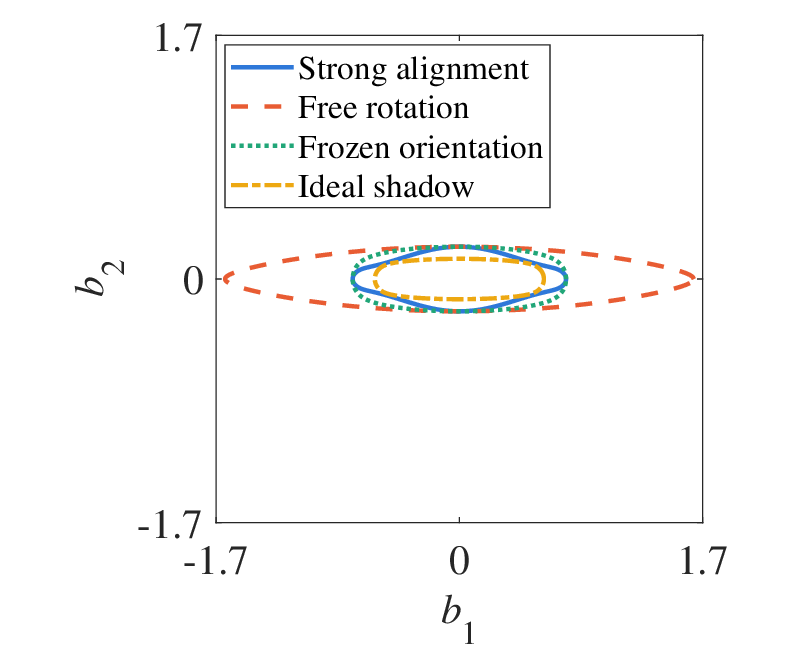}
  \caption{Capture sets of the inertial population under strong alignment, free rotation, frozen orientation on the starting plane at $Z=6L$ and its ideal shadow at broadside orientation, for $\kappa=10$ and $\gamma=1$: (\textit{a}) $\lambda=3$; (\textit{b}) $\lambda=5.5$.}
\label{fig:capture}
\end{figure}

\begin{table}
  \begin{center}
\def~{\hphantom{0}}\small\setlength{\tabcolsep}{3.2pt}
  \begin{tabular}{rcccccccccc}
      & & & \multicolumn{4}{c}{$E$} & & \multicolumn{3}{c}{$(K_{12}/K_S)\times10^{3}$}\\[2pt]
      $\kappa$ & $\lambda$ & $\lambda/\lpe$ & frozen & strong & free & Stokes & & strong & free & Stokes\\[3pt]
5 & 1.5 & 0.436 & 0.288 & 0.221 & 0.115 & 0.123 & & 79.1 & 41.2 & 37.8\\
 & 2.0 & 0.581 & 0.395 & 0.303 & 0.216 & 0.224 & & 33.3 & 23.8 & 20.6\\
 & 2.5 & 0.726 & 0.582 & 0.443 & 0.492 & 0.481 & & 18.2 & 20.2 & 16.1\\
 & 2.924$^{\dagger}$ & 0.849 & 1.000 & 0.747 & 1.964 & -- & & 11.9 & 31.2 & --\\
 & 3.2 & 0.929 & 2.047 & 1.490 & -- & -- & & 9.12 & -- & --\\[3pt]
10 & 1.5 & 0.243 & 0.304 & 0.239 & 0.098 & 0.111 & & 91.1 & 37.5 & 36.2\\
 & 2.0 & 0.324 & 0.381 & 0.301 & 0.153 & 0.171 & & 39.8 & 20.2 & 19.1\\
 & 3.0 & 0.486 & 0.527 & 0.424 & 0.292 & 0.319 & & 14.6 & 10.1 & 9.03\\
 & 4.0 & 0.648 & 0.740 & 0.598 & 0.546 & 0.564 & & 7.19 & 6.56 & 5.46\\
 & 4.642$^{\dagger}$ & 0.752 & 1.000 & 0.805 & 0.950 & 0.960 & & 4.90 & 5.79 & 4.64\\
 & 5.0 & 0.810 & 1.266 & 1.015 & 1.530 & -- & & 4.03 & 6.07 & --\\
 & 5.5 & 0.891 & 2.107 & 1.664 & 5.608 & -- & & 3.10 & 10.4 & --\\[3pt]
20 & 2.0 & 0.177 & 0.386 & 0.308 & 0.148 & 0.174 & & 41.5 & 20.0 & 19.8\\
 & 4.0 & 0.355 & 0.588 & 0.489 & 0.380 & 0.409 & & 7.89 & 6.13 & 5.38\\
 & 7.368$^{\dagger}$ & 0.653 & 0.999 & 0.853 & 0.923 & 0.876 & & 1.72 & 1.86 & 1.41\\
 & 9.0 & 0.798 & 1.573 & 1.336 & 1.903 & -- & & 1.01 & 1.43 & --\\
  \end{tabular}
  \caption{Interception efficiency and collision rate normalised by (\ref{eq:KS}), for $\gamma=1$ for inertial population with frozen orientation, under strong alignment and with free rotation, and Stokes population. The free-rotation values are extrapolated from release at $24L$ by (\ref{eq:release}) and omitted where $\Lambda>2.5$; the Stokes population is omitted near and between the two loci. The superscript symbol $^{\dagger}$ represents $\lambda=\lambda_w$, where the frozen efficiency is unity exactly.}
  \label{tab:eff}
  \end{center}
\end{table}

Figure~\ref{fig:eff_a} shows the interception efficiency $E$ at $\kappa=10$ as a function of the size ratio, for the inertial population under the three encounter models of \S\,\ref{sec:orientation} and for the Stokes population. Table~\ref{tab:eff} lists the corresponding values for $\kappa=5$, $10$ and $20$. The table also lists the collision rate itself, normalised by the Smoluchowski kernel (\ref{eq:KS}) of a pair of spheres, $K_{12}/K_S$. Since each efficiency is defined relative to the ideal kernel of its own population, the efficiencies of different populations cannot be compared directly, whereas $K_{12}/K_S$ measures all the rates against a common reference. It is given for the populations that can be realised physically; frozen orientation serves only to isolate the translational interaction, and only its efficiency is listed. The capture sets from which these efficiencies are obtained are shown in figure~\ref{fig:capture} for $\kappa=10$ and $\gamma=1$ at two size ratios, $\lambda=3$, well below the extinction locus, and $\lambda=5.5$, close to it. Each panel shows the capture sets on the starting plane at $Z=6L$, with the coordinate $b_1$ along the projection of $\bp_0$ and $b_2$ across it; the sets are computed in one quadrant and reflected about the two mirror planes of the pair. Each curve bounds a set of impact parameters whose trajectories reach contact: the ideal shadow (\ref{eq:steiner}), and the capture sets under frozen orientation, under strong alignment and with free rotation from broadside. Since the relative velocity on the starting plane differs from $\dU(\bp_0)$ only by the decaying hydrodynamic interaction, a boundary lying well inside the shadow indicates $E<1$ and one lying well outside it indicates $E>1$; the efficiencies themselves are obtained from the flux (\ref{eq:Kflux}), which accounts for the difference. The distortion of each boundary relative to the shadow shows in which direction the interactions act most strongly. Size ratios above the extinction locus are excluded here, since the efficiency, normalised by an ideal kernel that vanishes on the extinction locus, ceases to be a useful measure there; that region is examined in \S\,\ref{sec:locusHI} in terms of the collision rate itself.

\emph{Frozen orientation.} With the orientation held fixed at a broadside orientation, the interactions act only through the translational term $\delta\Wf$ in (\ref{eq:rdot}), whose sign is that of the difference between the buoyant weights of the two particles. Its effect can be clearly seen when the two particles are far apart, where the relative velocity reduces to the uniform relative settling velocity $\bpe\bg$, which acts as a stream carrying the spheroid past the sphere, plus a Stokeslet of strength proportional to $\delta$ (as discussed in \S\ref{sec:sbt}). For $\lambda<\lambda_w$ the sphere is the heavier particle, $\delta>0$, and the Stokeslet is directed against the stream. The relative trajectories then diverge around the contact surface, resulting in a slower approach, and trajectories that would graze the spheroid in the absence of interactions are deflected past it, so that $E<1$, as for spheres of unequal size \citep{Davis1984}. In physical terms, the flow generated by the heavier sphere carries the spheroid away from it more strongly than the flow generated by the spheroid carries the sphere towards it. The reduction is largest at small size ratios, where the sphere is heaviest relative to the spheroid, and at $\kappa=10$ the efficiency falls to approximately $0.30$ at $\lambda=1.5$ (table~\ref{tab:eff}). At $\lambda=\lambda_w$ the two weights are equal, the translational interaction vanishes identically, the trajectories are straight, and $E=1$ exactly, as shown in figure~\ref{fig:eff_a}. For $\lambda_w<\lambda<\lpe$ the spheroid is the heavier particle, $\delta<0$, and the Stokeslet is directed along the stream. The trajectories then converge towards the contact surface, the interaction draws the particles together, and the collision efficiency exceeds unity, i.e., $E>1$. This enhancement grows steeply towards the extinction locus. Above the locus ($\lambda>\lpe$) the stream reverses, $\bpe>0$, while $\delta$ remains negative, so that the Stokeslet again opposes the stream and the interaction once again hinders the approach (\S\,\ref{sec:locusHI}).


The capture sets of figure~\ref{fig:capture} show that the translational interaction does not simply rescale the shadow. Below $\lambda_w$ the boundary of the capture set for the frozen alignment lies inside the shadow and is contracted more strongly across $\bp_0$ than along it (figure~\ref{fig:capture_a}). Close to the locus the boundary of the capture set lies outside the shadow and is again distorted most across $\bp_0$ (figure~\ref{fig:capture_b}). The lateral displacement of the trajectories by the interaction is of the same order in the two directions, but the shadow is significantly narrower across the axis of the spheroid than along it, so that the relative change of the capture set, and hence its contribution to the change of the efficiency, is largest across the axis.

\emph{Strong alignment.} When the alignment is strong, the inertial torque holds the spheroid broadside but does not prevent its rotation about the vertical, so that the sphere can still turn the spheroid in azimuth. Let $\phi$ denote the angle between the axis $\bp$ and the horizontal offset of the pair. For a sphere above a broadside spheroid, the far field orientation rate due to the hydrodynamic interactions (\ref{eq:pdoth}) gives an azimuthal rotation rate $\dot\phi\propto(\br\bcdot\bg)\sin2\phi$, which turns the axis towards the direction perpendicular to the offset. As the spheroid turns, its end swings away from the path of the sphere, so that it extends towards that path by its thickness rather than its half-length, and a pair that would just collide under frozen orientation now misses. The spheroid thus behaves like a weathervane, orienting itself so as to present the least extent to the approaching sphere. On the two mirror planes, where $\sin2\phi=0$, the azimuthal torque vanishes and the capture boundary coincides with that under frozen orientation. When the azimuthal orientation of the spheroid stays between the two mirror planes, the boundary of the capture set lies inside the frozen one (figure~\ref{fig:capture}). Figure~\ref{fig:eff_b} shows the efficiency relative to its value under frozen orientation, $E/E_{fr}$, for $\kappa=5$, $10$ and $20$. Under strong alignment this ratio is nearly independent of the size ratio and lies between about $0.7$ and $0.85$, the reduction being weaker for more slender spheroids. Weathervaning therefore reduces the efficiency by an almost constant factor, while its variation with the size ratio is set, as under frozen orientation, by the translational interaction. The collision rate scaled by the Smoluchowski kernel, $K_{12}/K_S$, under strong alignment decreases monotonically with the size ratio for each aspect ratio (table~\ref{tab:eff}). It can be expressed as the product of the collision efficiency ($E$) and the ideal kernel of the aligned population ($K_b^{(0)}$), and the increase of the former towards the locus is outweighed by the decrease of the latter to zero.

\emph{Free rotation.} When the inertial torque is negligible, the sphere also tilts the spheroid out of the horizontal, and two competing effects arise. First, the disturbance flow of the sphere, which is directed downwards and is strongest close to the sphere, pushes the nearer end of the spheroid down more than the farther end. The axis therefore tilts towards the vertical, and the cross-section that the spheroid presents to the approaching sphere decreases: the spheroid feathers. Second, a tilted spheroid glides along its axis with the velocity $(\Gpa-\Gpe)\mu\,\bp$ towards the end that was pushed down and hence towards the vertical through the sphere. Feathering reduces the efficiency and gliding increases it, and their balance depends on the size ratio. Far below the extinction locus the approach is fast, the spheroid has little time to glide, and feathering dominates. At $\kappa=10$ and $\lambda=1.5$ the efficiency is about a third of its value under frozen orientation (figure~\ref{fig:eff_a}). The capture set of figure~\ref{fig:capture_a} shows the effect of the tilt directly. Along $\bp_0$ the free-rotation boundary is contracted well inside the frozen one, whereas across $\bp_0$, on the mirror plane (where no tilt is induced), it coincides with the frozen boundary. Near the extinction locus, by contrast, the difference of settling speeds $\bpe$ vanishes while the glide does not, so that gliding dominates and the capture set is drawn out along $\bp_0$ (figure~\ref{fig:capture_b}). At $\lambda=5.5$ and $\kappa = 10$ the collision efficiency is more than three times that under strong alignment (table~\ref{tab:eff}). This value, like the end of the free-rotation curve in figure~\ref{fig:eff_a}, excludes the pairs that the model holds in a steady configuration with the spheroid settling vertically above the sphere; this configuration, the fraction of pairs it traps and the criterion used to identify them are discussed in \S\,\ref{sec:accuracy} and Appendix~\ref{app:numerics}. Free rotation and strong alignment give the same efficiency a little below $\lambda_w$, where their curves cross in figure~\ref{fig:eff_b}. Feathering and gliding themselves balance where the free-rotation curve crosses $E/E_{fr}=1$, close to $\lambda_w$. Since the tilt that leads to gliding is acquired far upstream, the free-rotation efficiencies depend on the distance at which the spheroid is released from broadside, and the values in figure~\ref{fig:eff_a} and table~\ref{tab:eff} are corrected to release at infinity by (\ref{eq:release}). The competition between feathering and gliding is also apparent in the scaled collision rate $K_{12}/K_S$ (table~\ref{tab:eff}). Far below the extinction locus, the freely rotating spheroids collide significantly less compared to the strongly aligned ones. From about $\lambda_w$ onwards, however, they collide more often, and for $\kappa=5$ and $10$ the rate passes through a minimum and rises again towards the locus.

\emph{Stokes population.} In the Stokes population every spheroid rotates freely, as in the previous case, but it approaches the sphere from an orientation drawn at random far upstream rather than from broadside. Away from the two loci its efficiency lies close to the free-rotation value of the inertial population, from which it differs by less than a fifth (table~\ref{tab:eff}). The spheroids that start nearly broadside behave as the inertial population with free rotation, whereas those that start closer to the vertical collide more often. The disturbance flow of the sphere pushes the nearer end of a nearly vertical spheroid sideways, tilting it, and the spheroid then glides along its axis towards the sphere, so that the efficiency increases steadily from the broadside to the vertical members of the population. The interactions therefore modify the ratio of the collision rates of the two populations, which equals $\Rr$ (as discussed in \S\,\ref{sec:compare}) in their absence. Under strong alignment the ratio increases, since the aligned spheroids cannot feather, whereas the freely rotating spheroids of the Stokes population do. With free rotation the ratio of the collision rate of the inertial population to that of the Stokes population decreases, since both populations now feather, but only the Stokes population contains the nearly vertical spheroids whose glide towards the sphere raises its efficiency. Near and between the two loci, by contrast, some members of the Stokes population move very slowly relative to the sphere and are rotated appreciably long before they reach any finite starting plane. The computed kernel then depends on $Z$, and since the release law (\ref{eq:release}) applies only to spheroids released from broadside, it cannot be corrected; the Stokes population is therefore not computed there (table~\ref{tab:eff}). In terms of the collision rate itself, the Stokes population collides less often than the inertial population, under either strong alignment or free rotation, at every size ratio at which it is computed (table~\ref{tab:eff}). Its efficiency may exceed the free-rotation value, but its ideal kernel is smaller by the factor $\Rr$, and the latter effect prevails.

\subsection{Finite alignment rate}\label{sec:finiteRe}

At small but finite $Re_p$ the inertial torque acts throughout the approach of the spheroid to the sphere. Far from the sphere, where the interaction is weak, the tilt of the spheroid is then governed by two competing effects: the disturbance flow of the sphere tilts the spheroid at the rate (\ref{eq:tiltrate}) derived in Appendix~\ref{app:release}, whereas the inertial torque relaxes the tilt back to broadside at the rate $a_I$ of (\ref{eq:aI}). For a spheroid whose centroid lies at the vertical distance $z\gg1$ below the sphere and at the distance $b_1$ from the vertical through the sphere, measured along its axis (with the sense of $\bp$ chosen as in Appendix~\ref{app:release}), the two effects combine into
\begin{equation}
\dot\mu+a_I\,\mu=\frac{3\gamma b_1}{2\lambda^{3}z^{3}},
\label{eq:relax}
\end{equation}
with $\dot z=-|\bpe|$. The right-hand side of \eqref{eq:relax} is the forcing by the sphere, which grows as the spheroid approaches, and the term $a_I\mu$ is the damping by the inertial torque. Equation (\ref{eq:relax}) describes a forced relaxation, of the kind that \citet{ShaqfehKoch1990} solved along trajectories for fibres in an extensional flow. Its governing parameter is $kz$, with $k=a_I/|\bpe|$, the ratio of the time $z/|\bpe|$ needed to travel the distance $z$ to the relaxation time $1/a_I$; equivalently, $1/k$ is the distance travelled during one relaxation time. Forcing that acted more than a distance of order $1/k$ upstream has been damped out by the time the spheroid arrives, whereas forcing that acted closer has not. The release law (\ref{eq:release}) corresponds to an idealisation of this process, in which the tilt is suppressed entirely beyond the starting plane and not at all within it.

Since (\ref{eq:relax}) is linear in $\mu$, its solution for a spheroid that is broadside far upstream is the forcing accumulated along the approach, each contribution being reduced by the damping over the time elapsed since it acted. We measure this time by the distance $u=|\bpe|(t-t')$ travelled since then, so that the forcing acted at the distance $(z+u)$ and has since decayed by the factor $\mathrm{e}^{-ku}$. Now holding $b_1$ fixed in the forcing, we get
\begin{equation}
\mu(z)=\frac{3\gamma b_1}{2\lambda^{3}|\bpe|}\int_0^{\infty}\frac{\mathrm{e}^{-ku}}{(z+u)^{3}}\,\dd u ,
\label{eq:murelax}
\end{equation}
which reduces for $a_I=0$ to the free-rotation tilt $3\gamma b_1/(4\lambda^{3}|\bpe|z^{2})$ of Appendix~\ref{app:release}. The damping thus discounts the forcing that acted far upstream. The tilted spheroid glides towards the vertical through the sphere at the speed $(\Gpa-\Gpe)\mu$, as in (\ref{eq:glide}). Integrating the glide from infinity down to the distance $\ell$ below which the far-field form of the forcing in (\ref{eq:relax}) ceases to be valid, and comparing with free rotation, gives
\begin{equation}
\ln\frac{E(\infty)}{E(a_I)}=\Lambda\int_{\ell}^{\infty}\!\!\int_0^{\infty}\frac{1-\mathrm{e}^{-ku}}{(z+u)^{3}}\,\dd u\,\dd z=\Lambda\,k\,\mathcal{J}(k\ell),
\label{eq:relaxlaw}
\end{equation}
where
\begin{equation}
\mathcal{J}(k\ell)=\int_0^{\infty}\frac{x-1+\mathrm{e}^{-x}}{(k\ell+x)^{3}}\,\dd x=\tfrac12\,\mathrm{e}^{k\ell}E_1(k\ell).
\label{eq:mathcal_J}
\end{equation}
Here $E(a_I)$ is the collision efficiency at the alignment rate $a_I$, $E(\infty)$ is its free-rotation limit, $a_I\to0$, for release at infinity, and $E_1(k\ell)=\int_{k\ell}^{\infty}y^{-1}\mathrm{e}^{-y}\,\dd y$ is the exponential integral. In the double integral, the factor $(1-\mathrm{e}^{-ku})$ is the fraction of the forcing removed by the damping. The single integral $\mathcal{J}(k\ell)$ follows from the double integral on changing the variables to the distance $v=(z+u)$ at which the forcing acted and integrating first over $u$ at fixed $v$. The damping reduces the glide that brings the spheroid towards the vertical through the sphere, so that the width of the capture set along $\bp_0$, and with it the efficiency, decrease by the factor $\exp[-\Lambda k\mathcal{J}(k\ell)]$, while the width across $\bp_0$ is unchanged.

Now, two limits of the forced-relaxation law (\ref{eq:relaxlaw}) are considered. For weak damping, $k\ell\ll1$, which leads to $\mathcal{J}\simeq\tfrac12[\ln(1/k\ell)-\gamma_E]$, with $\gamma_E=0.5772$ as Euler's constant, so that $E(a_I)\to E(\infty)$ as $a_I\to0$. The logarithm reflects the continuous nature of the damping: the torque removes a fraction of order $kz$ of the tilt acquired at every distance $z$ between $\ell$ and $1/k$, and each of these scales contributes equally. A sharp release at the distance $Z$, by contrast, gives $\ln[E(\infty)/E]=\Lambda L/Z$, which equals $\Lambda k/2$ for $Z=2L/k$, without the logarithm. On the other hand, for strong damping, $k\ell\gg1$, which leads to $\mathcal{J}\simeq1/(2k\ell)$ and $\ln[E(\infty)/E(a_I)]\to\Lambda/(2\ell)$, which is the whole of the far-field glide of a freely rotating spheroid, (\ref{eq:binf}) with $z_n=\ell$. Therefore, in the strong damping limit, the whole of the tilt acquired in the far field gets suppressed. Like the release law (\ref{eq:release}), (\ref{eq:relaxlaw}) is first order in the glide, since $b_1$ is held fixed in the forcing, and it depends logarithmically on the cutoff $\ell$, which therefore plays the role of an effective parameter. We take $\ell$ equal to the length of the spheroid, $\ell=1$, the distance at which the far-field expansion of (\ref{eq:pdoth}) fails, and test the law against full computations below.

\begin{figure}
    \vspace{5mm}
  \centerline{\includegraphics[width=0.6\textwidth]{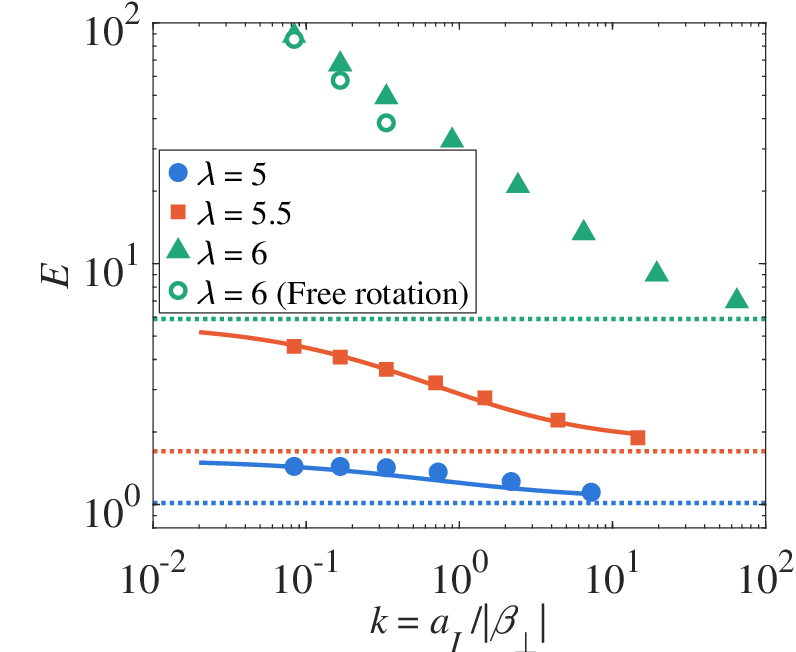}}
  \caption{The variation of collision efficiency of the inertial population at a finite alignment rate, for $\kappa=10$ and $\gamma=1$ at $\lambda=5$, $5.5$, and $6$, against $k=a_I/|\bpe|$. The computations with the full inertial rotation (\ref{eq:omegaI}) are represented by filled symbols, and the open symbols represent the collision efficiency for the free-rotation population released from broadside at $Z=2L/k$. Note that at $\lambda=6$, where $\Lambda=34$, the law does not apply and the free-rotation limit is not available. The solid lines represent the collision efficiencies obtained from the forced-relaxation law (\ref{eq:relaxlaw}) with $\ell=1$, starting from the free-rotation limit, at $\lambda=5$ and $5.5$. The dotted lines represent the collision efficiencies at the strong alignment limit for $\lambda=5$, $\lambda=5.5$, and $\lambda=6$. The range of $k$ differs between the size ratios because the same range of alignment rates, $1.3\times10^{-4}\le a_I\le0.1$, is divided by different values of $|\bpe|$.}
\label{fig:inertial}
\end{figure}

The law (\ref{eq:relaxlaw}) was tested against computations of the collision efficiency with the full inertial rotation (\ref{eq:omegaI}) added to the hydrodynamic one, for $\kappa=10$ at $\lambda=5$, $5.5$ and $6$ and for alignment rates between $a_I=1.3\times10^{-4}$ and $0.1$ (figure~\ref{fig:inertial}). Since the same range of $a_I$ is divided by a difference of settling speeds $|\bpe|$ that decreases towards the locus, it corresponds to larger values of $k$ at larger $\lambda$. In these computations the spheroid is released from broadside on a starting plane at $Z=6L/k$, restricted to $12L\le Z\le72L$. Where the upper limit is not reached, the memory of the starting orientation has decayed by the factor $\mathrm{e}^{-6}$ or less by the time the spheroid reaches the sphere. For $k<1/12$ the plane lies at $72L$ and the memory decays only by $\mathrm{e}^{-kZ/L}$; at $a_I=1.3\times10^{-4}$, for example, $kZ/L=0.68$ at $\lambda=5$ and $1.37$ at $\lambda=5.5$. Since the inertial torque can only reduce the tilt acquired upstream of the starting plane, the error incurred by releasing the spheroid at a finite distance cannot exceed that for free rotation. By the release law (\ref{eq:release}), the efficiencies computed at these points therefore differ from those for release at infinity by no more than the factor $\exp(-\Lambda L/Z)$, that is, by at most $1.0\,\%$ at $\lambda=5$ and $3.1\,\%$ at $\lambda=5.5$. The largest rates, $a_I\gtrsim10^{-2}$, correspond to $Re_p\gtrsim2$, beyond the range of validity of the small-$Re_p$ inertial torque; they test the model rather than the physics.

The law \eqref{eq:relaxlaw} agrees closely with the computations as long as the glide remains moderate. At $\lambda=5$, where $\Lambda=0.73$, the error of the law relative to the computed efficiency is at most $1.1\,\%$ for $k\le0.17$, and at $\lambda=5.5$, where $\Lambda=2.2$, it is at most $0.5\,\%$ for $k\le0.33$. These efficiencies exclude the pairs that end in a steady configuration of the model (\S\,\ref{sec:accuracy}). At larger $k$ the damping is no longer confined to the far field, since $k\ell$ is no longer small, and it modifies the near-field rotation that the law does not describe. The error then grows to $8\,\%$ and $4\,\%$ at $\lambda=5$ and $\lambda=5.5$, respectively, and the efficiency approaches its strong-alignment value only slowly as $k$ increases as shown in figure~\ref{fig:inertial}.

Closer to the extinction locus $(\lambda_\perp^*)$ the glide is no longer a small correction, and the law ceases to apply. At $\lambda=6$, $\Lambda=34$, so that the far-field glide alone would change the distance of the spheroid from the vertical through the sphere by a factor of order $\mathrm{e}^{\Lambda/2}$, and the first-order treatment fails. Moreover, the free-rotation limit from which the law starts is not available there (\S\,\ref{sec:awayHI}). The computed efficiency is instead close to that of free rotation released from broadside at the distance $Z=2L/k$, which represents the damping by a sharp release. For $Re_p\lesssim0.1$ the computed efficiencies remain very large (around $50$ to $90$ as shown in the figure~\ref{fig:inertial}), and the release model reproduces them with an error of at most about $30\,\%$. The release model slightly underestimates the computed efficiencies, and the underestimate increases with $k$. At larger alignment rates the efficiency decreases steadily towards its strong-alignment value of $5.9$, which it approaches only slowly, being still about $20\,\%$ above it at the largest rate computed. Figure~\ref{fig:locus_HI_a} shows the collision rate normalised by the Smoluchowski kernel, $K_{12}/K_S$, against the size ratio near the extinction locus for $\kappa=10$, for the ideal kernels of both populations and for the inertial population under strong alignment, with free rotation and at finite alignment rates. At $\lambda=6$, the efficiencies computed for $Re_p\lesssim0.1$ correspond to $K_{12}/K_S$ between $0.020$ and $0.036$, the larger values being attained at the smaller alignment rates. These rates are $8$ to $15$ times the strong-alignment result at the same size ratio, and $40$ to $80$ times the ideal isotropic kernel. Near the locus, therefore, a slow alignment fails to suppress the glide of the spheroid, and the glide raises the collision rate far above the strong-alignment result and the ideal isotropic kernel.

Again, the forced-relaxation law (\ref{eq:relaxlaw}) also delimits the range in which the free-rotation results apply near the locus. Free rotation is a good approximation only while the damping leaves the glide essentially intact, that is, while $\Lambda k\ln(1/k)\ll1$. Near the locus both $\Lambda$ and $k$ are governed by the vanishing difference of settling speeds, $\bpe\propto(\lpe-\lambda)$, so that $\Lambda\propto(\lpe-\lambda)^{-2}$ and $k\propto(\lpe-\lambda)^{-1}$. According to this weak-glide estimate, the condition therefore fails in a band below the locus whose width scales as $(\lpe-\lambda)\propto[a_I\ln(1/a_I)]^{1/3}$, that is, as $Re_p^{1/3}$ up to a logarithm. Outside this band the free-rotation results of \S\,\ref{sec:awayHI} apply. Inside it the efficiency depends on $Re_p$, and the free-rotation results no longer apply. The free-rotation efficiency itself would diverge at the extinction locus if the far-field glide law of Appendix~\ref{app:release} held up to it, since by \eqref{eq:binf} the capture set is drawn out along $\bp_0$ by the factor $\exp[\Lambda/(2z_n)]$, with $\Lambda\propto(\lpe-\lambda)^{-2}$. This divergence is a property of the far-field law, which is first order in the glide and assumes $b_1\ll z$. Once the glide displaces the spheroid by a distance comparable with its distance from the sphere, the tilting rate no longer scales as $b_1/z^{3}$, and the growth of the efficiency must slow. Finally, the strong-alignment limit, in which the spheroid is held broadside throughout the encounter, is not attained at small $Re_p$, however slow the approach to the sphere. In addition to suppressing the tilt acquired far upstream, it requires the inertial torque to dominate the hydrodynamic torque near contact, i.e., it requires $a_I\gg\Omega_h$, where the sphere tilts the spheroid at the rate $\Omega_h$. Since $a_I/\Omega_h<0.5\,Re_p$ (\S\,\ref{sec:orientation}), this condition cannot be met when $Re_p$ is small, and strong alignment is therefore a limiting model rather than a regime that small particles reach.

\subsection{Collision efficiencies near and beyond the extinction locus}\label{sec:locusHI}

\begin{figure}
  \vspace{5mm}
  \centering
  \panel[1.1]{fig:locus_HI_a}{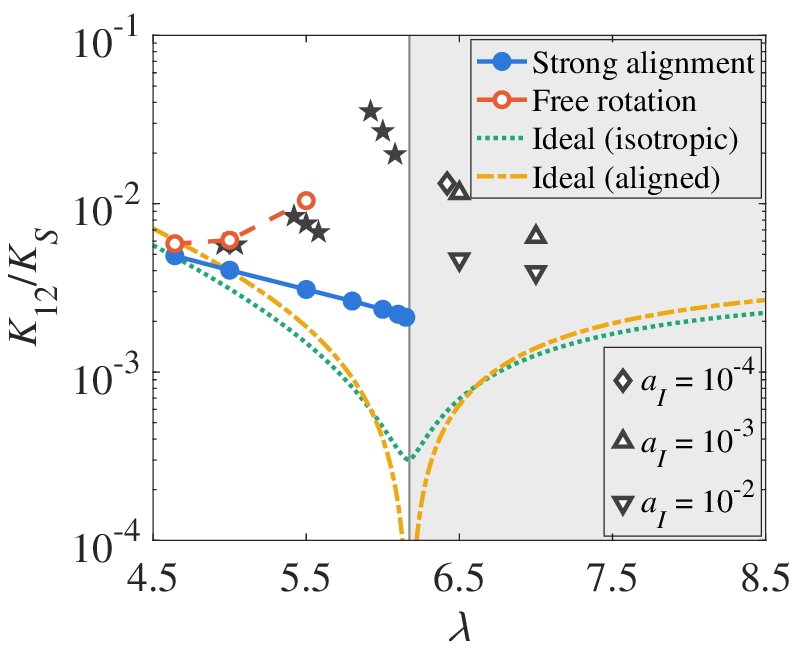}\hfill
  \panel[1.1]{fig:locus_HI_b}{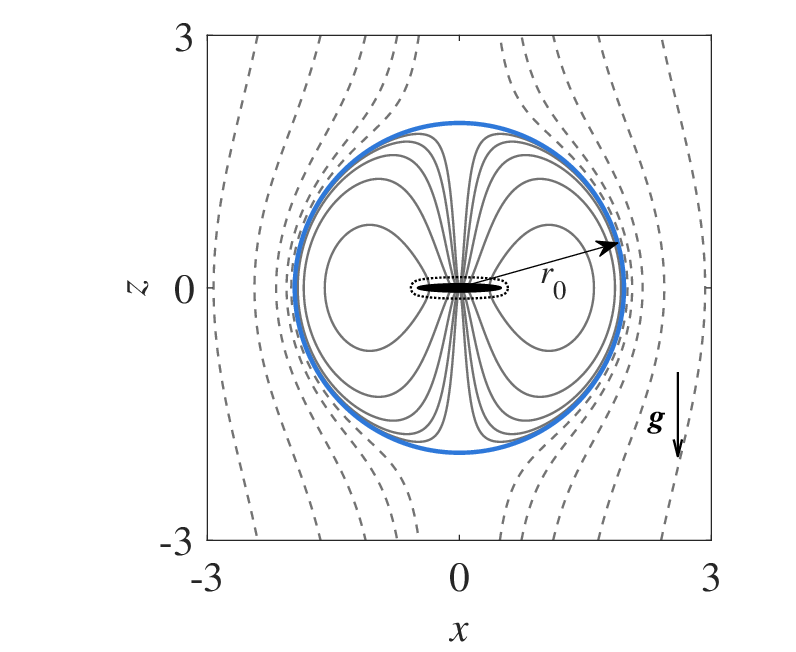}
  \caption{(\textit{a}) The variation of collision rate normalised by the Smoluchowski kernel \eqref{eq:KS} with respect to the size ratio for the ideal kernels of the isotropic and broadside populations, under strong alignment, and free rotation population (where $\Lambda\le2.5$) near the extinction locus when $\kappa=10$, and $\gamma=1$. The star-shaped symbols represent the normalised collision rate below the extinction locus for finite alignment rates corresponding to $k=0.083$, $0.17$, and $0.33$; where $a_I\le3\times10^{-3}$, that is $Re_p\lesssim0.6$. Above the extinction locus (shaded) no collisions are resolved under strong alignment up to $\lambda=9$; open diamond, open triangles, and open inverted triangles represent the normalised collision rates at finite alignment rates $a_I=10^{-4}$ ($Re_p\approx0.02$), $a_I=10^{-3}$ ($Re_p\approx0.2$), and $10^{-2}$ ($Re_p\approx2$), respectively. (\textit{b}) Streamlines of the far-field relative flow (\ref{eq:Psi}) in the vertical plane through $\bp_0$ at $\lambda=6.48$ (where, $\lambda/\lpe=1.05$). The dividing sphere $r=r_0=1.96L$ encloses the pair. Outside the dividing sphere, the streamlines are open (shown as dashed lines). The closed streamlines inside $r_0$ (shown as solid lines) are those of the far-field form and are indicative only. The spheroid (black) and the contact surface (dotted) are drawn to scale; since (\ref{eq:farfield}) is even in $\br$, the streamlines are also the paths of the centre of the sphere relative to the spheroid.}
\label{fig:locus_HI}
\end{figure}

On the extinction locus $(\lpe)$ the difference of settling speeds of the inertial population vanishes, but the interaction does not (\S\,\ref{sec:locus}). Far from the spheroid (\S\,\ref{sec:sbt}) the relative velocity (\ref{eq:rdot}) of a broadside pair reduces to a uniform stream plus a Stokeslet,
\begin{equation}
\dot{\br}\to\bpe\,\bg+\delta\,\tJ(\br)\bcdot\bg,
\label{eq:farfield}
\end{equation}
both axisymmetric about the vertical through the sphere. With $\Theta$ the angle between $\br$ and $-\bg$ and $r=|\br|$, its Stokes stream function is
\begin{equation}
\Psi=-r\sin^{2}\Theta\left(\tfrac12\bpe\,r+\delta\right).
\label{eq:Psi}
\end{equation}
When $\bpe$ and $\delta$ have opposite signs, $\Psi$ also vanishes on the sphere
\begin{equation}
r=r_0=\frac{2|\delta|}{|\bpe|},
\label{eq:r0}
\end{equation}
which is then a stream surface of the far-field flow (figure~\ref{fig:locus_HI_b}). The structure of the far-field flow near the locus differs qualitatively from that for a pair of spheres. For two spheres of equal density, equal settling speeds imply equal weights, so that the difference of settling speeds and the interaction vanish together; $r_0$ then remains of the order of the particle size, within the near field. For the spheroid--sphere pair, by contrast, the two zeros are separated: the difference of settling speeds of a broadside pair vanishes on the extinction locus $\lpe$, whereas the interaction vanishes at the equal-weight ratio $\lambda_w<\lpe$. Just below the locus, for $\lambda_w<\lambda<\lpe$, both $\bpe$ and $\delta$ are negative, the Stokeslet assists the stream, and every streamline from infinity is open. Just beyond the extinction locus, $\bpe>0>\delta$, the Stokeslet opposes the stream, and the dividing sphere (\ref{eq:r0}) appears. Since $\bpe$ vanishes on the locus whereas $\delta$ does not, its radius $r_0$ grows without bound as $\lambda\downarrow\lpe$.

This far-field picture holds only as far as the Stokes equations, and the description of an isolated pair, remain valid. At small but finite $Re_p$, fluid inertia modifies the Stokeslet beyond the Oseen length $\ell_O\sim L/Re_p$, and beyond the mean spacing $n_2^{-1/3}$, the distance at which a third particle is typically encountered, other particles intervene and the pair description no longer applies. Since $r_0$ is measured in units of $L$, the dividing sphere is therefore meaningful only if its dimensional radius lies well within both distances, $r_0L\ll\ell_O$ and $r_0L\ll n_2^{-1/3}$. Near the locus the difference of settling speeds varies linearly with the distance from it. Expanding $\bpe=(\Gpe-\gamma/\lambda^{2})$ about $\lpe$, where $\gamma/{\lpe}^{2}=\Gpe$, gives $|\bpe|\simeq2\Gpe|\lambda-\lpe|/\lpe$, so that the radius of the dividing sphere diverges as
\begin{equation*}
r_0\simeq\frac{|\delta|\,\lpe}{\Gpe\,|\lambda-\lpe|}.
\end{equation*}
The first condition, $r_0\ll\ell_O/L\sim Re_p^{-1}$, holds far from the locus but fails close to it, where $r_0$ becomes comparable with $Re_p^{-1}$. Setting $r_0\sim Re_p^{-1}$ shows that the dividing sphere of the Stokes problem is not realised within the narrow band
\begin{equation}
\lpe<\lambda\lesssim\lpe\left(1+Re_p\,\frac{|\delta|}{\Gpe}\right),
\label{eq:Oseen_band}
\end{equation}
above the locus, whose relative width $Re_p|\delta|/\Gpe$ is $0.08\,Re_p$ at $\kappa=10$. Within this band fluid inertia modifies the far-field flow, and the dividing sphere of the Stokes problem is not realised. However, this band, \eqref{eq:Oseen_band}, is narrow. For $Re_p\lesssim0.2$ it is several times narrower than the band in which the alignment rate controls the efficiency (\S\,\ref{sec:finiteRe}), so that the Oseen correction does not affect the conclusions drawn below.

\emph{Below the extinction locus.} As $\lambda\uparrow\lpe$ the difference of settling speeds of a broadside pair vanishes, but the interaction does not, since the weight difference $\delta$ vanishes only at $\lambda_w(<\lpe)$. The relative velocity (\ref{eq:rdot}) of a broadside spheroid therefore tends to $\delta_*\Wf$, where $\delta_*=3(\gamma{\lpe}^{-3}-\kappa^{-2})/8$ is the value of $\delta$ on the locus. The particles then still approach each other, driven entirely by the disturbance flow that each induces at the position of the other. Far from the pair this relative flow reduces to a Stokeslet, whose streamlines are open: they originate at infinity, approach the sphere and return to infinity, and those that intersect the contact surface carry pairs from infinity into contact. The flux into the contact surface (\ref{eq:K12flux}) therefore remains finite. Since it is linear in the relative velocity, the factor $|\delta_*|$ may be taken out, and
\begin{equation}
K_{12}\to n_1n_2\,U^{*}L^{2}\,|\delta_*|\,\mathcal{G}(\kappa),
\label{eq:Kfloor}
\end{equation}
where $\mathcal{G}$ is the inward flux of $\Wf$ alone through the contact surface of a broadside spheroid at $\lambda=\lpe$, along trajectories from infinity. It depends only on the geometry of the pair at the locus and on the encounter model, and it is of order unity, since $\Wf$ and the contact surface are both of order unity in units of $L$. The kernel thus keeps a floor on the locus, sustained by the weight difference of the pair rather than by the difference of their settling speeds. Since the ideal kernel of the aligned population vanishes there, the efficiency $E$ diverges as $\lambda\uparrow\lpe$, which is why the collision rate itself, rather than $E$, is used near the locus.
Under strong alignment at $\kappa=10$ the computed collision rate decreases steadily towards the locus, and its extrapolation to the locus gives $\mathcal{G}\approx1$ (table~\ref{tab:floor}). The floor decreases as the spheroid becomes more slender, and at every $\kappa$ it is higher with frozen orientation, for which the spheroid cannot weathervane. Although the ideal kernel of the aligned population vanishes on the locus, the aligned floor exceeds the isotropic ideal floor (\ref{eq:Kisolocus}) there, by a factor of about three at $\kappa=5$, seven at $\kappa=10$ and fifteen at $\kappa=20$. This ratio grows roughly in proportion to $\kappa$, because the two floors are set by different cross-sections of the spheroid: the isotropic floor by its end-on cross-section, of order $a^{2}$, and the aligned floor by the flux of the Stokeslet through its broadside section, of order $La$.

\begin{table}
  \begin{center}
    \begin{tabular}{l@{\hspace{2em}}ccc@{\hspace{2em}}ccc}
    & \multicolumn{3}{c}{$\mathcal{G}$} & \multicolumn{3}{c}{$(K_{12}/K_S)\times10^{3}$}\\[2pt]
    & $\kappa=5$ & $\kappa=10$ & $\kappa=20$ & $\kappa=5$ & $\kappa=10$ & $\kappa=20$\\[3pt]
    Strong alignment & 1.43 & 0.99 & 0.69 & 6.94 & 2.07 & 0.50\\
    Frozen orientation & 2.13 & 1.38 & 0.88 & 10.3 & 2.88 & 0.64\\
    Frozen orientation, estimate (\ref{eq:Gest}) & 2.28 & 1.46 & 0.92 & 11.1 & 3.05 & 0.67\\[3pt]
    Ideal kernel, isotropic population & 0.45 & 0.14 & 0.046 & 2.19 & 0.30 & 0.034\\
  \end{tabular}
  \caption{The floor of the collision rate below the extinction locus, for $\gamma=1$, computed under strong alignment and with frozen orientation, and estimated from (\ref{eq:Gest}); the last row gives, for comparison, the ideal floor (\ref{eq:Kisolocus}) of the isotropic population. The floor is given both as $\mathcal{G}=\lim_{\lambda\uparrow\lpe}K_{12}/(n_1n_2U^{*}L^{2}|\delta_*|)$ of (\ref{eq:Kfloor}) and as the collision rate normalised by the Smoluchowski kernel (\ref{eq:KS}) on the locus, $K_{12}/K_S=\mathcal{G}\,|\delta_*|/\Kbar_S$, where $\Kbar_S=K_S/(n_1n_2U^{*}L^{2})$ is evaluated at $\lambda=\lpe$.}
  \label{tab:floor}
  \end{center}
\end{table}

On the horizontal plane through the centre of the sphere, the interaction velocity of a broadside spheroid is vertical. If the flow enters the upstream half of the contact surface everywhere, its flux into that half equals, by continuity, its flux through the horizontal section $\mathcal{A}$ of the contact surface. By the symmetry of a broadside pair about this plane, $\mathcal{A}$ coincides with the broadside shadow of \S\,\ref{sec:ideal}, of area $A_b$ given in (\ref{eq:Ab_ICR}). On this plane $\br\bcdot\bg=0$ and $\mu=0$, so that, by (\ref{eq:Wclosed}), the Stokeslet part of $\Wf$ reduces to $\Iint01\bg$, with $\Iint01$ the moment defined in (\ref{eq:Imn}). Retaining only this part gives, for frozen orientation,
\begin{equation}
\mathcal{G}\approx\iint_{\mathcal{A}}\Iint01\,\dd A .
\label{eq:Gest}
\end{equation}
This estimate exceeds the computed values under frozen orientation by only a few per cent (table~\ref{tab:floor}), which confirms that the floor is carried by the far-field Stokeslet. Under strong alignment, weathervaning turns the axis of the spheroid away from the path of the sphere and lowers the floor further, by a third at $\kappa=5$ and by a fifth at $\kappa=20$.

\emph{Beyond the extinction locus.} Here the dividing sphere (\ref{eq:r0}) is a barrier for the far-field flow at fixed orientation. We now examine whether it also blocks the full dynamics, in which the near field and the rotation of the spheroid act as well. Under strong alignment, we mapped the closest approach $h_{min}$ of trajectories from infinity over the impact plane, refining the map around its smallest values and then minimising locally (Appendix~\ref{app:numerics}). At $\kappa=10$ no trajectory reaches contact for $\lambda\le9$, and at $\kappa=5$ none reaches contact at the two size ratios computed, $\lambda=3.5$ and $4.5$ ($\lambda/\lpe=1.02$ and $1.31$). Close to the extinction locus the closest approach agrees with the gap left between the dividing sphere and the tips of the spheroid, $(r_0-c-r_s)$, which confirms that the far-field barrier controls the encounter. As $\lambda$ increases the dividing sphere shrinks, and the closest approach decreases with it. It becomes smallest on the mirror plane that contains $\bp_0$, where the azimuthal torque vanishes and the spheroid cannot weathervane; even slightly off this plane, the closest approach is markedly larger. A capture set could in principle hide between the points of the map. Where $h_{min}$ changes abruptly between neighbouring points, the trajectories divide around a stagnation point of the relative flow, and refining these jumps by bisection to a width of order $10^{-10}L$ reveals no smaller approach. The only trajectory that reaches contact is the exact head-on one, a set of measure zero that carries no flux: it runs along the vertical through the sphere into the front stagnation point of the dividing sphere, from which numerical error carries it around the dividing sphere to the rear stagnation point and into contact. Collisions resume only once the sphere is no thicker than the spheroid, $\lambda\gtrsim\kappa$, and then only within a thin layer about the mirror plane that contains $\bp_0$, on which the azimuthal torque vanishes by symmetry; the resulting efficiencies are of order $10^{-2}$ at most, and such layers persist up to the largest size ratio computed. This upper edge of the blocked band is set by the near field, where the first reflection is least accurate (\S\,\ref{sec:accuracy}).

With free rotation, or at a finite alignment rate, the barrier fails. Outside the dividing sphere the sphere tilts the spheroid at a rate of order $\tfrac38\gamma\lambda^{-3}r^{-2}$, and the time spent in this region grows as $1/\bpe$, while its extent $r_0$ grows in the same proportion. The tilt accumulated during the approach,
\begin{equation*}
\int_{r_0}^{\infty}\frac{3\gamma}{8\lambda^{3}}\,\frac{\dd r}{r^{2}\bpe}=\frac{1}{2\left[\lambda^{3}/(\gamma\kappa^{2})-1\right]},
\end{equation*}
is therefore independent of $\bpe$ and remains of order unity near the extinction locus. The spheroid thus reaches the dividing sphere appreciably tilted, and its glide along its axis, which does not vanish with $\bpe$, overwhelms the vanishing difference of settling speeds. With free rotation, a grid search accordingly finds collisions, from irregular capture sets that cover a substantial fraction of the impact plane. At a finite alignment rate the capture sets are also irregular, but their flux is well defined. For $\kappa=10$, $\gamma=1$ and $a_I=10^{-3}$, which corresponds to $Re_p\approx0.2$, the collision rate at $\lambda/\lpe=1.05$ and $1.13$ is about six and three times the floor (\ref{eq:Kfloor}), respectively, that strong alignment sustains just below the locus. It also exceeds by more than an order of magnitude the isotropic floor (\ref{eq:Kisolocus}), that is, the ideal collision rate of the isotropic population on the extinction locus, which remains finite there because tilted spheroids still settle at speeds different from that of the sphere. At $a_I=10^{-3}$ the capture sets extend to $4.8L$ from the vertical through the sphere at $\lambda/\lpe=1.05$ and to $1.8L$ at $\lambda/\lpe=1.13$, well beyond the dividing sphere. The tilted spheroid glides along its axis and thereby draws in spheres that are offset in that direction. A smaller alignment rate raises the collision rate further. At $a_I=10^{-4}$ ($Re_p\approx0.02$), which is close to the alignment rate of the example in \S\,\ref{sec:example}, the rate at $\lambda/\lpe=1.05$ reaches $K_{12}/K_S=1.3\times10^{-2}$, or $6.4$ times the floor, and the capture set extends to $8.3L$. A larger alignment rate lowers it. At $a_I=10^{-2}$ ($Re_p\approx2$) the rates are lower than at $a_I=10^{-3}$ by factors of $2.5$ at $\lambda/\lpe=1.05$ and $1.6$ at $\lambda/\lpe=1.13$, and both lie at about twice the strongly aligned floor (figure~\ref{fig:locus_HI_a}). At $a_I=10^{-1}$ they fall below that floor. The blocking of collisions above the locus is therefore a consequence of the strong-alignment limit, in which the tilt of the spheroid is suppressed throughout the encounter. At small $Re_p$ the alignment rate $a_I$ is small, the tilt induced by the sphere persists, and collisions continue above the locus. For $a_I\gtrsim10^{-3}$ their rate decreases as $a_I$ increases and the spheroid is held closer to the broadside orientation; at smaller $a_I$ it changes by less than $15\,\%$.

These results hold at a fixed distance from the locus. The estimate of the tilt given above integrates the hydrodynamic rotation without damping. At a finite alignment rate, the spheroid spends a time of order $r_0/|\bpe|$ outside the dividing sphere, and the alignment accumulated during this outer approach is measured by
\begin{equation*}
D_{out}=\frac{a_I\,r_0}{|\bpe|}=\frac{2a_I|\delta|}{\bpe^{2}}.
\end{equation*}
For $D_{out}\ll1$ the spheroid reaches the dividing sphere with the tilt of free rotation. For $D_{out}\gg1$, alignment reduces that tilt by a factor of order $D_{out}^{-1}$, even though $a_I\ll\Omega_h$ near contact. The computations of figure~\ref{fig:locus_HI_a} span both regimes: $D_{out}$ is small at the smallest alignment rates and exceeds unity at $a_I=10^{-2}$ for both size ratios. The collision rate falls as $D_{out}$ increases, but collisions persist up to $D_{out}\approx72$ ($a_I=10^{-1}$ at $\lambda/\lpe=1.05$), where alignment also dominates the hydrodynamic rotation near contact ($a_I/\Omega_h=6$). At a fixed $Re_p$, however, $D_{out}$ grows without bound as $\lambda\to\lpe$, because $|\bpe|$ vanishes on the locus while $\delta$ does not. With $|\bpe|\simeq2\Gpe(\lambda/\lpe-1)$ and $a_I=F_b\Gpe Re_p$, at $\kappa=10$, $D_{out}$ exceeds unity for
\begin{equation*}
\frac{\lambda}{\lpe}-1\lesssim\left(\frac{F_b\,|\delta^{*}|\,Re_p}{2\,\Gpe}\right)^{1/2}\approx0.09\,Re_p^{1/2},
\end{equation*}
where $\delta^{*}$ is the value of $\delta$ on the extinction locus. At $Re_p=0.02$ this band, $(\lambda/\lpe-1)\lesssim0.013$, is wider than the Oseen band (\ref{eq:Oseen_band}). All the points computed at small $Re_p$ lie outside it, and whether collisions continue within it is not established here.

\subsection{Accuracy of the model and the effect of the gap}\label{sec:accuracy}

The efficiencies above are interception efficiencies of a first-reflection slender-body model. We now test the model and add the physics of the gap.

\emph{Slender-body truncation.} The slender-body closure was tested by replacing it with exact results for a spheroid, while the first-reflection approximation was retained. We replaced the line of Stokeslets of the spheroid by its exact singularity solution: uniform Stokeslets and parabolically weighted potential doublets on the focal segment \citep{ChwangWu1975}. We also replaced the slender-body law for the translation of the spheroid, the first of (\ref{eq:sbtresponse_u}), by the exact Fax\'en law of a spheroid \citep{Kim1985,ClaeysBrady1993},
\begin{equation}
\boldsymbol{U}=\frac{1}{2c_f}\int_{-c_f}^{c_f}\left[1+\frac{1-e^{2}}{4e^{2}}\left(c_f^{2}-\xi^{2}\right)\nabla^{2}\right]\boldsymbol{u}(\br+\xi\bp)\,\dd\xi,\qquad c_f=ec,
\label{eq:faxen}
\end{equation}
which keeps the weight-difference form (\ref{eq:wrel}) exactly, by reciprocity. The resulting changes in the collision efficiencies are small. At frozen orientation and $\kappa=10$ they range from $+0.3$ to $+0.8\,\%$ below $\lambda_w$ and from $-0.4$ to $-1.6\,\%$ above it, and they do not exceed $2.3\,\%$ for $\kappa=5$ and $0.4\,\%$ for $\kappa=20$. Under strong alignment the changes are the same to within $0.1\,\%$. The slender-body closure of the translational interaction is therefore accurate within the first-reflection model, even for $\kappa=5$. The slender-body law (\ref{eq:sbtresponse_p}) for the rotation was retained. Its error is also of relative order $\kappa^{-2}$: in a linear flow, the rate at which the strain rotates an exact spheroid is smaller than its slender-body value by the Bretherton constant $\mathcal{B}$ of \S\,\ref{sec:sbt}, which is $0.92$ for $\kappa=5$ and $0.98$ for $\kappa=10$.

\begin{figure}
  \vspace{5mm}
  \centering
  \panel[1.1]{fig:accuracy_a}{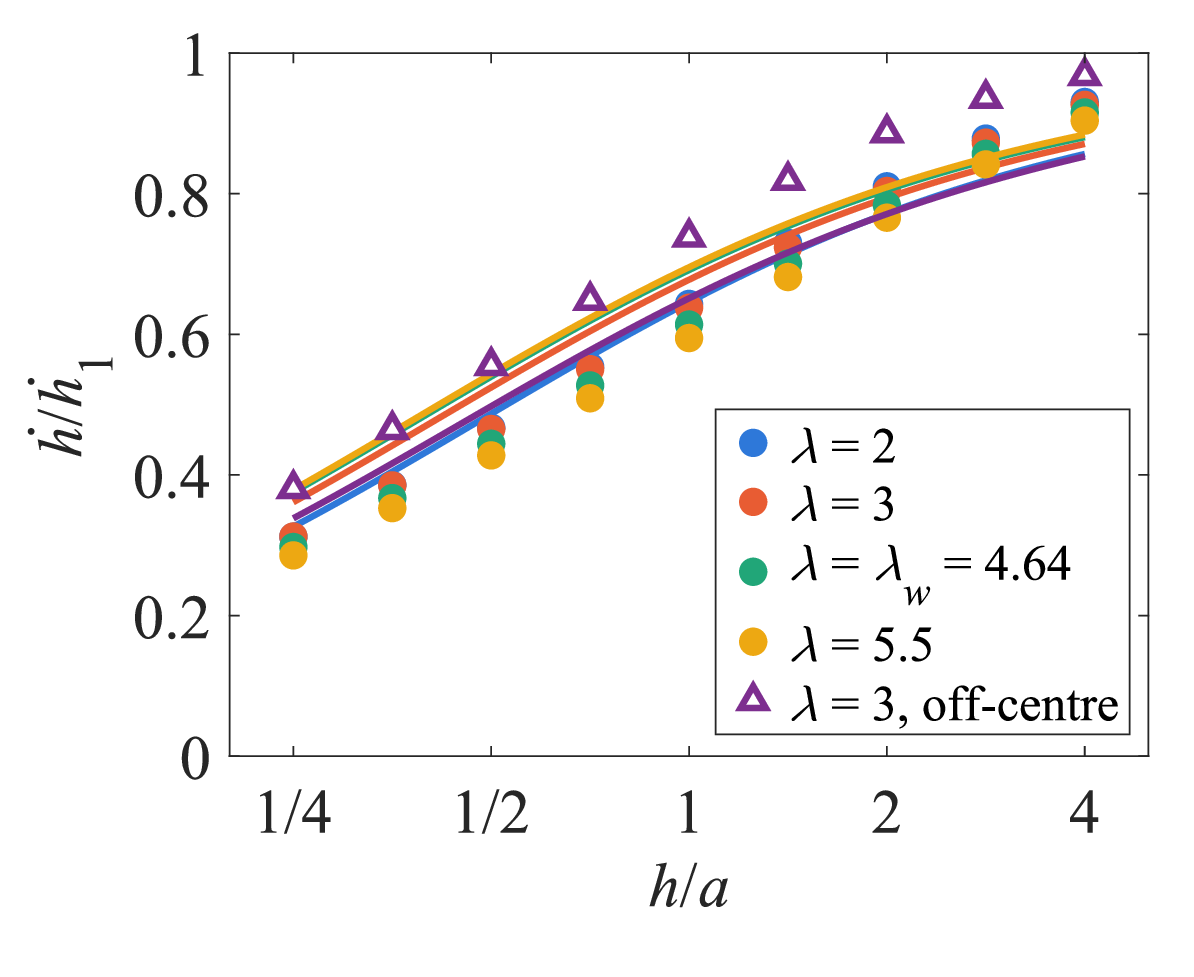}\hfill
  \panel[1.1]{fig:accuracy_b}{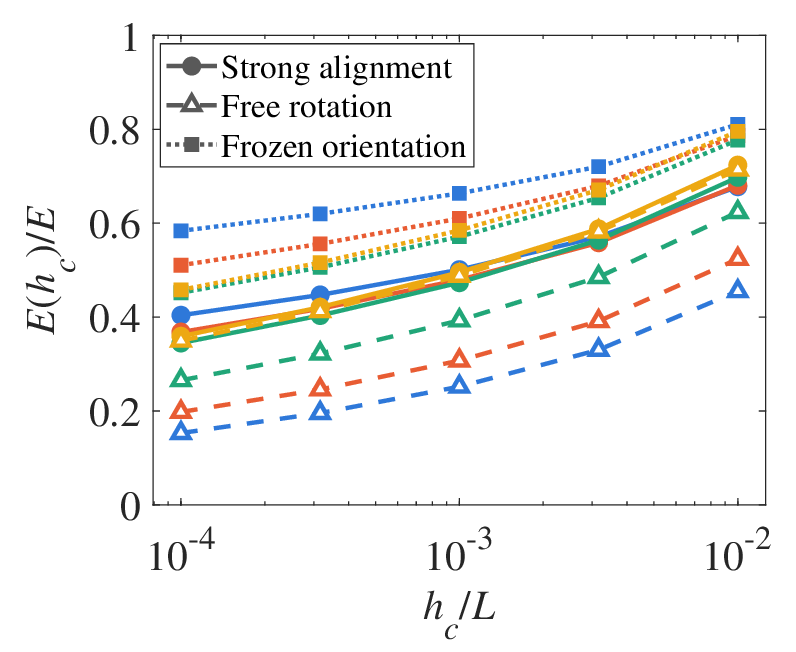}
  \caption{Accuracy of the model and effect of the gap, $\kappa=10$ and $\gamma=1$. (\textit{a}) Rate at which the gap closes $\dot{h}$, divided by its first-reflection value $\dot h_1$, against the gap $h$ in units of the transverse semi-axis $a$, for a sphere approaching the middle of a broadside spheroid head-on (filled circles; $\lambda=2$, $3$, $\lambda_w=4.64$ and $5.5$) and the point midway between the centre and a tip (open triangles; $\lambda=3$). The boundary-element solutions are represented by symbols and the solutions with first-reflection model with the lubrication film (\ref{eq:hstar}) are represented by solid lines. (\textit{b}) Efficiency with the film and a capture gap $h_c$, divided by the interception efficiency, for frozen orientation (squares), strong alignment (circles) and free rotation released at $12L$ (open triangles), with the colours of (\textit{a}).}
\label{fig:accuracy}
\end{figure}

\emph{Reflection truncation.} To test the first-reflection approximation itself, the full Stokes problem for the pair was solved by a boundary-element method (Appendix~\ref{app:bem}), which reproduces the exact squeeze mobility of two equal spheres \citep{Brenner1961} to within $0.5\,\%$ at a gap of $0.1$ radii. The comparison is first made for a sphere approaching the middle of a broadside spheroid head-on, at $\kappa=10$ and $\gamma=1$. Figure~\ref{fig:accuracy_a} shows the rate at which the gap $h$ closes, normalised by the closing rate $\dot h_1$ of the first-reflection model. Far from contact the first reflection is accurate: at $h=4a$ it overestimates the closing rate by $8$ to $11\,\%$ for $2\le\lambda\le5.5$. Near contact it is not, because it contains no lubrication resistance. The overestimate grows to $23$ to $31\,\%$ at $h=2a$ and to $56$ to $68\,\%$ at $h=a$, and it reaches a factor of $3.2$ to $3.5$ at $h=a/4$. This factor depends mainly on $h/a$, and only weakly on $\lambda$. At $\lambda=\lambda_w$ the first-reflection interaction vanishes, but the exact interaction, although weak, does not: it lowers the closing rate by $8.5\,\%$ at $h=4a$.

A second configuration tests the rotation. At $\lambda=3$ the sphere is placed above the point midway between the centre and a tip of the spheroid, so that it also tilts the spheroid. The closing rate is again overestimated near contact, by $13\,\%$ at $h=2a$ and by a factor of $2.6$ at $h=a/4$. The rate at which the sphere tilts the spheroid, however, agrees with the boundary-element value to within $3\,\%$ down to $h=a/2$, and to within $4.7\,\%$ at $h=a/4$. In this configuration, at least, the model therefore describes well the rotation that drives feathering and gliding.

\emph{Steady pairs.} Between the two loci a vertical spheroid settles faster than the sphere (\S\,\ref{sec:locus}). The flow of the sphere can turn a passing spheroid vertical just above the sphere, and the spheroid then tends to close the gap. For $\lambda>\lambda_w$, however, the spheroid is the heavier particle: its flow pushes the sphere ahead more strongly than the flow of the sphere draws the spheroid along, and the gap can stop closing. In the model the pair then settles together, with the spheroid vertical on the axis above the sphere. With free rotation this configuration is stable, and at $\lambda=5.5$ it remains stable for $a_I\le1.5\times10^{-2}$ ($k\le2.3$); at larger $a_I$ the model has no stable steady configuration. Trajectories that end in it are counted as misses. For each aspect ratio considered, the steady configuration exists in a band of size ratios that begins just above the axial locus and ends short of the extinction locus; the band is narrow at $\kappa=5$ and widens as $\kappa$ increases. Across the band the gap between the lower tip and the sphere decreases from about two spheroid lengths to nearly zero. The boundary-element solution confirms this equilibrium at $\lambda=5.3$ and $5.4$, with gaps of about $0.57L$ and $0.19L$, respectively, compared with $0.61L$ and $0.29L$, respectively, in the model. At $\lambda=5.5$, however, it finds no equilibrium. Over the whole range of gaps examined, from $0.4L$ down to $0.015L$, the gap continues to close, at between a tenth and a third of the difference between the settling speeds of the two particles in isolation. The pairs that the model holds in a steady configuration at this size ratio would therefore collide. To quantify them, each trajectory from the starting plane is assigned one outcome: contact, trapping when it comes within $10^{-7}$ of a stable steady configuration, or escape (Appendix~\ref{app:numerics}). At $\lambda=5.5$, with free rotation released at $24L$, $14\,\%$ of the pairs that come within $c+r_s+0.35L$ of each other collide, $2\,\%$ are trapped and $84\,\%$ escape. The trapped pairs start in a sector a few degrees wide about $\bp_0$, just outside the capture set, and their flux is $16\,\%$ of the colliding flux. This fraction decreases steadily as the alignment rate increases, and no pair is trapped once the steady configuration loses its stability ($k>2.3$), under strong alignment or at frozen orientation. Counting the trapped pairs as collisions would raise the corresponding efficiencies by these fractions. Closer to $\lambda_w$ the trapped flux is larger, reaching $94\,\%$ of the colliding flux at $\lambda=5.2$ for free rotation, whereas at $\kappa=20$ no trajectory from infinity is found to reach the steady configuration. The efficiencies reported here are single-encounter rates, which exclude the trapped pairs. In a suspension these pairs would accumulate until they are released, for example by Brownian rotation or turbulence, or until they collide; their contribution to the collision rate would require a population balance, which is beyond the scope of this work.

\emph{The lubrication film.} As the gap $h$ closes, the film of fluid between the surfaces resists the motion with the force $R_{lub}\dot h$. For surfaces whose relative principal radii of curvature at the point of closest approach are $R_x$ and $R_y$,
\begin{equation}
R_{lub}=\frac{12\upi\mu_f\,(R_xR_y)^{3/2}}{(R_x+R_y)\,h},
\label{eq:Rlub}
\end{equation}
as discussed in Appendix~\ref{app:lub}. At the equator of a broadside spheroid, $1/R_x=a/c^{2}+1/r_s$ and $1/R_y=1/a+1/r_s$. For $r_s\gg a$ these give $R_{lub}=12\upi\mu_f r_s^{1/2}a^{3/2}/h$, the sphere--fibre result of \citet{Harlen1999}, which is much smaller than the resistance $6\upi\mu_f r_s^{2}/h$ of a sphere approaching a plane. This limit is approached only slowly: for $r_s/a=1.6$, as on the extinction locus at $\kappa=10$, (\ref{eq:Rlub}) is $0.35$ times the sphere--fibre value. \citet{Harlen1999} showed that, in the time the sphere takes to move through one radius, the gap shrinks by the factor $\exp[-\tfrac12(r_s/a)^{3/2}]$, so that the film is negligible when $r_s\gg a$. Near the extinction locus, however, $r_s/a=(\gamma Y)^{-1/2}$ is only about $1.5$ to $1.8$ for $5\le\kappa\le20$, and the film is then important, as they also noted.

The film is therefore added to the first-reflection model as a resistance in series. The gap then closes at the rate
\begin{equation}
\dot h=\dot h_1\,\frac{h}{h+h^{*}},\qquad h^{*}=m\,R_{lub}\,h,
\label{eq:hstar}
\end{equation}
where $\dot h_1$ is the closing rate of the first-reflection model, given by (\ref{eq:hdot}), and $m$ is the relative mobility of the pair in that model for equal and opposite forces applied at the contact point, namely the sum of the isolated mobilities of the two particles minus twice their mutual mobility. Since $R_{lub}\propto h^{-1}$, the film length $h^{*}$ is independent of $h$, and with this choice of $m$, (\ref{eq:hstar}) reproduces the exact lubrication limit for two spheres. The film force is shared between the translation of both particles and the rotation of the spheroid in proportion to their mobilities. At $\kappa=10$, $h^{*}$ lies between $0.40a$ and $0.51a$ near contact, and decreases slowly as the sphere becomes smaller.

The model is tested against the boundary-element rates in figure~\ref{fig:accuracy_a}. At $\lambda=2$ it agrees with them to within $5\,\%$ down to $h=a/4$. For smaller spheres the film of the model is too weak; at $h=a/4$ it overestimates the closing rate by $16\,\%$, $26\,\%$ and $33\,\%$ at $\lambda=3$, $\lambda=\lambda_w=4.64$ and $5.5$, respectively. For the off-centre approach at $\lambda=3$, by contrast, it underestimates the rate by $10$ to $11\,\%$. For head-on approach, the boundary-element rates near contact follow the first-reflection rate multiplied by $h/(h+h^{*}_e)$, with an effective film length $h^{*}_e$ between $0.56a$ and $0.67a$. The film length of the model is shorter than $h^{*}_e$ by a factor that increases from $1.1$ at $\lambda=2$ to $1.6$ at $\lambda=5.5$, whereas for the off-centre approach it is longer. To assess the sensitivity of the results to this discrepancy, the efficiencies were also computed with $h^{*}$ multiplied by $(1+0.09\lambda)$, which reproduces the head-on factors to within $8\,\%$.

With the film, the gap closes exponentially in time and never vanishes, so that contact requires a non-hydrodynamic mechanism. In a gas, continuum lubrication fails once the gap is comparable to the mean free path \citep{Sundararajakumar1996}, and van der Waals attraction acts at similar distances \citep{Dhanasekaran2021a,Patra2022}. These effects are modelled by counting a collision when $h$ falls below a capture gap $h_c$. Figure~\ref{fig:accuracy_b} shows the resulting efficiencies divided by the interception efficiencies. At $\kappa=10$ and $h_c=10^{-3}L$ this factor is $0.57$ to $0.66$ for frozen orientation, $0.47$ to $0.50$ under strong alignment and $0.25$ to $0.49$ with free rotation. Therefore, freely rotating spheroids lose efficiency most far below the extinction locus, where feathering already lowers the efficiency, and least near the extinction locus. The factor decreases slowly with $h_c$, by between $0.2$ and $0.4$ from $h_c=10^{-2}L$ to $10^{-4}L$. Near the extinction locus, at $\lambda=6$ and $6.1$, it is about $0.6$ for frozen orientation and $0.5$ under strong alignment, so that the floor of the collision rate survives at about half its interception value. Lengthening the film by the factor $(1+0.09\lambda)$ lowers these factors by a further $12$ to $30\,\%$ at $\lambda=3$ and $5.5$.

\section{Parametrisations for collision models}\label{sec:practical}

This section collects the results in a form that can be used in collision models, for example in bin or particle-based models of riming and aggregation. Lengths are in units of $L$ and velocities in units of $U^{*}$ (\ref{eq:Ustar}). The formulas assume that translation obeys the Stokes equations, that particle inertia and Brownian motion are negligible (\S\S\,\ref{sec:isolated} and \ref{sec:orientation}), and that the suspension is dilute.

\subsection{Ideal kernels}\label{sec:fitsideal}

The ideal kernels depend only on the three dimensionless parameters of the pair: the aspect ratio $\kappa$ of the spheroid, the size ratio $\lambda$ and the density ratio $\gamma$. For a given $\kappa$, the axial and transverse resistance functions $X$ and $Y$ follow from (\ref{eq:XY}), and with them the settling speeds of a vertical and of a broadside spheroid, $\Gpa=1/(\kappa^{2}X)$ and $\Gpe=1/(\kappa^{2}Y)$. The sphere settles with the speed $S=\gamma/\lambda^{2}$. The speed mismatches $\bpa=\Gpa-S$ and $\bpe=\Gpe-S$ between the sphere and a vertical and a broadside spheroid, respectively, then fix the relative velocity (\ref{eq:dU}) at every orientation.

For the broadside-aligned population, the ideal kernel is given in closed form by (\ref{eq:Kb}), with the broadside cross-section (\ref{eq:Ab_ICR}). The only special function in this cross-section is the complete elliptic integral $\mathcal{E}(e)$, which sets the perimeter of the broadside shadow of the spheroid. For $\kappa\ge3$ it may be replaced by the leading terms of its expansion for a slender ellipse,
\begin{equation}
\mathcal{E}(e)\simeq1+\frac{\ln4\kappa-\tfrac12}{2\kappa^{2}},
\label{eq:Eslender}
\end{equation}
whose relative error is largest at $\kappa=3$, where it is $0.31\,\%$, and decreases rapidly as the spheroid becomes more slender. For the isotropic population, the ideal kernel is (\ref{eq:Kiso}), with the three contributions $T_1$, $T_2$ and $T_3$ given by (\ref{eq:T123}). The terms $T_1$ and $T_3$, which arise from the area of the projected ellipse and from the area of the disc, follow in closed form from the function $\Phi$ (\ref{eq:Phi}). Only the perimeter term $T_2$ requires numerical quadrature in $\mu$, with the eccentricity $e_\perp(\mu)$ of the projected ellipse obtained from (\ref{eq:ae}) and the angle $\chi$ from (\ref{eq:dUmag_Orientation}). Gauss--Legendre quadrature with $16$ nodes on $0\le\mu\le1$ evaluates $T_2$ with a relative error below $10^{-4}$ for all $\kappa\ge2$ and $\lambda\ge1.5$. The quadrature is exact on the extinction locus, where the integrand is linear in $\mu$, and its error is largest on the axial locus, where the integrand varies as $(1-\mu^{2})^{1/2}$, so that its slope becomes infinite at $\mu=1$. Finally, the positions of the extinction and axial loci follow from (\ref{eq:lperp}) and (\ref{eq:lpar}), respectively, or for a slender spheroid from (\ref{eq:lperpslender}) and (\ref{eq:lparslender}), respectively, and the floor of the isotropic kernel on the extinction locus follows from (\ref{eq:Kisolocus}).

\subsection{Interception efficiencies}\label{sec:fitsE}

\begin{figure}
  \centerline{\includegraphics[width=0.6\textwidth]{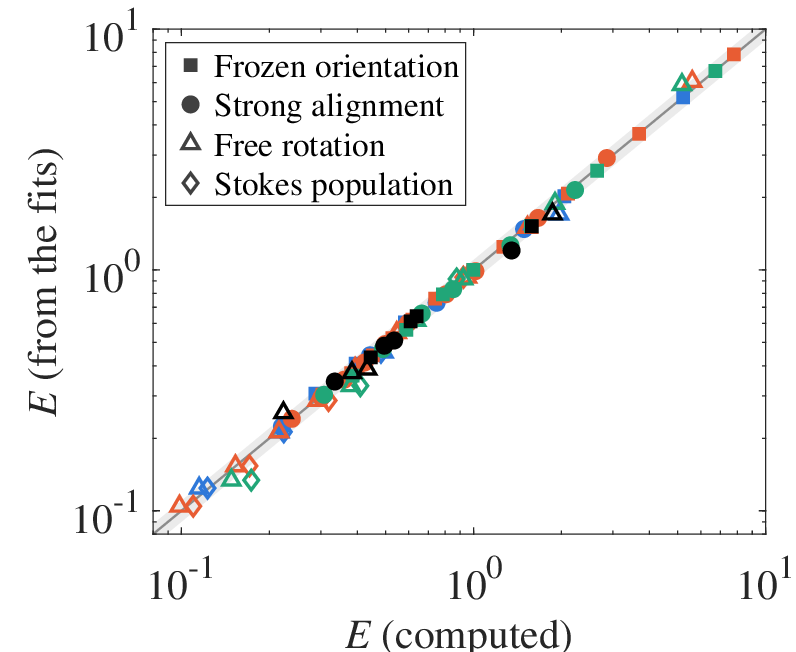}}
  \caption{Interception efficiencies from the fits (\ref{eq:Efit})--(\ref{eq:freefit}) against the computed values, for frozen orientation, under strong alignment, free rotation and the Stokes population (compared with (\ref{eq:freefit})), for $\kappa=5$, $10$ and $20$ with $\gamma=1$ (colours as in figure~\ref{fig:eff_b}), and for $\kappa=10$ with $\gamma=0.5$, $1.09$ and $2$ (black). The shaded band is $\pm10\,\%$.}
\label{fig:fits}
\end{figure}

The collision efficiencies are fitted in terms of three parameters. The first compares the strength $\delta$ of the interaction (\ref{eq:delta}) with the difference $|\bpe|$ of the settling speeds, relative to the size of the broadside cross-section \eqref{eq:Ab_ICR},
\begin{equation}
\sigma=\frac{\delta}{|\bpe|},\qquad \sigma_A=\frac{\sigma}{A_b^{1/2}}.
\label{eq:sigmaA}
\end{equation}
Since the interaction velocity decays as $\delta/r$ far from the pair, $|\sigma|$ is the distance at which it becomes comparable to the difference of the settling speeds, and $\sigma_A$ measures this distance in units of the linear size $A_b^{1/2}$ of the broadside cross-section (\ref{eq:Ab_ICR}). The second parameter is the glide parameter $\Lambda$ of the release law (\ref{eq:release}), which measures the far-field glide of a freely rotating spheroid. The third is the ratio $\Pi=\Omega_h/|\bpe|=9\gamma/(2\lambda^{3}|\bpe|)$ of the rate $\Omega_h$ at which the sphere rotates the spheroid (\S\,\ref{sec:orientation}) to the rate at which the pair approaches. The fits were constructed from the computations for $\kappa=5$, $10$ and $20$, with $\gamma=1$ and $1.5\le\lambda<\lpe$, and were tested at $\kappa=10$ for $\gamma=0.5$, $1.09$ and $2$ (figure~\ref{fig:fits}).

\emph{Frozen orientation.} The efficiency is fitted by
\begin{equation}
E_{fr}=
\begin{cases}
\exp\left(-1.68\,\sigma_A-13.1\,\sigma_A^{2}\right), & \lambda\le\lambda_w,\\[1ex]
1+\mathcal{G}_{fr}\,|\sigma|/A_b, & \lambda_w\le\lambda<\lpe,
\end{cases}
\label{eq:Efit}
\end{equation}
where
\begin{equation}
    \mathcal{G}_{fr}=2.93\,(a+r_{s\perp}^{*})\ln4\kappa.
    \label{eq:Fitt_G_fr}
\end{equation}
Here $r_{s\perp}^{*}$ is the matched radius (\ref{eq:lperp}). For $\gamma=1$, the errors of the two branches are below $7\,\%$ and $3.3\,\%$, respectively, and the fit remains accurate to about $4\,\%$ at the other density ratios tested. Below $\lambda_w$, where $\sigma_A>0$, the interaction lowers the efficiency below unity, and as $\sigma_A$ increases, the efficiency reduces further. Above $\lambda_w$ it raises the efficiency above unity, and the second branch is an additive law, $K_{12}\simeq\Kb+n_1n_2U^{*}L^{2}|\delta|\,\mathcal{G}_{fr}$, in which the flux due to the difference of the settling speeds and the flux due to the Stokeslet interaction simply add.

\emph{Strong alignment.} Weathervaning reduces the collision efficiency by a nearly constant factor compared to the frozen value,
\begin{equation}
E_{s}=W\,E_{fr},\qquad W=1-\frac{0.62}{\ln2\kappa}.
\label{eq:Wfit}
\end{equation}
With (\ref{eq:Efit}) for $E_{fr}$, this fit agrees with the computations to within $5.2\,\%$ for $\lambda/\lpe\le0.95$ and $\gamma=1$, and to within $11\,\%$ for different density ratios. Closer to the extinction locus the computed ratio $E_s/E_{fr}$ falls slowly below $W$. 

\emph{Free rotation.} The combined effect of feathering and gliding is fitted by
\begin{equation}
E_{free}=E_{fr}\exp\left(1.69\,\Lambda^{1/2}+\frac{2.61}{1+\Pi}-1.98\right),
\label{eq:freefit}
\end{equation}
with $E_{fr}$ from (\ref{eq:Efit}). The first term in the exponent increases with the glide, whereas the second decreases as the rotation by the sphere becomes faster than the approach, and hence as feathering strengthens. The fit agrees with the computations to within $15\,\%$ where $\Lambda\le2.5$, for all the values of $\gamma$ tested. The Stokes population also follows (\ref{eq:freefit}), to within $12\,\%$ for $\kappa=5$ and $10$; for $\kappa=20$ its efficiency is up to $30\,\%$ higher than (\ref{eq:freefit}) below $\lambda_w$.

\emph{Inertial population at small $Re_p$.} Since $a_I/\Omega_h<0.5\,Re_p$ (\S\,\ref{sec:orientation}), the spheroid rotates almost freely during an encounter, and the inertial torque acts mainly on the tilt acquired far upstream. Where $\Lambda\le2.5$, this effect is accounted for by correcting the free-rotation fit with the forced-relaxation law (\ref{eq:relaxlaw}), taken with $\ell=1$,
\begin{equation}
E=E_{free}\exp\left[-\Lambda\,k\,\mathcal{J}(k)\right],\qquad k=\frac{a_I}{|\bpe|},
\label{eq:finitefit}
\end{equation}
where $a_I$ is given by (\ref{eq:aI}) and $\mathcal{J}$ by \eqref{eq:mathcal_J}. For $\kappa=10$, the correction factor reproduces the computed ratio $E/E_{free}$ to within $1.1\,\%$ for $k\le0.17$. The discrepancy grows with the alignment rate, reaching $5.2\,\%$ for $k\le0.35$ and $8.2\,\%$ for $k\le0.73$. At $\lambda=5$ the last bound corresponds to $Re_p\lesssim2$. The error of (\ref{eq:freefit}) adds to these values. Far from the locus the correction is negligible. Within about $0.1\lpe$ of the locus, however, $\Lambda$ exceeds $2.5$ and the efficiency depends strongly on $Re_p$. There it is close to the efficiency of free rotation released from broadside at $Z=2L/k$ (figure~\ref{fig:inertial}), and the collision rate can reach a few times $10^{-2}K_S$. The strong-alignment fit (\ref{eq:Wfit}) does not apply to the inertial population at small $Re_p$, since suppressing the tilt induced near contact would require $a_I\gg\Omega_h$.

\subsection{The gap}\label{sec:fitsgap}

With the lubrication film of \S\,\ref{sec:accuracy} and a capture gap $h_c$, the efficiency is $NE$, where $E$ is the interception efficiency of \S\,\ref{sec:fitsE} and $N$ is the near-contact factor of table~\ref{tab:gap}. The factor depends on the encounter model and on $\kappa$, and falls slowly as $h_c$ decreases, so it can be interpolated linearly in $\ln (h_c)$. At $h_c=10^{-3}L$ it lies between $0.18$ and $0.82$. It is largest for frozen orientation and for thin spheroids, and smallest for free rotation far below the locus. For the inertial population at small $Re_p$, which rotates almost freely during an encounter, the free-rotation values apply. The uncertainty of the film model, measured by the test of \S\,\ref{sec:accuracy}, is up to about $30\,\%$ in $N$.

\begin{table}
  \begin{center}
\def~{\hphantom{0}}\small\setlength{\tabcolsep}{3.4pt}
  \begin{tabular}{rcccccccccccc}
      & & \multicolumn{3}{c}{$h_c/L$ (frozen)} & & \multicolumn{3}{c}{$h_c/L$ (strong)} & & \multicolumn{3}{c}{$h_c/L$ (free)}\\[2pt]
      $\kappa$ & $\lambda$ & $10^{-2}$ & $10^{-3}$ & $10^{-4}$ & & $10^{-2}$ & $10^{-3}$ & $10^{-4}$ & & $10^{-2}$ & $10^{-3}$ & $10^{-4}$\\[3pt]
        5 & 2 & 0.63 & 0.46 & 0.36 & & 0.48 & 0.29 & 0.20 & & 0.37 & 0.18 & 0.10\\
        & 2.92$^{\dagger}$ & 0.68 & 0.50 & 0.39 & & 0.56 & 0.37 & 0.25 & & 0.59 & 0.40 & 0.29\\[3pt]
        10 & 2 & 0.81 & 0.66 & 0.58 & & 0.68 & 0.50 & 0.40 & & 0.46 & 0.25 & 0.15\\
        & 3 & 0.79 & 0.61 & 0.51 & & 0.68 & 0.48 & 0.37 & & 0.52 & 0.31 & 0.20\\
        & 4.64$^{\dagger}$ & 0.78 & 0.57 & 0.45 & & 0.70 & 0.47 & 0.34 & & 0.62 & 0.39 & 0.27\\
        & 5.5 & 0.80 & 0.58 & 0.46 & & 0.72 & 0.50 & 0.36 & & 0.71 & 0.49 & 0.35\\[3pt]
        20 & 2 & 0.94 & 0.82 & 0.76 & & 0.85 & 0.71 & 0.63 & & 0.65 & 0.44 & 0.32\\
        & 7.37$^{\dagger}$ & 0.89 & 0.64 & 0.51 & & 0.84 & 0.58 & 0.44 & & 0.82 & 0.55 & 0.41\\
  \end{tabular}
  \caption{Near-contact factor $N=E(h_c)/E$: the efficiency with the lubrication film and a capture gap $h_c$, divided by the interception efficiency, for $h_c/L=10^{-2}$, $10^{-3}$ and $10^{-4}$ and $\gamma=1$. Free rotation is released from broadside at $12L$. The superscript symbol $^{\dagger}$ represents $\lambda=\lambda_w$, where the frozen efficiency is unity exactly.}
  \label{tab:gap}
  \end{center}
\end{table}

\subsection{An example}\label{sec:example}

As an illustration, consider ice columns of length $L=70$\,\textmu m and aspect ratio $\kappa=10$ settling in air at $-10\,^{\circ}$C and $800$\,hPa, where $\rho_f=1.06$\,kg\,m$^{-3}$ and $\mu_f=1.67\times10^{-5}$\,Pa\,s, together with water droplets, so that $\gamma=1.09$. The velocity scale (\ref{eq:Ustar}) is then $U^{*}=0.146$\,m\,s$^{-1}$, a broadside column settles at $U_\perp=3.8$\,mm\,s$^{-1}$, and the particle Reynolds number is $Re_p=0.017$. The droplet that settles at the same speed as a broadside column, and hence lies on the extinction locus, has the radius $r_{s\perp}^{*}=5.4$\,\textmu m ($\lpe=6.45$). The droplet whose buoyant weight equals that of the column ($\lambda=\lambda_w$) has the radius $7.3$\,\textmu m, and the droplet that settles at the same speed as a vertical column, on the axial locus, has the radius $6.5$\,\textmu m.

With $F_b=0.189$, (\ref{eq:aI}) gives the alignment rate $a_I=0.18$\,s$^{-1}$, or $8.5\times10^{-5}$ in units of $U^{*}/L$. The columns are therefore kept broadside between encounters (\S\,\ref{sec:orientation}), since alignment is fast compared with rotational diffusion, with turbulent shear and with the interval between encounters. First, rotational diffusion is negligible: $D_r/a_I=0.01$, which corresponds to an r.m.s.\ tilt of $6^{\circ}$. Note that near the extinction locus this residual tilt is not negligible: the axial drift $(\Gpa-\Gpe)\sin(6^{\circ})$ that it produces is about $0.7|\bpe|$ for a droplet of radius $5.6$\,\textmu m, and exceeds $0.2|\bpe|$ within about $10\,\%$ of $r_{s\perp}^{*}$, so that the broadside idealisation (\ref{eq:psib}) is approximate there. Second, turbulence is negligible provided that its shear rate $(\epsilon/\nu)^{1/2}$ is small compared with $a_I$, that is, provided that $\epsilon\ll5\times10^{-7}$\,m$^{2}$\,s$^{-3}$. Third, alignment is fast compared with the interval between encounters: for $100$ droplets per cm$^{3}$ the ratio of the two times (\S\,\ref{sec:orientation}) is about $10^{-2}$. In such calm air the columns therefore fall broadside between encounters and form the inertial population. During an encounter, however, alignment is slow. The ratio $k=a_I/|\bpe|$ is below $0.03$ unless the droplet radius lies within $6\,\%$ of $r_{s\perp}^{*}$, so that the column rotates almost freely and $E\simeq E_{free}$. Moreover, $a_I/\Omega_h<0.02$ for $\lambda\le10$, so that the tilt induced near contact is never suppressed, and the band above the locus in which collisions are blocked under strong alignment (\S\,\ref{sec:locusHI}) is not expected to arise, except possibly within about $1\,\%$ of $r_{s\perp}^{*}$, where alignment acts before the spheroid reaches the dividing sphere.

For a droplet of radius $7$\,\textmu m ($\lambda=5$), the fits (\ref{eq:Efit}) and (\ref{eq:freefit}) give $E_{fr}=1.12$ and $E_{free}=1.14$, and the correction (\ref{eq:finitefit}) lowers $E_{free}$ by only $0.6\,\%$. Since $\Kb/(n_1n_2)=3.9\times10^{-12}$\,m$^{3}$\,s$^{-1}$, the interception kernel is $K_{12}/(n_1n_2)=4.4\times10^{-12}$\,m$^{3}$\,s$^{-1}$. Continuum lubrication fails once the gap becomes comparable to the mean free path of the air (\S\,\ref{sec:accuracy}), which under these conditions is about $0.07$\,\textmu m, or $10^{-3}L$. Taking the capture gap $h_c$ equal to this mean free path, interpolation in table~\ref{tab:gap} gives the near-contact factor $N\approx0.43$ for free rotation, which reduces the kernel to $K_{12}/(n_1n_2)\approx1.9\times10^{-12}$\,m$^{3}$\,s$^{-1}$.

Droplets of radii between $5.4$ and $6$\,\textmu m lie within about $0.1\lpe$ of the locus, where the efficiency depends on $Re_p$ (\S\,\ref{sec:finiteRe}). For $\lambda/\lpe\approx0.97$, a droplet of radius $5.6$\,\textmu m, the computations of \S\,\ref{sec:finiteRe} for $\gamma=1$ give $K_{12}/K_S\approx0.036$ at $k=0.083$ ($Re_p\approx0.026$). Taking this value as an estimate for the present example, in which $k\approx0.05$, gives $K_{12}/(n_1n_2)\approx2.7\times10^{-11}$\,m$^{3}$\,s$^{-1}$ before the near-contact factor is applied, about six times the value for the droplet of radius $7$\,\textmu m. On the locus itself, the strongly aligned floor (\ref{eq:Kfloor}) gives the estimate $1.6\times10^{-12}$\,m$^{3}$\,s$^{-1}$ for the rate sustained by the weight difference alone; at the alignment rate of this example the rate is expected to be higher (\S\,\ref{sec:locusHI}), but it has not been computed.

Two conditions limit the validity of this example. First, the droplets have inertia. Their Stokes number $\tau_sU_\perp/a$, based on the relaxation time $\tau_s=2\rho_sr_s^{2}/(9\mu_f)$ of a droplet, the broadside settling speed and the semi-minor axis of the column, is $0.43$ at $r_{s\perp}^{*}$ and grows as $r_s^{2}$; the interception efficiencies of the larger droplets therefore require inertial corrections. The column itself is not free of inertia. Its Stokes number based on the length, $St_c=\tau_cU_\perp/L$ (\S\,\ref{sec:isolated}), is only $0.03$, but that based on the transverse semi-axis, $\tau_cU_\perp/a$, is $0.6$. The latter is the relevant measure near contact, where the relative motion of the pair varies over distances comparable to the thickness of the column rather than its length. These Stokes numbers, like that of the droplets, scale as $L^{3}$ at fixed $\kappa$, so that a shorter column would reduce them, but only at the expense of alignment, since the r.m.s.\ tilt grows as $L^{-7/2}$. Inertial corrections, for the columns as well as for the droplets, are therefore required. Second, the far-field construction of \S\,\ref{sec:locusHI} requires the dividing sphere to be small compared with the mean spacing of the droplets, $n_2^{-1/3}\approx30L$ at $100$ droplets per cm$^{3}$. This holds only for $|\lambda-\lpe|\gtrsim0.01\lpe$, that is, for droplet radii that differ from $r_{s\perp}^{*}$ by more than about $1\,\%$.

\section{Conclusions}\label{sec:conclusions}

We have computed the collision rate of a prolate spheroid and a sphere that settle through a quiescent fluid, for a randomly oriented population of spheroids and for a broadside-aligned population, resulting from weak fluid inertia.

Without interactions the collision cross-section is the shadow of the Minkowski sum of the two particles: an ellipse fattened by a disc, whose area follows from Steiner's formula. This gives the kernel of the aligned population in closed form and that of the isotropic population as a one-dimensional quadrature. Alignment raises the ideal collision rate by at most $4/\upi$, a bound reached where the settling speed of a vertical spheroid matches the sphere. It extinguishes the rate on the locus $\lambda= \lpe =\kappa(\gamma Y)^{1/2}$, where the settling speed of a broadside spheroid matches the sphere. The isotropic kernel keeps a small closed-form floor there, because every spheroid then approaches the sphere end-on. The sharp minimum of the isotropic kernel is a remnant of this locus rather than a coincidence of mean settling speeds; the same argument is consistent with the dip found by \citet{Naso2018} for thin plates.

In a first-reflection slender-body model the relative translational interaction is proportional to the difference of the buoyant weights, so that it vanishes at $\lambda_w=\gamma^{1/3}\kappa^{2/3}$. The rotation of the spheroid is set by the sphere alone. With the orientation held fixed, the efficiency is below unity when the sphere is heavier, and above unity only for $\lambda_w<\lambda<\lpe$. Under strong alignment the spheroid weathervanes and the efficiency falls by $15$ to $30\,\%$. With free rotation the spheroid feathers, which lowers the efficiency by up to two thirds far from the locus, and glides, which raises it near the locus. The Stokes population behaves like the inertial population with free rotation. At a finite alignment rate the inertial torque damps the tilt over the whole approach. A model of forced relaxation captures this, and shows that the alignment rate controls the efficiency in a band below the locus whose width grows as $Re_p^{1/3}$ up to a logarithm. In this band the glide that slow alignment fails to suppress raises the collision rate far above the strong alignment result and the ideal isotropic kernel: at $\kappa=10$ and $Re_p\approx0.03$, $3\,\%$ below the locus, it is $15$ times the former and $80$ times the latter. Near the extinction locus the far-field relative flow is a uniform stream plus a Stokeslet. Below the locus it sustains a floor of the collision rate, set by the weight difference. Above it the two oppose and a dividing stream surface forms. Under strong alignment, no trajectory from infinity then reaches contact in a band above the extinction locus. With free rotation, and at every finite alignment rate computed, collisions nevertheless continue there, at rates several times the strongly aligned floor just below the locus. Strong alignment requires the inertial torque to suppress the tilt induced near contact, $a_I\gg\Omega_h$, whereas $a_I/\Omega_h<0.5\,Re_p$, so that this limit is not attained at small $Re_p$. Whether collisions continue arbitrarily close to the locus has not been established. Within a narrow band, $(\lambda/\lpe-1)\lesssim0.09\,Re_p^{1/2}$, alignment damps the tilt before the spheroid reaches the dividing sphere, and this band lies outside the range computed here.

The efficiencies above are interception efficiencies of the first-reflection model. The slender-body closure of the translational interaction is accurate to within $2.3\,\%$. In the configuration tested, the rotation predicted by the model agrees with a boundary-element solution of the full pair problem to within $4.7\,\%$. Near contact, however, the first reflection overestimates the rate at which the gap closes, by a factor of $3.2$ to $3.5$ at $h=a/4$. A lubrication film was therefore added in series with the model, based on the weak sphere--spheroid resistance identified by \citet{Harlen1999}. It reproduces the boundary-element rates to within $16\,\%$ for $\lambda\le3$, and to within a third for smaller spheres. With a capture gap of $10^{-3}L$, comparable to the mean free path of air for columns of length $70$\,\textmu m, the film reduces the efficiencies to between $20$ and $80\,\%$ of their interception values, the reduction being largest for freely rotating spheroids far below the extinction locus. Fits for the ideal kernels, the interception efficiencies and this near-contact factor are given in \S\,\ref{sec:practical}. The indicative example of \S\,\ref{sec:example} illustrates the regime of small ice columns in calm air: the columns fall broadside between encounters but rotate almost freely during them, so that the free-rotation efficiencies apply except close to the locus. Two effects neglected here limit this picture. The first is particle inertia. The Stokes number of the droplets is about $0.4$ at the matched size and grows with their size. That of the columns is only $0.03$ when based on their length, but about $0.6$ when based on their transverse semi-axis, which is the relevant scale near contact. Inertial corrections are therefore required for both particles. The second is turbulence, which randomises the orientations of such small columns unless it is very weak. The same formulation can be extended to oblate spheroids, pairs of spheroids, branched crystals, particle inertia, turbulence and electric charge.

\begin{bmhead}[Funding.]
A.C. was supported by an Institute Postdoctoral Fellowship of the Indian Institute of Technology Madras.
\end{bmhead}

\begin{bmhead}[Declaration of interests.]
The authors report no conflict of interest.
\end{bmhead}

\begin{bmhead}[Author ORCIDs.] \\
Arnab Choudhury \href{https://orcid.org/0000-0002-6366-2752}{https://orcid.org/0000-0002-6366-2752}; \\
Anubhab Roy \href{https://orcid.org/0000-0002-0049-2653}{https://orcid.org/0000-0002-0049-2653}.
\end{bmhead}

\begin{appen}

\section{The line integrals of slender-body theory}\label{app:lineint}

Expanding the dyad in (\ref{eq:W}) and (\ref{eq:pdoth}), with $r_g=\br\bcdot\bg$ and $\mu=\bp\bcdot\bg$, reduces every integral to the moments
\begin{equation}
\Iint{m}{n}(\br,\bp)=\int_{-1/2}^{1/2}\frac{s^{m}\,\dd s}{|\br+s\bp|^{n}},\qquad m \in \{0,1,2\}, \quad n \in \{1,3,5\}.
\label{eq:Imn}
\end{equation}
The expression of $\Wf$ in \eqref{eq:W} can be written as
\begin{eqnarray}
\nonumber \Wf&=&\Iint01\bg+\Iint03r_g\br+\Iint13(\mu\br+r_g\bp)+\Iint23\mu\bp\\
&&+\frac{r_s^{2}}{3}\left[\Iint03\bg-3\Iint05r_g\br-3\Iint15(\mu\br+r_g\bp)-3\Iint25\mu\bp\right].
\label{eq:Wclosed}
\end{eqnarray}
In (\ref{eq:pdoth}) the additional weight $s$ raises the order $m$ of every moment by one. The terms that would then require the cubic moments $\int s^{3}|\bx|^{-n}\,\dd s$ are all parallel to $\bp$, and they are annihilated by the projector $\tI-\bp\bp$, so that only the moments (\ref{eq:Imn}) are needed. These are evaluated in closed form. With $\zeta=\br\bcdot\bp$ the component of $\br$ along the axis and $\rho^{2}=|\br|^{2}-\zeta^{2}$ the square of its distance from the axis, the substitution $w=s+\zeta$ gives
\begin{equation*}
|\br+s\bp|^{2}=w^{2}+\rho^{2}\equiv R^{2},
\end{equation*}
with limits $w_\pm=\zeta\pm\tfrac12$ and $R_\pm=(w_\pm^{2}+\rho^{2})^{1/2}$. The required primitives,
\begin{equation}
\int\frac{\dd w}{R}=\sinh^{-1}\frac{w}{\rho},\qquad
\int\frac{\dd w}{R^{3}}=\frac{w}{\rho^{2}R},\qquad
\int\frac{\dd w}{R^{5}}=\frac{w}{3\rho^{2}R^{3}}+\frac{2w}{3\rho^{4}R},\qquad
\int\frac{w\,\dd w}{R^{n}}=\frac{R^{2-n}}{2-n},
\label{eq:primitives}
\end{equation}
may be verified by differentiation. Writing $s=w-\zeta$, so that $s^{2}=w^{2}-2\zeta w+\zeta^{2}$, and using $w^{2}=R^{2}-\rho^{2}$ to reduce the even powers of $w$, every moment follows from (\ref{eq:primitives}):
\begin{gather}
\Iint01=\sinh^{-1}\left(\frac{w_+}{\rho}\right)-\sinh^{-1}\left(\frac{w_-}{\rho}\right),\qquad
\Iint11=R_+-R_--\zeta\,\Iint01,\nonumber\\
\Iint03=\frac{1}{\rho^{2}}\left(\frac{w_+}{R_+}-\frac{w_-}{R_-}\right),\qquad
\Iint13=\frac{1}{R_-}-\frac{1}{R_+}-\zeta\,\Iint03,\nonumber\\
\Iint23=\Iint01-\rho^{2}\Iint03+2\zeta\left(\frac{1}{R_+}-\frac{1}{R_-}\right)+\zeta^{2}\Iint03,\nonumber\\
\Iint05=\frac{1}{3\rho^{2}}\left(\frac{w_+}{R_+^{3}}-\frac{w_-}{R_-^{3}}\right)+\frac{2}{3\rho^{2}}\Iint03,\qquad
\Iint15=\frac13\left(\frac{1}{R_-^{3}}-\frac{1}{R_+^{3}}\right)-\zeta\,\Iint05,\nonumber\\
\Iint25=\Iint03-\rho^{2}\Iint05+\frac{2\zeta}{3}\left(\frac{1}{R_+^{3}}-\frac{1}{R_-^{3}}\right)+\zeta^{2}\Iint05.
\label{eq:Iclosed}
\end{gather}
Two numerical precautions are needed. First, $\Iint01$ is written in terms of $\sinh^{-1}$ rather than the equivalent logarithm $\ln[(w_++R_+)/(w_-+R_-)]$, which suffers from cancellation when $w_\pm<0$. Second, close to the extension of the axis, where $w_-w_+>0$ and $\rho\ll|w_\pm|$, the differences in $\Iint03$ and $\Iint05$ cancel almost completely; at $\rho=10^{-7}|w_-|$, for example, the closed form of $\Iint05$ loses ten significant digits. In this region, taken as $\rho<0.2\min|w_\pm|$, the moments are instead evaluated by $48$-point Gauss--Legendre quadrature in $\ln|w|$. This is possible because $w$ does not change sign on the interval, and the integrands are then smooth. The moments so obtained agree with adaptive quadrature to within $10^{-10}$ everywhere, including on the axis.

\section{Near-sphere expansions of the ideal kernels}\label{app:nearsphere}

Consider the limit $e\to0$ at fixed $L$ and $\gamma=1$, in which the spheroid tends to a sphere of diameter $L$. The settling speed of the sphere is then $S=\lambda^{-2}$, which is less than unity for $\lambda>1$. The two factors of the ideal kernel, the cross-section and the relative velocity, are expanded in turn.

For the cross-section, $a=c(1-e^{2})^{1/2}$, and (\ref{eq:ae}) gives the semi-axis and eccentricity of the projected ellipse, respectively, as
\begin{equation*}
a_e=c\left(1-\tfrac12e^{2}\cos^{2}\chi\right)+O(e^{4}),\qquad e_\perp^{2}=e^{2}\sin^{2}\chi+O(e^{4}).
\end{equation*}
Substituting these, together with $\mathcal{E}(e_\perp)=\tfrac{\upi}{2}\left[1-\tfrac14e_\perp^{2}+O(e_\perp^{4})\right]$, into the Steiner formula (\ref{eq:steiner}) gives
\begin{equation}
\frac{A_\perp}{\upi(c+r_s)^{2}}=1-\frac{e^{2}}{2}\,(1+\cos^{2}\chi)\,\frac{\lambda}{\lambda+1}+O(e^{4}),
\label{eq:Anearsphere}
\end{equation}
where $\upi(c+r_s)^{2}$ is the cross-section of the equivalent pair of spheres.

For the relative velocity, the settling speeds of a nearly spherical spheroid follow from (\ref{eq:XY}) as $\Gpa=1-\tfrac35e^{2}+O(e^{4})$ and $\Gpe=1-\tfrac{7}{10}e^{2}+O(e^{4})$. Equation (\ref{eq:dUmag}) then gives
\begin{equation}
\frac{|\dU|}{1-S}=1-\frac{e^{2}\lambda^{2}}{\lambda^{2}-1}\left(\tfrac35\cos^{2}\theta+\tfrac{7}{10}\sin^{2}\theta\right)+O(e^{4}).
\label{eq:dUnearsphere}
\end{equation}
At $e=0$ the relative velocity is vertical, so $\chi=\theta+O(e^{2})$. Since neither $|\dU|$ nor $A_\perp$ depends on $\chi$ at leading order, $\chi$ may be replaced by $\theta$ in the $O(e^{2})$ terms.

The normalised kernel is the product of (\ref{eq:Anearsphere}) and (\ref{eq:dUnearsphere}). For the isotropic population it is averaged over orientations with $\avg{\cos^{2}\theta}=\tfrac13$, and for the aligned population it is evaluated at $\theta=\upi/2$. This gives
\begin{equation}
\Kbar=1-\frac23\,e^{2}\,\frac{\lambda(2\lambda-1)}{\lambda^{2}-1}+O(e^{4}),\qquad
\Kbar_b=1-e^{2}\left[\frac{7}{10}\,\frac{\lambda^{2}}{\lambda^{2}-1}+\frac{\lambda}{2(\lambda+1)}\right]+O(e^{4}),
\label{eq:Kbarnearsphere}
\end{equation}
whose ratio is (\ref{eq:Rnearsphere}). The deficit in the isotropic kernel has two parts. The kinematic part, $\tfrac23e^{2}\lambda^{2}/(\lambda^{2}-1)$, arises from the lower mean settling speed of the spheroid, and the geometric part, $\tfrac23e^{2}\lambda/(\lambda+1)$, from the thinning of its shadow. The coefficients were confirmed to six significant figures by Richardson extrapolation in $e^{2}$ of the exact kernels. The expansions are not uniform as $\lambda\to1$, where both loci approach the point $\lambda=1$ of two equal spheres and $(1-S)$ vanishes.

\section{The shadow of the spheroid and the function $\Phi$}\label{app:shadow}

The shadow of the contact body is obtained from the support function. For a convex body $\mathcal{K}$, the support function
\begin{equation*}
h_{\mathcal{K}}(\hat{\boldsymbol{m}})=\max_{\bx\in\mathcal{K}}\bx\bcdot\hat{\boldsymbol{m}}
\end{equation*}
is the distance from the origin to the supporting plane of $\mathcal{K}$ with outward normal $\hat{\boldsymbol{m}}$, and it determines the body uniquely \citep{Schneider2014}. Moreover, the support function of the projection of $\mathcal{K}$ onto a plane coincides with that of $\mathcal{K}$ for every direction $\hat{\boldsymbol{m}}$ lying in that plane. The spheroid $\mathcal{E}$ is the image of the unit ball under the symmetric linear map $\mathsfbi{M}=c\,\bp\bp+a\,(\tI-\bp\bp)$, that is, $\bx=\mathsfbi{M}\boldsymbol{y}$ with $|\boldsymbol{y}|\le1$. Hence $h_{\mathcal{E}}(\hat{\boldsymbol{m}})=\max_{|\boldsymbol{y}|\le1}\boldsymbol{y}\bcdot\mathsfbi{M}\hat{\boldsymbol{m}}=|\mathsfbi{M}\hat{\boldsymbol{m}}|$, so that
\begin{equation*}
h_{\mathcal{E}}^{2}=c^{2}(\hat{\boldsymbol{m}}\bcdot\bp)^{2}+a^{2}\left[1-(\hat{\boldsymbol{m}}\bcdot\bp)^{2}\right].
\end{equation*}
Let $\hat{\boldsymbol{u}}$ be the direction of projection and $(\hat{\boldsymbol{e}}_1,\hat{\boldsymbol{e}}_2)$ an orthonormal basis of the plane normal to it, chosen so that $\bp=\cos\chi\,\hat{\boldsymbol{u}}+\sin\chi\,\hat{\boldsymbol{e}}_1$. For a direction $\hat{\boldsymbol{m}}=\cos\varphi\,\hat{\boldsymbol{e}}_1+\sin\varphi\,\hat{\boldsymbol{e}}_2$ in this plane, $\hat{\boldsymbol{m}}\bcdot\bp=\sin\chi\cos\varphi$, and
\begin{equation*}
h_{\mathcal{E}}^{2}=\left(c^{2}\sin^{2}\chi+a^{2}\cos^{2}\chi\right)\cos^{2}\varphi+a^{2}\sin^{2}\varphi .
\end{equation*}
This is the support function of an ellipse with semi-axes $a_e$ and $a$, as given by (\ref{eq:ae}), so the shadow of the spheroid is this ellipse. Since projection commutes with Minkowski addition, the shadow of the contact body is the ellipse enlarged by a disc of radius $r_s$. Its area follows from the Steiner formula (\ref{eq:steiner}). The curve parallel to the ellipse at a distance $t$ has length $P_e+2\upi t$, so the enlargement adds the area
\begin{equation*}
\int_0^{r_s}\left(P_e+2\upi t\right)\dd t=P_er_s+\upi r_s^{2}
\end{equation*}
to that of the ellipse.

The function $\Phi$ appearing in (\ref{eq:T123}) is defined, for $\beta\ge0$ and $\beta^{2}+m\ge0$, by
\begin{equation}
\Phi(\beta^{2},m)=\int_0^{1}\left(\beta^{2}+m\mu^{2}\right)^{1/2}\dd\mu=\frac{\left(\beta^{2}+m\right)^{1/2}}{2}+
\begin{cases}
\dfrac{\beta^{2}}{2m^{1/2}}\,\sinh^{-1}\dfrac{m^{1/2}}{\beta}, & m>0,\\[2ex]
\dfrac{\beta^{2}}{2|m|^{1/2}}\,\sin^{-1}\dfrac{|m|^{1/2}}{\beta}, & m<0,
\end{cases}
\label{eq:Phi}
\end{equation}
with $\Phi(\beta^{2},0)=\beta$. For $m>0$, (\ref{eq:Phi}) follows from the primitive
\begin{equation*}
\int\left(\beta^{2}+m\mu^{2}\right)^{1/2}\dd\mu=\frac{\mu}{2}\left(\beta^{2}+m\mu^{2}\right)^{1/2}+\frac{\beta^{2}}{2m^{1/2}}\sinh^{-1}\frac{m^{1/2}\mu}{\beta}.
\end{equation*}
For $m<0$ the same primitive holds with the inverse hyperbolic sine replaced by an inverse sine and $m^{1/2}$ by $|m|^{1/2}$. Both branches tend to $\Phi(\beta^{2},0)=\beta$ as $m\to0$.

\section{Numerical method and convergence}\label{app:numerics}

\emph{Integration.} Trajectories are integrated with an adaptive Runge--Kutta scheme of relative tolerance $10^{-7}$, in two stages: with steps bounded by $0.5L/V$, where $V=|\dU|+\tfrac34(\gamma\lambda^{-3}+\kappa^{-2})$ includes the scale of the interaction velocity, until the separation enters a sphere of radius $c+r_s+0.35L$, and with steps $25$ times smaller inside it. Contact is detected on the exact gap $h$, computed from the nearest point of the spheroid by Newton iteration. A trajectory ends in trapping when it comes within $10^{-7}$, in $\br$ and $\bp$ together, of a linearly stable steady configuration (\S\,\ref{sec:accuracy}), and in escape when it passes the plane at $Z+L$ downstream. A trajectory that has entered the sphere of radius $c+r_s+0.35L$ eight times without any of these, or that has run for a time $80Z/|\dU|$, is stopped and counted as a miss. Contact and trapping occur well within these limits, within $Z/|\dU|$ and $5Z/|\dU|$ of the release, respectively, in the cases of \S\,\ref{sec:accuracy}. Above the locus the trajectories stopped by the limits circle the spheroid in bound orbits: at $(\lambda,a_I)=(6.5,10^{-4})$ they carry $11\,\%$ of the colliding flux, and with $100$ entries and ten times the time, $47$ of $51$ sampled ones still circle and the collision rate changes by $0.3\,\%$. Below the locus, with free rotation, they instead separate sideways: the spheroid leaves the encounter tilted close to $\theta_v$ (\S\,\ref{sec:locus}), at which it settles as fast as the sphere, and glides away at the speed $(\Gpa-\Gpe)\sin\theta_v\cos\theta_v$; at $\lambda=5.5$ they are $2\,\%$ of the pairs that enter the encounter sphere, with $16\,\%$ of the colliding flux, and followed to $800Z/|\dU|$ all of them are more than $10^{4}L$ apart.

\emph{Rays.} For a broadside spheroid $33$ rays span a quadrant, equally spaced for frozen orientation and strong alignment, and clustered towards $\bp_0$ at the angles $\tfrac{\upi}{2}(j/32)^{2}$ for free rotation, whose capture sets are drawn out along $\bp_0$ by gliding. Tracings on $9$ equally spaced rays ($17$ for $\kappa=20$) differ from those reported by at most $1.7\,\%$ for frozen orientation and strong alignment, and tracings on $17$ clustered rays by at most $1.9\,\%$ for free rotation; near the locus, $9$ equally spaced rays overestimate the free-rotation efficiency by up to a factor of $2.2$, which is why clustered rays are used there. For an inclined spheroid, which has one mirror plane, $17$ equally spaced rays span a half-plane ($33$ for $\kappa=20$); doubling them changes the efficiency of the Stokes population at $(10,3)$ by $0.1\,\%$. Free rotation can produce capture sets with holes: near the axial locus, a nearly vertical spheroid approached head-on is turned to the vertical, where it settles at almost the speed of the sphere, and the sphere drifts past it. The outcome is therefore sampled at $25$ (broadside) or $17$ (inclined) points along each ray, out to $2.2(c+r_s)$ and beyond while collisions persist, and every change is refined by bisection; for star-shaped sets this agrees with bisection to $0.13\,\%$. With the lubrication film (\S\,\ref{sec:accuracy}) the capture sets under frozen orientation and strong alignment are no longer star-shaped; they develop a narrow lobe along $\bp_0$, a few degrees wide, in which the sphere approaches a tip of the spheroid, where the film is weakest. The near-contact factors of table~\ref{tab:gap} and figure~\ref{fig:accuracy_b} are therefore traced on $257$ equally spaced rays per quadrant, every ray being scanned; $129$ rays give the same values to within $0.003$, and a tolerance of $10^{-9}$ or four times smaller steps near the sphere change them by at most $0.001$. Above the locus at a finite alignment rate the capture sets have parts detached from the inner part, out to $4.8L$ at $(\lambda,a_I)=(6.5,10^{-3})$ and $8.3L$ at $(6.5,10^{-4})$. There the outcome is sampled every $0.01L$ out to $5L$ ($10L$ at $a_I=10^{-4}$), on $192$ rays per quadrant placed at the midpoints of equal sectors, so that none lies on a mirror plane, where trajectories stall; half as many rays and samples change the collision rate by at most $0.9\,\%$, and maps of the outcome out to $14L$ ($16L$ at $a_I=10^{-4}$) show no collision beyond the ranges scanned.

\emph{Starting plane.} At frozen orientation, where the flux is exactly independent of $Z$, this change measures the numerical error; under strong alignment the azimuthal rotation upstream of the plane contributes in addition a change of relative order $(L/Z)^{3}$ (\S\,\ref{sec:method}), which is smaller still. At the points nearest the locus, where $Z=2r_0$ reaches $42L$, moving the starting plane to $12L$ changes $E$ by at most $0.13\,\%$. Above the extinction locus, at a finite alignment rate, the starting plane is placed at $Z=\max(6L/k,\,12L,\,2r_0)$, without the upper limit of $72L$ used in \S\,\ref{sec:finiteRe}. At $a_I=10^{-3}$ and $\lambda=6.5$, where this gives $Z\approx15L$, the collision rate increases by $3\,\%$ when the distance of the plane is doubled and by a further $1.2\,\%$ when it is doubled again, so that $Z=62L$ is used there. At $a_I=10^{-4}$, where $Z=6L/k=154L$, and at all the other points above the locus, doubling the distance of the plane changes the rate by less than $1\,\%$.

\emph{Closest approach.} Above the locus, the minimum gap $h_{min}$ along each trajectory is recorded on a $17\times9$ grid over the quadrant of impact parameters, starting at $Z=\max(6L,2r_0)$. The ten cells with the smallest corner values are then split into four, four times, which brings the finest cells to $0.003L$ ($0.004L$ across $\bp_0$ for $\kappa=5$). The head-on point is excluded from this ranking (\S\,\ref{sec:locusHI}). Finally $h_{min}$ is minimised locally by the simplex method from the best point; at $(\kappa,\lambda)=(10,6.3)$ the grid minimum is confirmed instead by the bisections of \S\,\ref{sec:locusHI}.

\section{The release law for the free rotation population}\label{app:release}

With free rotation, the efficiencies of the inertial population are computed for spheroids released from broadside on a starting plane at the finite upstream distance $Z$ (\S\,\ref{sec:method}). This appendix derives the release law (\ref{eq:release}), which relates them to the efficiency for release at infinity. The derivation uses the far field of the pair dynamics (\ref{eq:rdot})--(\ref{eq:pdoth}), in which the tilt induced by the sphere and the resulting glide of the spheroid along its axis are both small.

Below the extinction locus the sphere settles faster than a broadside spheroid, $\bpe<0$, so that the sphere approaches the spheroid from above. Since $\bp$ and $-\bp$ describe the same spheroid, the sense of $\bp$ is chosen to point from the centroid of the spheroid towards the vertical through the centre of the sphere, and the separation is written as $\br=z\bg-b_1\bp$. Here $z>0$ is the vertical distance of the centroid below the centre of the sphere and $b_1>0$ is the distance of the centroid from the vertical through the sphere, measured along the axis, with $b_1\ll z$. With this choice, $\mu=\bp\bcdot\bg>0$ when the end of the spheroid nearer to the vertical through the sphere lies below the other end. An offset across the axis lies on the mirror plane $b_1=0$, on which no tilt is induced, and does not enter the derivation. As in the rest of the paper, $z$ and $b_1$ are measured in units of $L$, whereas the starting distance $Z$ is dimensional, as in \S\,\ref{sec:method}; the starting plane is therefore at $z=Z/L$.

For $z\gg1$, the Stokeslet in (\ref{eq:pdoth}) may be expanded as $\tJ(\br+s\bp)\simeq\tJ(\br)+s\,(\bp\bcdot\nabla)\tJ(\br)$, and the potential doublet, which decays faster, may be neglected. Since $\int_{-1/2}^{1/2}s\,\dd s=0$ and $\int_{-1/2}^{1/2}s^{2}\,\dd s=\tfrac{1}{12}$, this gives
\begin{equation}
\dot{\bp}_h\simeq\frac{3\gamma}{8\lambda^{3}}\,(\tI-\bp\bp)\bcdot(\bp\bcdot\nabla)\big[\tJ(\br)\bcdot\bg\big].
\label{eq:pdoth_far}
\end{equation}
With $\tJ(\br)\bcdot\bg=\bg/|\br|+\br\,(\br\bcdot\bg)/|\br|^{3}$, and $\bp\bcdot\bg=0$, $\br\bcdot\bg=z$ and $\bp\bcdot\br=-b_1$ for a broadside spheroid,
\begin{equation}
(\bp\bcdot\nabla)\big[\tJ(\br)\bcdot\bg\big]=\frac{b_1}{|\br|^{3}}\,\bg+\frac{z}{|\br|^{3}}\,\bp+\frac{3zb_1}{|\br|^{5}}\,\br .
\end{equation}
The projection $(\tI-\bp\bp)$ removes the components along $\bp$, and the component along $\bg$ gives, to leading order in $b_1/z$,
\begin{equation}
\dot\mu=\bg\bcdot\dot{\bp}_h\simeq\frac{3\gamma}{8\lambda^{3}}\left(\frac{b_1}{z^{3}}+\frac{3b_1}{z^{3}}\right)=\frac{3\gamma\,b_1}{2\lambda^{3}z^{3}} ,
\label{eq:tiltrate}
\end{equation}
which is the forcing of (\ref{eq:relax}). The sphere therefore pushes the nearer end of the spheroid downwards, and tilts its axis towards the vertical.

By (\ref{eq:dU}), a spheroid of tilt $\mu$ has the axial relative velocity $(\Gpa-\Gpe)\mu\,\bp$. Since $b_1=-\bp\bcdot\br$, its distance from the vertical through the sphere changes at the rate
\begin{equation}
\dot b_1=-(\Gpa-\Gpe)\,\mu .
\label{eq:glide}
\end{equation}
Since $\Gpa>\Gpe$ and $\mu>0$, the spheroid glides along its axis towards the end that has been pushed down, and hence towards the vertical through the sphere.

To leading order the spheroid approaches the sphere at the speed $|\bpe|$, so that $\dd z=-|\bpe|\,\dd t$. For a spheroid released from broadside at $z=Z/L$, $\mu(Z/L)=0$, and, since $b_1$ varies slowly compared with $\mu$, (\ref{eq:tiltrate}) integrates to
\begin{equation}
\mu(z)=\frac{3\gamma\,b_1}{2\lambda^{3}|\bpe|}\int_{z}^{Z/L}\frac{\dd z'}{z'^{3}}=\frac{3\gamma\,b_1}{4\lambda^{3}|\bpe|}\left(\frac{1}{z^{2}}-\frac{L^{2}}{Z^{2}}\right).
\end{equation}
Substituting in (\ref{eq:glide}), with $\dd b_1/\dd z=\dot b_1/\dot z=(\Gpa-\Gpe)\mu/|\bpe|$,
\begin{equation}
\frac{\dd\ln b_1}{\dd z}=\frac{\Lambda}{2}\left(\frac{1}{z^{2}}-\frac{L^{2}}{Z^{2}}\right),
\label{eq:dlnb}
\end{equation}
where $\Lambda$ is the dimensionless parameter of (\ref{eq:release}). Integrating (\ref{eq:dlnb}) from $z=Z/L$ to a distance $z_n\ll Z/L$ at which the near-field interactions take over gives
\begin{equation}
\ln\frac{b_1(z_n)}{b_1(Z)}=-\frac{\Lambda}{2}\left(\frac{1}{z_n}-\frac{2L}{Z}\right)+O\!\left(\frac{z_nL^{2}}{Z^{2}}\right),
\label{eq:bZ}
\end{equation}
whereas for release at infinity, for which the term in $L^{2}/Z^{2}$ is absent from (\ref{eq:dlnb}),
\begin{equation}
\ln\frac{b_1(z_n)}{b_1(\infty)}=-\frac{\Lambda}{2z_n} .
\label{eq:binf}
\end{equation}

Whether a trajectory reaches contact is decided in the near field, by the distance $b_1(z_n)$ on arrival there, and the critical value of this distance is the same for both releases, since their tilts at $z_n$ differ only by a relative amount of order $z_n^{2}L^{2}/Z^{2}$. It follows from (\ref{eq:bZ}) and (\ref{eq:binf}) that the critical impact parameter for release at a finite distance $Z$ is smaller than that for release at infinity by the factor $\exp(-\Lambda L/Z)$. This factor is independent of both $z_n$ and the critical value itself. Every point on the boundary of the capture set is therefore drawn towards the axis by the same factor in the direction of $\bp_0$, while its extent across $\bp_0$ is unchanged. The area of the capture set is thus reduced by $\exp(-\Lambda L/Z)$, and the kernel (\ref{eq:Kflux}) and the efficiency are reduced by the same factor, which gives the release law (\ref{eq:release}).

Note that the derivation neglects terms of relative order $b_1/z$ in the tilting rate (\ref{eq:tiltrate}), the variation of $b_1$ in the integration of the tilt and the change of the approach speed with the tilt, and it requires $\Lambda L/Z$ to be moderate. At $(\kappa,\lambda)=(10,5.5)$, where $\Lambda=2.2$, the limits $E(\infty)$ inferred from release at $12L$ and at $24L$ agree to $1.2\,\%$, and that inferred from release at $6L$ is $4\,\%$ higher, of the order of the neglected terms. The free-rotation efficiencies of table~\ref{tab:eff} are extrapolated to an infinitely distant release using the law (\ref{eq:release}), $E(\infty)=E(24L)\exp(\Lambda/24)$. The values computed at $Z=24L$ thus lie at most $10\,\%$ below their extrapolated limits. Entries with $\Lambda>2.5$ are omitted, since the correction is no longer small there.

\section{Boundary-element method}\label{app:bem}

For the benchmark of \S\,\ref{sec:accuracy}, the mobility problem for the two rigid particles is solved by a single-layer boundary-integral method \citep{Pozrikidis1992}. The particles are labelled by $i,j\in\{1,2\}$, with $1$ the spheroid and $2$ the sphere, and $S_j$ denotes the surface of particle $j$. The disturbance flow is represented by a distribution of Stokeslets over both surfaces, with density $\boldsymbol{f}$, the force per unit area exerted by each surface on the fluid. The no-slip condition then requires that, for every point $\bx$ on the surface $S_i$,
\begin{equation}
\sum_{j=1}^{2}\int_{S_j}\mathsfbi{G}(\bx-\boldsymbol{y})\bcdot\boldsymbol{f}(\boldsymbol{y})\,\dd S_y=\boldsymbol{U}_i+\boldsymbol{\Omega}_i\times(\bx-\bx_i),
\qquad
\mathsfbi{G}(\boldsymbol{\varrho})=\frac{1}{8\upi\mu_f}\left(\frac{\tI}{\varrho}+\frac{\boldsymbol{\varrho}\boldsymbol{\varrho}}{\varrho^{3}}\right).
\label{eq:bie}
\end{equation}
Here $\mathsfbi{G}$ is the Oseen tensor, $\boldsymbol{\varrho}=\bx-\boldsymbol{y}$ is the vector from the source point $\boldsymbol{y}$ to the field point $\bx$, $\varrho=|\boldsymbol{\varrho}|$, and $\bx_i$, $\boldsymbol{U}_i$ and $\boldsymbol{\Omega}_i$ are the centre, translational velocity and angular velocity of particle $i$, respectively. The left-hand side is the velocity induced at $\bx$ by both particles, and the right-hand side is the rigid-body velocity of the surface on which $\bx$ lies. The unknowns are $\boldsymbol{f}$ on both surfaces and the twelve components of $\boldsymbol{U}_i$ and $\boldsymbol{\Omega}_i$. Since the inertia of the particles is neglected, the system is closed by requiring each particle to be force- and torque-free: the force exerted on the fluid balances the buoyant weight $\boldsymbol{F}_i$ of the particle, and the torque vanishes,
\begin{equation}
\int_{S_i}\boldsymbol{f}\,\dd S=\boldsymbol{F}_i,\qquad
\int_{S_i}(\bx-\bx_i)\times\boldsymbol{f}\,\dd S=\boldsymbol{0}.
\label{eq:bieforce}
\end{equation}
For a closed surface, the single-layer operator has the outward unit normal $\hat{\boldsymbol{n}}$ as a null vector: a uniform normal traction is balanced by a uniform pressure inside the surface and drives no flow, so (\ref{eq:bie}) determines $\boldsymbol{f}$ only up to a multiple of $\hat{\boldsymbol{n}}$ on each surface. This multiple exerts no net force or torque either, since $\hat{\boldsymbol{n}}$ and $(\bx-\bx_i)\times\hat{\boldsymbol{n}}$ integrate to zero over a closed surface. The indeterminacy is removed by adding $\hat{\boldsymbol{n}}(\bx)\int_{S_i}\hat{\boldsymbol{n}}\bcdot\boldsymbol{f}\,\dd S$ to the left-hand side of (\ref{eq:bie}), which selects the solution with zero mean normal traction and leaves the velocities unchanged.

Both surfaces are discretised into flat triangular elements. The traction is constant on flat triangles and the equations are collocated at their centroids. The self-integral is done exactly in the radial direction in polar coordinates about the centroid, and nearby integrals by subdividing the source triangle. The sphere is meshed from a subdivided icosahedron ($5120$ triangles) and the spheroid by rings of equal arclength ($70$ rings of $24$ points), and both meshes are refined towards the point of closest approach until the triangles are smaller than $0.2$ times their distance from it, and smaller than $h/6$. A single sphere has a drag error of $0.08\,\%$ and a single spheroid of $0.3\,\%$ at $\kappa=10$. As a test, two equal spheres approaching each other under equal and opposite forces were considered, for which the exact solution of \citet{Brenner1961} is available. The relative mobility is first divided by the value that the same meshes give for the isolated particles, which removes most of the discretisation error. The normalised mobility then agrees with the exact solution to within $0.1\,\%$ at a gap of one radius and to within $0.5\,\%$ at a gap of $0.1$ radii. We apply the same normalisation to the sphere--spheroid rates. At the most demanding gap, $h=a/4$, the meshes have $12\,538$ triangles for $\lambda=3$; coarser meshes, with $1280$ triangles on the sphere, $50$ rings of $16$ points on the spheroid and refinement to $0.3$ times the distance ($5288$ triangles), change the closing rate by $1.0$ to $2.3\,\%$, whereas smaller triangles at contact ($h/10$) or a more accurate quadrature change it by less than $0.1\,\%$.

\section{Lubrication between a sphere and a spheroid}\label{app:lub}

Near the point of closest approach the gap has the local form
\begin{equation*}
h(x,y)=h+\frac{x^{2}}{2R_x}+\frac{y^{2}}{2R_y},
\end{equation*}
where $x$ and $y$ are measured along the principal directions of curvature of the spheroid, and the relative radii of curvature, defined by $1/R_x=k_m+1/r_s$ and $1/R_y=k_\varphi+1/r_s$, combine the curvatures of the two surfaces. At the point $\bx=c\cos\vartheta\,\bp+a\sin\vartheta\,\hat{\boldsymbol{e}}_\rho$ of the spheroid, with polar angle $\vartheta$, the meridional and azimuthal curvatures are
\begin{equation}
k_m=\frac{ca}{q^{3}},\qquad k_\varphi=\frac{c}{aq},\qquad q=\left(c^{2}\sin^{2}\vartheta+a^{2}\cos^{2}\vartheta\right)^{1/2}.
\end{equation}
The pressure in the film satisfies the Reynolds equation $\nabla\bcdot(h^{3}\nabla p)=12\mu_f\dot h$, which has the exact solution
\begin{equation*}
p=\frac{6\mu_f(-\dot h)\,R_xR_y}{(R_x+R_y)\,h(x,y)^{2}},
\end{equation*}
vanishing far from the point of closest approach. Integrating this pressure over the plane gives the force $R_{lub}(-\dot h)$, with $R_{lub}$ given by (\ref{eq:Rlub}). For $R_x=R_y=R$ it reduces to $6\upi\mu_f R^{2}/h$, the familiar result for a sphere of radius $R$ approaching a plane.

In (\ref{eq:hstar}), $m$ is the relative mobility of the pair along the normal $\bnu$ for equal and opposite unit forces applied at the contact point. It consists of an isolated part and a mutual part. The isolated part is
\begin{equation*}
m_{iso}=\frac{1}{6\upi\mu_f r_s}+\bnu\bcdot\mathsfbi{M}_c\bcdot\bnu+m_{rot},
\end{equation*}
in which the three terms are the translational mobilities of the sphere and of the spheroid and the contribution of the rotation of the spheroid. Here $\mathsfbi{M}_c=[X^{-1}\bp\bp+Y^{-1}(\tI-\bp\bp)]/(6\upi\mu_f c)$ is the translational mobility tensor of the spheroid, and
\begin{equation*}
m_{rot}=\frac{3\ln(2\kappa)}{\upi\mu_f L^{3}}\,|\boldsymbol{d}_\perp|^{2},
\end{equation*}
where $\boldsymbol{d}=(\bx^{*}-\br)\times\bnu$ is the torque arm of a unit normal force at the contact point and $\boldsymbol{d}_\perp$ its component normal to $\bp$. Under strong alignment only rotation about the vertical is admissible, so that only the corresponding component of $\boldsymbol{d}_\perp$ contributes, and at frozen orientation $m_{rot}=0$. The mutual part is the normal velocity induced at the contact point of the spheroid by a unit force on the sphere, computed from (\ref{eq:usphere}), (\ref{eq:sbtresponse_u}) and (\ref{eq:sbtresponse_p}). By reciprocity, it equals the velocity induced on the sphere by a unit force on the spheroid. The relative mobility is the isolated part minus twice the mutual part. Near contact at $\kappa=10$ the mutual part is $0.23$ to $0.31$ of the isolated mobility, so that $m$ is $0.39$ to $0.55$ of $m_{iso}$.

With this choice of $m$, the series combination (\ref{eq:hstar}) is an addition of resistances: writing $F=\dot h_1/m$ for the internal force of the first-reflection model, (\ref{eq:hstar}) becomes $\dot h=F/(1/m+R_{lub})$. Near contact $R_{lub}$ dominates and $\dot h\to F/R_{lub}$, which for two spheres is the exact lubrication limit. In the trajectory calculations the film is represented by the force
\begin{equation*}
F_{film}=-\left(1-\frac{h}{h+h^{*}}\right)\frac{\dot h_1}{m_{iso}},
\end{equation*}
applied at the contact point. The resulting changes in the translation of the sphere, and in the translation and admissible rotation of the spheroid, follow from their isolated mobilities, and together they reduce the closing rate from $\dot h_1$ to the value (\ref{eq:hstar}). As a check, the whole correction was instead applied to the relative translation alone. At $\kappa=10$ and $h_c=10^{-3}L$ this changes the efficiencies by less than $0.2\,\%$ for frozen orientation, $3\,\%$ with free rotation and $10\,\%$ under strong alignment.

\end{appen}\clearpage

\bibliographystyle{jfm}
\bibliography{sphere_spheroid}

@article{Smoluchowski1917,
  author  = {Smoluchowski, M. von},
  title   = {Versuch einer mathematischen {T}heorie der {K}oagulationskinetik kolloider {L}{\"o}sungen},
  journal = {Z. Phys. Chem.},
  volume  = {92},
  pages   = {129--168},
  year    = {1917}
}

@article{Davis1984,
  author  = {Davis, R. H.},
  title   = {The rate of coagulation of a dilute polydisperse system of sedimenting spheres},
  journal = {J. Fluid Mech.},
  volume  = {145},
  pages   = {179--199},
  year    = {1984}
}

@article{Batchelor1982,
  author  = {Batchelor, G. K.},
  title   = {Sedimentation in a dilute polydisperse system of interacting spheres. {P}art 1. {G}eneral theory},
  journal = {J. Fluid Mech.},
  volume  = {119},
  pages   = {379--408},
  year    = {1982}
}

@article{BatchelorWen1982,
  author  = {Batchelor, G. K. and Wen, C.-S.},
  title   = {Sedimentation in a dilute polydisperse system of interacting spheres. {P}art 2. {N}umerical results},
  journal = {J. Fluid Mech.},
  volume  = {124},
  pages   = {495--528},
  year    = {1982}
}

@article{Dhanasekaran2021a,
  author  = {Dhanasekaran, J. and Roy, A. and Koch, D. L.},
  title   = {Collision rate of bidisperse spheres settling in a compressional non-continuum gas flow},
  journal = {J. Fluid Mech.},
  volume  = {910},
  pages   = {A10},
  year    = {2021}
}

@article{Dhanasekaran2021b,
  author  = {Dhanasekaran, J. and Roy, A. and Koch, D. L.},
  title   = {Collision rate of bidisperse, hydrodynamically interacting spheres settling in a turbulent flow},
  journal = {J. Fluid Mech.},
  volume  = {912},
  pages   = {A5},
  year    = {2021}
}

@article{Patra2022,
  author  = {Patra, P. and Koch, D. L. and Roy, A.},
  title   = {Collision efficiency of non-{B}rownian spheres in a simple shear flow -- the role of non-continuum hydrodynamic interactions},
  journal = {J. Fluid Mech.},
  volume  = {950},
  pages   = {A18},
  year    = {2022}
}

@article{Patra2023,
  author  = {Patra, P. and Koch, D. L. and Roy, A.},
  title   = {Collision efficiency of like-charged spheres settling in a quiescent environment},
  journal = {J. Fluid Mech.},
  volume  = {968},
  pages   = {A22},
  year    = {2023}
}

@misc{Warrier2026,
  author       = {Warrier, S. and Roy, A. and Patra, P.},
  title        = {Electrostatic enhancement of particle collision rates in atmospheric flows},
  year         = {2025},
  howpublished = {arXiv:2512.24512}
}

@book{KimKarrila1991,
  author    = {Kim, S. and Karrila, S. J.},
  title     = {Microhydrodynamics: Principles and Selected Applications},
  publisher = {Butterworth--Heinemann},
  year      = {1991}
}

@article{ChwangWu1975,
  author  = {Chwang, A. T. and Wu, T. Y.},
  title   = {Hydromechanics of low-{R}eynolds-number flow. {P}art 2. {S}ingularity method for {S}tokes flows},
  journal = {J. Fluid Mech.},
  volume  = {67},
  pages   = {787--815},
  year    = {1975}
}

@article{Batchelor1970,
  author  = {Batchelor, G. K.},
  title   = {Slender-body theory for particles of arbitrary cross-section in {S}tokes flow},
  journal = {J. Fluid Mech.},
  volume  = {44},
  pages   = {419--440},
  year    = {1970}
}

@article{Cox1970,
  author  = {Cox, R. G.},
  title   = {The motion of long slender bodies in a viscous fluid. {P}art 1. {G}eneral theory},
  journal = {J. Fluid Mech.},
  volume  = {44},
  pages   = {791--810},
  year    = {1970}
}

@article{Cox1965,
  author  = {Cox, R. G.},
  title   = {The steady motion of a particle of arbitrary shape at small {R}eynolds numbers},
  journal = {J. Fluid Mech.},
  volume  = {23},
  pages   = {625--643},
  year    = {1965}
}

@article{ProudmanPearson1957,
  author  = {Proudman, I. and Pearson, J. R. A.},
  title   = {Expansions at small {R}eynolds numbers for the flow past a sphere and a circular cylinder},
  journal = {J. Fluid Mech.},
  volume  = {2},
  pages   = {237--262},
  year    = {1957}
}

@article{KhayatCox1989,
  author  = {Khayat, R. E. and Cox, R. G.},
  title   = {Inertia effects on the motion of long slender bodies},
  journal = {J. Fluid Mech.},
  volume  = {209},
  pages   = {435--462},
  year    = {1989}
}

@article{Dabade2015,
  author  = {Dabade, V. and Marath, N. K. and Subramanian, G.},
  title   = {Effects of inertia and viscoelasticity on sedimenting anisotropic particles},
  journal = {J. Fluid Mech.},
  volume  = {778},
  pages   = {133--188},
  year    = {2015}
}

@article{Jeffery1922,
  author  = {Jeffery, G. B.},
  title   = {The motion of ellipsoidal particles immersed in a viscous fluid},
  journal = {Proc. R. Soc. Lond. A},
  volume  = {102},
  pages   = {161--179},
  year    = {1922}
}

@article{JayaweeraMason1965,
  author  = {Jayaweera, K. O. L. F. and Mason, B. J.},
  title   = {The behaviour of freely falling cylinders and cones in a viscous fluid},
  journal = {J. Fluid Mech.},
  volume  = {22},
  pages   = {709--720},
  year    = {1965}
}

@article{Klett1995,
  author  = {Klett, J. D.},
  title   = {Orientation model for particles in turbulence},
  journal = {J. Atmos. Sci.},
  volume  = {52},
  pages   = {2276--2285},
  year    = {1995}
}

@book{PruppacherKlett1997,
  author    = {Pruppacher, H. R. and Klett, J. D.},
  title     = {Microphysics of Clouds and Precipitation},
  series    = {Atmospheric and Oceanographic Sciences Library},
  volume    = {18},
  publisher = {Kluwer Academic},
  year      = {1997}
}

@book{LambVerlinde2011,
  author    = {Lamb, D. and Verlinde, J.},
  title     = {Physics and Chemistry of Clouds},
  publisher = {Cambridge University Press},
  year      = {2011}
}

@article{GrabowskiWang2013,
  author  = {Grabowski, W. W. and Wang, L.-P.},
  title   = {Growth of cloud droplets in a turbulent environment},
  journal = {Annu. Rev. Fluid Mech.},
  volume  = {45},
  pages   = {293--324},
  year    = {2013}
}

@article{Sundararajakumar1996,
  author  = {Sundararajakumar, R. R. and Koch, D. L.},
  title   = {Non-continuum lubrication flows between particles colliding in a gas},
  journal = {J. Fluid Mech.},
  volume  = {313},
  pages   = {283--308},
  year    = {1996}
}

@article{Roy2019,
  author  = {Roy, A. and Hamati, R. J. and Tierney, L. and Koch, D. L. and Voth, G. A.},
  title   = {Inertial torques and a symmetry breaking orientational transition in the sedimentation of slender fibres},
  journal = {J. Fluid Mech.},
  volume  = {875},
  pages   = {576--596},
  year    = {2019}
}

@article{Gustavsson2019,
  author  = {Gustavsson, K. and Sheikh, M. Z. and Lopez, D. and Naso, A. and Pumir, A. and Mehlig, B.},
  title   = {Effect of fluid inertia on the orientation of a small prolate spheroid settling in turbulence},
  journal = {New J. Phys.},
  volume  = {21},
  pages   = {083008},
  year    = {2019}
}

@article{Sheikh2020,
  author  = {Sheikh, M. Z. and Gustavsson, K. and Lopez, D. and L{\'e}v{\^e}que, E. and Mehlig, B. and Pumir, A. and Naso, A.},
  title   = {Importance of fluid inertia for the orientation of spheroids settling in turbulent flow},
  journal = {J. Fluid Mech.},
  volume  = {886},
  pages   = {A9},
  year    = {2020}
}

@article{Naso2018,
  author  = {Naso, A. and Jucha, J. and L{\'e}v{\^e}que, E. and Pumir, A.},
  title   = {Collision rate of ice crystals with water droplets in turbulent flows},
  journal = {J. Fluid Mech.},
  volume  = {845},
  pages   = {615--641},
  year    = {2018}
}

@article{Jucha2018,
  author  = {Jucha, J. and Naso, A. and L{\'e}v{\^e}que, E. and Pumir, A.},
  title   = {Settling and collision between small ice crystals in turbulent flows},
  journal = {Phys. Rev. Fluids},
  volume  = {3},
  pages   = {014604},
  year    = {2018}
}

@article{WangJi2000,
  author  = {Wang, P. K. and Ji, W.},
  title   = {Collision efficiencies of ice crystals at low--intermediate {R}eynolds numbers colliding with supercooled cloud droplets: a numerical study},
  journal = {J. Atmos. Sci.},
  volume  = {57},
  pages   = {1001--1009},
  year    = {2000}
}

@article{Schlamp1975,
  author  = {Schlamp, R. J. and Pruppacher, H. R. and Hamielec, A. E.},
  title   = {A numerical investigation of the efficiency with which simple columnar ice crystals collide with supercooled water drops},
  journal = {J. Atmos. Sci.},
  volume  = {32},
  pages   = {2330--2337},
  year    = {1975}
}

@book{Schneider2014,
  author    = {Schneider, R.},
  title     = {Convex Bodies: The {B}runn--{M}inkowski Theory},
  edition   = {2nd},
  publisher = {Cambridge University Press},
  year      = {2014}
}

@article{SaffmanTurner1956,
  author  = {Saffman, P. G. and Turner, J. S.},
  title   = {On the collision of drops in turbulent clouds},
  journal = {J. Fluid Mech.},
  volume  = {1},
  pages   = {16--30},
  year    = {1956}
}

@article{Siewert2014,
  author  = {Siewert, C. and Kunnen, R. P. J. and Schr{\"o}der, W.},
  title   = {Collision rates of small ellipsoids settling in turbulence},
  journal = {J. Fluid Mech.},
  volume  = {758},
  pages   = {686--701},
  year    = {2014}
}

@article{Jiang2024,
  author  = {Jiang, X. and Xu, C. and Zhao, L.},
  title   = {Settling and collision of spheroidal particles with an offset mass centre in a quiescent fluid},
  journal = {J. Fluid Mech.},
  volume  = {984},
  pages   = {A40},
  year    = {2024}
}

@article{GavzeKhain2022,
  author  = {Gavze, E. and Khain, A.},
  title   = {Gravitational collision of small nonspherical particles: swept volumes of prolate and oblate spheroids in calm air},
  journal = {J. Atmos. Sci.},
  volume  = {79},
  pages   = {1493--1514},
  year    = {2022}
}

@article{SlomkaStocker2020a,
  author  = {S{\l}omka, J. and Stocker, R.},
  title   = {On the collision of rods in a quiescent fluid},
  journal = {Proc. Natl Acad. Sci. USA},
  volume  = {117},
  pages   = {3372--3374},
  year    = {2020}
}

@article{SlomkaStocker2020b,
  author  = {S{\l}omka, J. and Stocker, R.},
  title   = {Bursts characterize coagulation of rods in a quiescent fluid},
  journal = {Phys. Rev. Lett.},
  volume  = {124},
  pages   = {258001},
  year    = {2020}
}

@article{ClaeysBrady1993,
  author  = {Claeys, I. L. and Brady, J. F.},
  title   = {Suspensions of prolate spheroids in {S}tokes flow. {P}art 1. {D}ynamics of a finite number of particles in an unbounded fluid},
  journal = {J. Fluid Mech.},
  volume  = {251},
  pages   = {411--442},
  year    = {1993}
}

@article{Kim1985,
  author  = {Kim, S.},
  title   = {Sedimentation of two arbitrarily oriented spheroids in a viscous fluid},
  journal = {Int. J. Multiphase Flow},
  volume  = {11},
  pages   = {699--712},
  year    = {1985}
}

@article{Harlen1999,
  author  = {Harlen, O. G. and Sundararajakumar, R. R. and Koch, D. L.},
  title   = {Numerical simulations of a sphere settling through a suspension of neutrally buoyant fibres},
  journal = {J. Fluid Mech.},
  volume  = {388},
  pages   = {355--388},
  year    = {1999}
}

@article{KochShaqfeh1989,
  author  = {Koch, D. L. and Shaqfeh, E. S. G.},
  title   = {The instability of a dispersion of sedimenting spheroids},
  journal = {J. Fluid Mech.},
  volume  = {209},
  pages   = {521--542},
  year    = {1989}
}

@article{ShaqfehKoch1988,
  author  = {Shaqfeh, E. S. G. and Koch, D. L.},
  title   = {The effect of hydrodynamic interactions on the orientation of axisymmetric particles flowing through a fixed bed of spheres or fibers},
  journal = {Phys. Fluids},
  volume  = {31},
  pages   = {728--743},
  year    = {1988}
}

@article{ShaqfehKoch1990,
  author  = {Shaqfeh, E. S. G. and Koch, D. L.},
  title   = {Orientational dispersion of fibers in extensional flows},
  journal = {Phys. Fluids A},
  volume  = {2},
  pages   = {1077--1093},
  year    = {1990}
}

@article{Shin2009,
  author  = {Shin, M. and Koch, D. L. and Subramanian, G.},
  title   = {Structure and dynamics of dilute suspensions of finite-{R}eynolds-number settling fibers},
  journal = {Phys. Fluids},
  volume  = {21},
  pages   = {123304},
  year    = {2009}
}

@article{Sheikh2022,
  author  = {Sheikh, M. Z. and Gustavsson, K. and L{\'e}v{\^e}que, E. and Mehlig, B. and Pumir, A. and Naso, A.},
  title   = {Colliding ice crystals in turbulent clouds},
  journal = {J. Atmos. Sci.},
  volume  = {79},
  pages   = {2205--2218},
  year    = {2022}
}

@article{Brenner1961,
  author  = {Brenner, H.},
  title   = {The slow motion of a sphere through a viscous fluid towards a plane surface},
  journal = {Chem. Engng Sci.},
  volume  = {16},
  pages   = {242--251},
  year    = {1961}
}

@book{Pozrikidis1992,
  author    = {Pozrikidis, C.},
  title     = {Boundary Integral and Singularity Methods for Linearized Viscous Flow},
  publisher = {Cambridge University Press},
  year      = {1992}
}

@misc{Roy2026orientation,
  title={Orientation and slip statistics of spheroids settling in turbulence},
  author={Roy, Anubhab},
  note={{arXiv}, 2611.07990},
  doi={10.48550/arXiv.2611.07990},
  year={2026}
}

@article{Sheikh2024,
  author  = {Sheikh, M. Z. and Gustavsson, K. and L{\'e}v{\^e}que, E. and Mehlig, B. and Pumir, A. and Naso, A.},
  title   = {Effect of turbulence on the collision rate between settling ice crystals and droplets},
  journal = {J. Atmos. Sci.},
  volume  = {81},
  pages   = {887},
  year    = {2024}
}

@article{Jiang2021,
  author  = {Jiang, F. and Zhao, L. and Andersson, H. I. and Gustavsson, K. and Pumir, A. and Mehlig, B.},
  title   = {Inertial torque on a small spheroid in a stationary uniform flow},
  journal = {Phys. Rev. Fluids},
  volume  = {6},
  pages   = {024302},
  year    = {2021}
}

@misc{Patra2026,
  author       = {Patra, P. and Koch, D. L. and Roy, A.},
  title        = {Collision efficiency of rapidly settling particle pairs in a turbulent flow},
  year         = {2026},
  howpublished = {arXiv:2608.16528}
}

@article{Bretherton1962,
  author  = {Bretherton, F. P.},
  title   = {The motion of rigid particles in a shear flow at low {R}eynolds number},
  journal = {J. Fluid Mech.},
  volume  = {14},
  pages   = {284--304},
  year    = {1962}
}

\end{document}